\documentclass[a4paper,dvips,12pt]{book}

\usepackage{epsfig,latexsym,enumerate,lscape,amssymb,graphicx}

\newcommand{\mnras}{Monthly Not. RAS }

\newcommand{\prd}{Phys. Rev. D }                

\newcommand{\prl}{Phys. Rev. Lett. }

\newcommand{\nat}{Nature }

\newcommand{\cqg}{Class. Quant. Gravity }

\newcommand{\pramjp}{Pramana J. Phys. }

\newcommand{\aaa}{Astron. and Astrophys. }

\newcommand{\pp}{Preprint gr-qc/}

\begin{document}

\pagestyle{empty}

$$\left. \right. $$

\vspace{3.20cm}

\begin{center}Forwarded\end{center} 

\vspace{1.8cm}

\indent\indent\indent\indent\indent\indent\indent\indent \hspace{0.1cm}(Prof. D.C. Srivastava)\\
\indent\indent\indent\indent\indent\indent\indent\indent \hspace{0.1cm}Supervisor\\
\indent\indent\indent\indent\indent\indent\indent\indent \hspace{0.1cm}Department of Physics\\
\indent\indent\indent\indent\indent\indent\indent\indent \hspace{0.1cm}DDU Gorakhpur University\\
\indent\indent\indent\indent\indent\indent\indent\indent \hspace{0.1cm}Gorakhpur

\vspace{2.5cm}

\begin{center}Forwarded\end{center} 

\vspace{1cm}

\indent\indent\indent\indent\indent\indent\indent\indent \hspace{0.1cm}(Prof. C.P.D. Dwivedi) \\
\indent\indent\indent\indent\indent\indent\indent\indent \hspace{0.1cm}Head, Department of Physics\\
\indent\indent\indent\indent\indent\indent\indent\indent \hspace{0.1cm}DDU Gorakhpur University\\
\indent\indent\indent\indent\indent\indent\indent\indent \hspace{0.1cm}Gorakhpur

\baselineskip=0.83cm

\begin{center}{\Huge\bf STUDIES IN \\ GRAVITATIONAL WAVE \\
DATA ANALYSIS  \\}

\vspace{1.0cm}

\epsfig{file=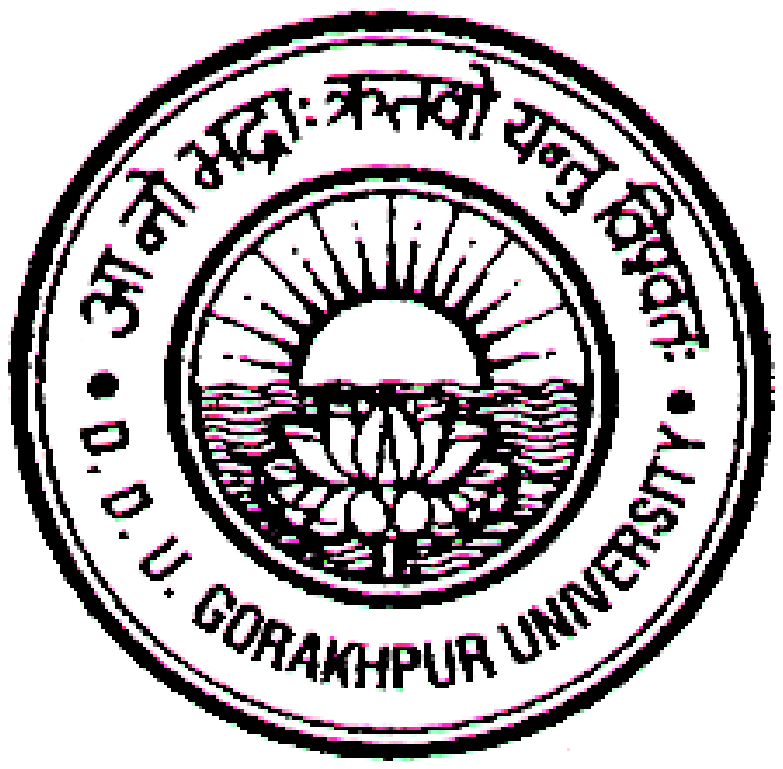,height=4cm}

\vspace{1.0cm}

\Large {\it A thesis submitted to the \/}\\
{\bf DDU GORAKHPUR UNIVERSITY\/}\\
for the degree of\\
{\bf Doctor of Philosophy}\\
IN PHYSICS

\vspace{1.90cm}

by \\ 

\vspace{0.5cm}

{\bf SANJAY KUMAR SAHAY\/} \\   

\vspace{1.0cm}

Department of Physics \\
DDU Gorakhpur University\\
Gorakhpur - 273009\\

\vspace{1.50cm}

\large December 2001
\end{center}
\clearpage
$$\left. \right. $$

\vspace{235pt}
\begin{center}{\Huge $\cal TO \hspace{5mm} MY \hspace{5mm} FATHER$ \\    

\vspace{5mm}

$\cal LATE \hspace{5mm} SHRI \hspace{5mm} KRISHNA \hspace{5mm} SAHAY$ }
\end{center}

\pagestyle{plain}

\pagenumbering{roman}

\setcounter{page}{1}

\tableofcontents

\listoftables

\listoffigures

\begin{center}{\LARGE\bf ACKNOWLEDGEMENT}
\end{center}
\addcontentsline{toc}{chapter}{\numberline{}Acknowledgement\/}
\vspace{1cm}

I would like to thank Prof. D.C. Srivastava, my thesis Supervisor
whose guidance and constant encouragement throughout the course of
the work has been a pleasant, rich and a rewarding experience. I am thankful
to my colleague Dr. S.S. Prasad, UNPG College, Padrauna for various
useful discussions and comments. A part of the
work has been carried out during my various visits at IUCAA,
Pune. I have been benefited by the discussions,
comments and suggestions of IUCCA Scientists notably of Prof. S.V. Dhurandhar.
I am indebted to all of them for their kindness and help. I thankfully
acknowledge the local hospitality, financial support, computer and library
facilities made available to me during the course of stay. I would like
to thank Prof. A.K. Kembhavi,
Dean Visitor Academic Programmes and  Mrs. S. Shankar,
Administrative Officer, Visitor Services, and staff members of IUCAA
for their kind cooperation and assistance during the course of
stay in IUCAA. In this connection, my heartfelt thanks are also to Prof. J.V.
Narlikar, Director, IUCAA for creating such
a stimulating research atmosphere in IUCAA.

\par I am thankful to friends and the staff members of Physics
Department for extending me moral support during the work. I would also like to
thank Prof. R.M. Misra, Vice-Chancellor, DDU Gorakhpur
University for providing requisite facilities for the research work. I am also 
thankful to the staff of the Computer Centre of the University for providing
me necessary facilities.

\par I would like to thank Drs. B.S. Sathyaprakash and S. Bose, Cardiff,
U.K. for various useful suggestions. I also like to thank Prof. A.R.
Tharafdar, Convener, Indian Institute of Technology, Kharagpur for local
hospitality and computational facility during my visit there.

\par My heartfelt special thanks are due to my mother and sisters for their
love and continuous support.

\par I gratefully acknowledge the grant of a fellowship from Department of
Science and Technology, New Delhi vide research grant number SP/S2/0-15/93.
\vspace{0.5cm}
\begin{flushleft}
\begin{tabular}{ll}
Place: Gorakhpur  \hspace{5.70cm}  & (Sanjay Kumar Sahay)\\
Date:     & 
\end{tabular}
\end{flushleft}

\newpage
\begin{center}{\LARGE\bf DECLARATION}
\end{center}
\addcontentsline{toc}{chapter}{\numberline{}Declaration\/}

\vspace{2.0cm}

\par  Certified that the work incorporated in the thesis
{\bf\em `Studies in Gravitational Wave Data Analysis'\/}
submitted by Mr. {\bf Sanjay Kumar Sahay\/} was carried out by the
candidate under my supervision. Such materials as have been obtained
from other sources has been duly acknowledged in the thesis.
\vspace{2cm}
\begin{flushleft}
\begin{tabular}{ll}
Place: Gorakhpur  \hspace{5.0cm}  & (Prof. D.C. Srivastava)\\
Date:     & Supervisor\\
& Department of Physics\\
& DDU Gorakhpur University\\
& Gorakhpur
\end{tabular}
\end{flushleft}

\newpage

\begin{center}{\LARGE\bf LIST OF PUBLICATIONS}
\end{center}
\addcontentsline{toc}{chapter}{\numberline{}List of Publications\/}
\vspace{1cm}

\begin{enumerate}

\item Srivastava, D.C and Sahay, S.K., {\it Data analysis of continuous 
gravitational wave: Fourier transform\/}, \pp 0009094 (2000).

\item Srivastava, D.C and Sahay, S.K., {\it Analysis of noise free response
of interferometric antenna to Gravitational Radiation I \& II}, Proceedings
of the Fourth International 
Conference on Gravitation and Cosmology, \pramjp {\bf 55}, 589 (2000).
 
\item Srivastava, D.C and Sahay, S.K., {\it Data analysis of continuous
gravitational wave: Fourier transform-I\/}, communicated
to \mnras (2001).

\item Srivastava, D.C and Sahay, S.K., {\it Data analysis of continuous
gravitational wave: Fourier transform-II\/}, communicated to \mnras (2001).

\item Srivastava, D.C and Sahay, S.K., {\it Data analysis of continuous
gravitational wave: All sky search and study of templates\/}, communicated
to \mnras (2001).
\end{enumerate}

\chapter*{Preface}
\addcontentsline{toc}{chapter}{\numberline{}Preface}
\indent This thesis is devoted to the investigations of gravitational wave data
analysis from a continuous source e.g. a pulsar, a binary star system. The
research in the field of gravitational radiation started after its
formulation by 
Einstein in (1916) as propagating gravitational disturbance described by
the linearised field limit of his General Theory of Relativity (GR) but has 
received serious impetus towards its detection after the announcement of its
detection by Weber in 1969 using aluminium bar detectors. This field has
now emerged
and established itself with General Relativity (GR), Astrophysics and Numerical
analysis as its equally important facets. Ofcourse, the technological advancements
being employed in the construction of detectors with day by day improving sensitivity
have played the crucial role. To date, although the results of Weber could not
be confirmed and we do not have as yet any direct detection of gravitational wave (GW)
yet it is not a matter of concern. Because on one hand the sensitivity required
for the announcement of definite  
detection of GW bathing earth is yet to be achieved by the detectors whereas on
the other
hand we have an indirect evidence of the existence of GW observed in 1974 as  
the slowing down of the binary pulsar PSR 1913 + 16 arising because of back
reaction of GW emission. GW Scientists all over the globe
are putting more persuasive arguments regarding the feasibility of GW detection
in `near future' and the advantages to be
achieved once the ``Gravitational Wave Astronomy'' as they call it, is established.
A huge amount of money is involved in these projects to the extent that many
of the detectors are built in collaboration e.g. American Laser Interferometric
Gravitational Wave Observatory (LIGO), Italian French Gravitational
Wave Observatory (VIRGO), British-German Observatory (GE0600). As a consequence
the literature is full of update reviews on gravitational wave astronomy
notably by Thorne (1987), Blair (1991), Schutz (1999), Grishchuk et.al
(2000), where the related issues viz., the fall outs, pre-requisites and
prospects are discussed and scrutinised with rigour and minute detail.

\par The thesis starts with two introductory Chapters
dealing with gravitational wave and data analysis concepts, respectively. The
matter covered in these Chapters are restricted to the extent they are supposed
to provide continuity and coherence to the investigations presented in Chapters
3-5. Further, the material presented in these chapters depend heavily on the
reviews by Grishchuk (2000), Schutz (1999), Sathyaprakash (1999). Our job have
been, frankly speaking, to cut and paste. Let us hope this has not defaced the
truth. For the portion describing GR we have adopted notations and conventions
of Schutz (1989). The source codes of the numerical computations are relegated
to the appendices A-C.

\par The term ``Pulsar'' used in the thesis refers to spinning neutron
stars. A Pulsar will emit GW signal over extended
period of time only
when it has long-lived asymmetry. Several mechanisms have been given
for such an asymmetry to arise. Some Pulsars emit GW almost monochromatically
and are remarkably good clocks as its periods have been measured upto
13 significant digits. However, the GW signals from
Pulsars are very weak $(\le 10^{-25}$) and will be buried in the broadband
noise of the detector. In order to detect
the signal from the dominant noise one has to analyse the long time
observation data of months/years. The output of detector is highly involved
function of many initial parameters. It is usually not possible to obtain the
Fourier Transform (FT) analytically. Hence, FT has to be obtained via numerical
methods. But it appears to be computationally
demanding even for the standard computers expected to be available in 
near future. Therefore, one will have to work with efficient data
analysis techniques, efficient not only in picking weak signal from
the noisy data but also in terms of computing-cost requirements. Chapter 2
is on the problem and technique for the data analysis. The
noises and sensitivity of the detector has been briefly described. The 
{\it Matched Filtering\/}, a technique of the optimal method for detecting unknown
signal and which describes drop in signal-to-noise ratio in terms of 
{\it Fitting Factor\/} $(FF)$, is discussed. The computational cost and detection criteria are also
discussed.

\par The output of detector will be Doppler modulated (both in
frequency and in amplitude) due to the motions of Earth and star. To work in
Fourier space, one usually
employs the numerical method called Fast Fourier Transform (FFT).
Fortunately, we have been able to develop analytical FT. We have employed
the results to obtain efficient data analysis algorithms for continuous
gravitational wave (CGW).

\par In Chapter 3 we have studied the noise free
complete response of Laser Interferometer detector for CGW for its arbitrary location 
and that of the source. We have taken into account the rotational motion
of Earth about its spin axis as well as its orbital motion around Sun. We 
have developed analytical FT of Frequency modulated output for one
day observation time. In Chapter 4 this analysis has been
generalised for (i) one year observation
time and for (ii) any arbitrary duration of observation data set. The
emission of GW from pulsars as such may contain two or more frequencies.
Hence, we finally generalise the transform for N-component signal. We
also give the method to account for spin down of CGW. The Frequency
modulation (FM) smears out
a monochromatic signal into a small bandwidth
around the signal frequency. The FT contains
double series of Bessel functions and the computational cost depends 
primarily because of Bessel functions which in
turn depends on source location and frequency.

\par The Amplitude modulation (AM) arises due to the anisotropic
response of the detector i.e. the detector possesses a quadrupole antenna
pattern. The effect of AM is to split the signal frequency $f_o$ into five
lines corresponding to frequencies $f_o \pm 2f_{rot}$,
$f_o \pm f_{rot}$ and $f_o$
where $f_{rot}$ represents the rotational frequency of Earth. Hence, for optimal detection
of Amplitude modulated signal one will need four additional linear filters. For a 2-component pulsar
signal the number of filters will increase further by a factor of $2$.
The cases of Doppler modulation investigated by us reveal that as a result of
the AM the most of the power lies in the frequency band $f + 2f_{rot}$ 
and the least in the frequency $f - 2f_{rot}.$ 

\par An important advantage of our analytical results over FFT arises when one
looks for the resolution of the FT. The resolution offered by the FFT 
is inversely proportional to the observation time, $T_o$ whereas, the
analytical FT is independent of $T_o$.
This means that for short observation time the resolution offered by FFT is small
whereas for long observation time there would be high computational
demands. We have developed, using our analytical FT, a semi-analytical formula
to calculate the Doppler shift for one year observation data for any
source location and frequency. 
A monochromatic signal of one kHz frequency from a source 
located at $\theta = \pi /2$ is found to spread into approximately 3.13
millions Fourier bins. We observe that the number of Fourier 
bins increases with the source location $\theta$ and frequency $f_o.$

\par The strategy for the detection of GW signal is to make use of FT to dig
out the signal from the noisy output of detectors. To achieve this, one
constructs the templates which are best educated guesses of the expected
signal waveform. In Chapter 5 we have applied our results on FT obtained in chapter 4.  
Applying the technique of {\it Matched
Filtering\/} we have computed the
number of templates required for all sky search problem 
for a pre-assigned $(FF)$. It is estimated that one
would require about $1.44 \times 10^{10}$, $3.5 \times 10^{10}$ and $5.5
\times 10^{10}$ number of templates for respective data analysis of 30, 120
and 365 days for a FF of 0.97. We have studied the variation of the
estimated number of templates with frequency. It is found that
the numbers of templates required are approximately $1.22 \times 10^{10}$,
$2.16 \times 10^{10}$ and $5 \times 10^{10}$ for respective signal
frequencies 20, 50 and 100 Hz for analysing data of 120
days with a FF of 0.97. In this Chapter we have also studied the variation
of the FF with source location $[\theta ,
\phi ]$ and have found a marked symmetry. The FF for any
arbitrarily chosen $\theta$, is the same as corresponding to $\pi - \theta$.
Similar symmetry have also been observed for $\phi$. This
symmetry property will reduce the computation burden by a factor of four for
all sky search problem. It is not clear whether this symmetry property can be
established analytically as well. However the source location, because of this
symmetry, becomes uncertain and its exact location determination will require
t
he use of other methods.

\vspace{6.9cm}

\baselineskip=0.1cm

\noindent ------------------------------------------------------------------------------------------------------   \\
The thesis has been typeset by the candidate in \LaTeX.

\cleardoublepage
\baselineskip=0.83cm

\pagenumbering{arabic}

\setcounter{page}{1}

\chapter{Introduction: Gravitational Waves, Detectors and Sources}
\begin{figure}[hbt]
\centering
\epsfig{file=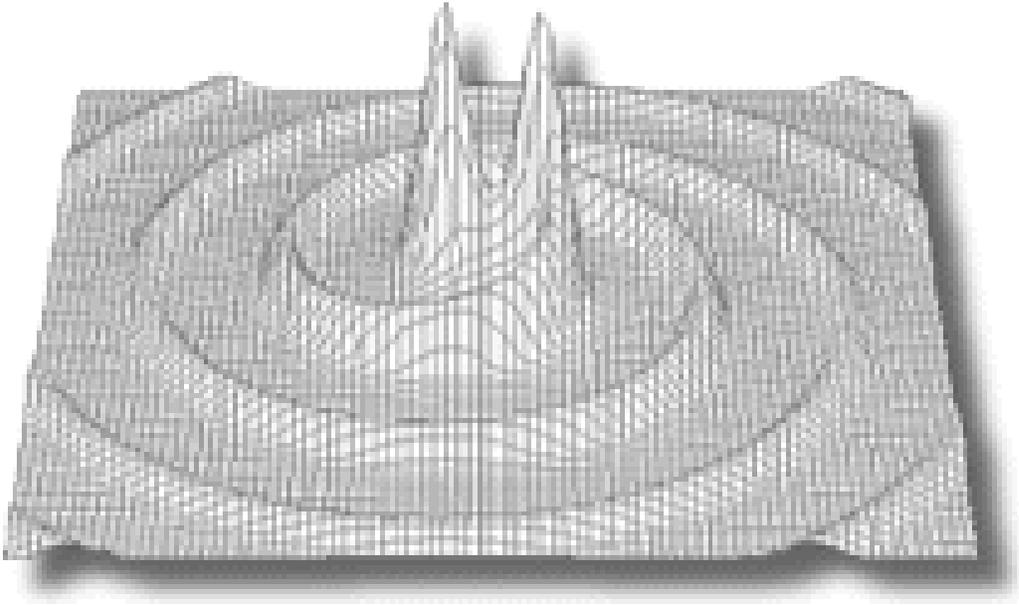,height=8.0cm}
\caption{Gravitational wave from binary.}
\label{fig:gwbinary}
\end{figure}

Gravitational waves (GW) like any other type of waves are  
propagating perturbations of some flat background space-time.
These are identified as small ripples rolling across space-time in the same
manner as water waves
are on an otherwise flat ocean. These waves originate from the most energetic
events in our Universe such as rotating neutron stars, colliding neutron
star binaries, supernovae explosions and gravitational collapse in 
Black Holes. They manifest themselves
as strains in space-time that periodically stretch and 
compress matter. The GW emanating from a binary may be 
represented as in Fig.~(\ref{fig:gwbinary}).

\par Newtonian gravitation does not have the provision for GW. In several
Lorentz covariant gravitational theories e.g. scalar, vector, tensor theories
the GW arise as the spreading out gravitational influence. The basis of most
current thinking of GW is Einstein's theory of General Relativity (GR). In fact,
GW was first studied by Einstein in 1916 by applying linear field approximation
to GR. However, he was misled to the result
that an accelerated spherical mass would emit GW. He corrected his mistake in
1918 and showed that the first order term of GW was quadrupolar. We discuss
the properties of GW in the next section. 

\par The strength of GW is so small that there is no hope of its detection
in the manner Hertz demonstrated the existence of electromagnetic waves. The
reason may be attributed to the extremely small value
of the universal constant of gravitation  
($\mathrm{G} = 6.7 \times 10^{-11}\mbox{Nm}^2\mbox{kg}^{-2}$). The dimensionless 
amplitude of typical GW reaching the earth is only of the
order of $10^{-17}$. This put in other words means that such a GW will put
a rod of one meter length into a oscillation with an amplitude of one millionth 
of the radius of a hydrogen atom. Any chance to observe the effects of GW
would require acceleration of astrophysical size masses at relativistic speeds.
Several GW-detectors are being built at present,
or have been proposed for the future, that hope to detect these very 
small vibrations. The best known of these are the {\em Laser 
Interferometer Ground Observatory}; {\sc LIGO}, and the {\em Laser 
Interferometer Space Antenna}; {\sc LISA}. As the names suggest, the 
former is a terrestrial observatory and the latter a space based one. 
Needless to say, what enormous engineering achievements these 
detectors require to detect the oscillations mentioned above.

\par The interaction of GW with matter is also very small and this leads to
some profound benefits for astrophysics. The wave will not scatter, so they
emanate undisturbed from the regions of their origin viz., 
densest regions of the Universe, inner cores of imploding
stars, the earliest instants of the primeval Universe, and the
formation of Black Holes. 
Hence, they will provide information
on the dynamics in these regions about which there are no other means.

\par A piece of indirect evidence affirming the existence
of GW came in 1974 when Joseph Taylor and Russell
Hulse studied the pulsar PSR 1913+16 which is a binary partner of compact
neutron star and is orbiting around the other in a slowly
decreasing eight hour period. The orbital frequency acceleration was
observed to be in perfect agreement with what expected from the energy
loss arising due to their calculated GW
emission (Damour and Taylor, 1992). For their accomplishment, Hulse and Taylor were awarded the
1993 Nobel Prize. 

\par Unfortunately PSR 1913 + 16 is presently emitting GW
too feeble for their direct detection on Earth and such a condition will
continue until the end of its pas-de-deux inspiral some 350 million
years hence. Interestingly, the density of observed binaries in our galaxy
is such that a few late stage 
inspiral events are expected per century to occur. Hence, the
study of the orbits of the binary neutron star system 
and their coalescence have played important role in the 
setting design criteria for some of the instruments to come into operation
in the next few years. To the date we have only the indirect evidence of
GW. Yet this has not been of much concern for theoretical physicists because
the strength of the postulated GW signals are
below the detection threshold currently available.

\par The quest to detect GW started in its earnest as early as in 1965
with the pioneering work of J. Weber on resonating bars. These are
basically high quality {\it bells\/} designed to be rung by transient
GW. With this beginning a small GW
detection community has arisen and thrived continuing to improve on
the original idea. The main effort is directed towards noise
reduction by introducing ever lower cryogenic temperature and ever
more sensitive displacement sensor to improve detector sensitivity.
Many more bars; progressively sophisticated ones have been 
built. The  {\it Explorer\/} which is quietly
functioning at CERN since 1989, is one such detector worth mention. We
discuss GW detectors in section 2.

\par There is an excellent prospect of detection of 
GW in the near future. There will emerge what has been called
``Gravitational Wave Astronomy''. This will provide 
another window for observing Universe. The expectation is that
it will uncover new phenomena as well as add new insights into
phenomena already observed in electromagnetic part of the spectrum.
GW emanate due to the accelerated motions of mass in the
interior of objects. As remarked earlier, these regions are otherwise obscure in
electromagnetic, and possibly, even in neutrino astronomy. The GW arise due to the
coherent effects of masses moving together
rather than individual motions of smaller constituents such as atoms
or charged particles which generate the electromagnetic 
radiation.

\vspace{0.65cm}

\section{Gravitational waves}
\label{sec:gr}
The gravitational waves like electromagnetic waves may be defined as time
varying gravitational fields in absence of its sources. The gravitation theory,  
widely believed, is the one proposed by Einstein and famous as GR.
In its mathematical essence it is expressed via the equation known by his name viz.
\begin{equation}
\mathrm{G}_{\alpha\beta} = \mathrm{R}_{\alpha\beta} - 
\frac{1}{2} g_{\alpha\beta} \mathrm{R} = 8\pi \mathrm{T}_{\alpha\beta}\footnote{For notations,
conventions and basic definitions the reader may refer to Schutz (1989)} 
\label{einstein}
\end{equation}                    

\vspace{0.1cm}

\noindent The conceptual meaning of this equation is that a localized density
distribution characterised by the energy momentum tensor, $T_{\alpha\beta}$
curves the space-time around it. The
curvature and other geometrical properties of the space-time are
characterised by Ricci curvature tensor $R_{\alpha\beta}$ and the underlying metric tensor
$g_{\alpha\beta}$. The curved space-time forces a free mass particle or light
to follow the geodesics. These geodesics measure the effect of gravitation.
The GR in its weak field limit yields the Newton's theory of gravitation. In this
approximation the gravitational field is considered to be represented by a
metric with a linear modification of the background Lorentz space-time metric,
$\eta_{\alpha\beta}$. It is important to remark that many of the basic concepts of GW 
theory can be understood in this approximation and were introduced and
developed by Einstein himself. At present GW theory; its issues and problems,
are dealt in detail in research text books of GR. 
Due to the limited scope of the thesis we base our discussion on 
linear field approximation and incorporate few aspects of GW required for
understanding of the subject matter presented in the thesis.

\vspace{0.65cm}

\subsection{Linear field approximation}

The metric of the
space-time may be expressed as
\begin{equation}
g_{\alpha\beta} = \eta_{\alpha\beta} + h_{\alpha\beta}\; , \qquad |h_{\alpha\beta}| \ll 1
\label{eq:gr2}
\end{equation}

\vspace{0.1cm}

\noindent It is straight
forward to compute Einstein tensor, $G_{\alpha\beta}$ and one obtains
\begin{equation}
G_{\alpha\beta} = - \frac{1}{2}
\left[ \bar{h}_{\alpha\beta ,\, \mu}^{\ \ \ \ \ ,\, \mu} +
\eta_{\alpha\beta}\bar{h}_{\mu\nu}^{\  \  \ ,\, \mu\nu} - \bar{h}_{\alpha\mu ,\, \beta}^{\  \  \  \  \ , \, \mu}
- \bar{h}_{\beta\mu ,\, \alpha}^{ \  \  \  \  \ , \, \mu} + O \left( h^2_{\alpha\beta}\right)\right]
\end{equation}

\noindent where
\begin{equation}
\bar{h}^{\alpha\beta} = h^{\alpha\beta} - \frac{1}{2}\eta^{\alpha\beta}h\, ; \qquad
h = h^\alpha_{ \ \alpha} = - \bar{h}^\alpha_{ \ \alpha}
\end{equation}

\vspace{0.1cm}

\noindent We raise and lower the tensor indices using the Lorentz metric. Let us note
that the expression for $G_{\alpha\beta}$ would simplify considerably if
we could require
\begin{equation}
\bar{h}^{\mu\nu}_{\ \ \ , \, \nu} = 0
\label{eq:gr5}
\end{equation}

\vspace{0.1cm}

\noindent In fact, there is a gauge freedom available as infinitesimal
coordinate transformation defined as
\begin{equation}
x^\alpha \longrightarrow x'^{\,\alpha} = x^\alpha + \xi^\alpha \, (x^\beta)\; ,
\quad \left|\xi^\alpha \right| \ll 1
\label{eq:gr6}
\end{equation}

\vspace{0.1cm}

\noindent where $\xi^\alpha$ is arbitrary. This gauge transformation
preserves~(\ref{eq:gr2}). It
can be shown that, in order to achieve the condition~(\ref{eq:gr5}), 
$\xi^\alpha$ has to be chosen as to satisfy 
\begin{equation}
\Box \xi^\mu = \xi^{\mu\nu}_{\ \ \ , \,\nu} = \bar{h}^{(old) \mu\nu}_{ \ \ \ \ \ \ \ , \,\nu}
\label{eq:gr7}
\end{equation}

\vspace{0.1cm}

\noindent where the symbol $\Box$ is used for the four dimensional Laplacian:
\begin{equation}
\Box f = f^{\, ,\, \mu}_{\ \ \ ,\, \mu} = \eta^{\mu\nu}f_{,\, \mu\nu} = \left( - \frac{\partial^2}{
\partial t^2} + \bigtriangledown^2\right) f
\end{equation}

\vspace{0.1cm}

The gauge condition defined via~(\ref{eq:gr5}) is called Lorentz gauge.
We adopt this gauge. In
literature harmonic gauge and de Donder gauge are other names for this gauge. 
The Einstein tensor, to the first order in $h_{\alpha\beta}$ becomes 
\begin{equation}
\mathrm{G}^{\alpha\beta} = - \frac{1}{2}\Box\bar{h}^{\alpha\beta}
\end{equation}

\vspace{0.1cm}

\noindent Thus the weak-field Einstein equations take the form
\begin{equation}
\Box\bar{h}^{\mu\nu} = - 16\pi \mathrm{T}^{\mu\nu}
\label{eq:10gr}
\end{equation}

\noindent In the Newtonian limit where the gravitational fields are too weak to
produce velocities near the speed of light,
\begin{equation}
\left|\mathrm{T}^{oo}\right| \gg \left|\mathrm{T}^{oj}\right| \gg \left|\mathrm{T}^{ij}\right|\, ; \quad \mathrm{T}^{oo}
\simeq \varrho\, , \quad \Box^2 \simeq \bigtriangledown^2
\end{equation}

\noindent Equation~(\ref{eq:10gr}) now results into
\begin{equation}
\bigtriangledown^2\bar{h}^{oo} = - 16\pi\varrho
\end{equation}

\noindent This equation may be compared to Newtonian equation for gravitational
potential, $\varphi$ i.e.
\begin{equation}
\bigtriangledown^2\varphi = -4\pi\varrho
\end{equation}

\noindent One obtains, after some calculations
\begin{equation}
h^{oo} = 2\varphi\, , \quad h^{xx} = h^{yy} = h^{zz} = -2\varphi
\end{equation}

\noindent and accordingly the space-time metric for a Newtonian gravitational
field is represented via
\begin{equation}
ds^2 = - (1 + 2\varphi) dt^2 + (1 - 2\varphi )( dx^2 + dy^2 + dz^2 )
\end{equation}

\vspace{0.65cm}

\subsection{Propagation of gravitational waves}

The GW in the regions far away from its sources may be described in the weak
field approximation. The Einstein equations, in absence of matter and
electromagnetic fields $(T_{\alpha\beta} = 0 )$ take the form
\begin{equation}
\Box\bar{h}^{\mu\nu} = \left( -  \frac{\partial^2}{\partial t^2} + \bigtriangledown^2 \right)
\bar{h}^{\alpha\beta} = 0
\label{eq:gr10}
\end{equation}

\vspace{0.1cm}

\noindent This is the three dimensional wave equation and its solution may be taken as \\
\begin{equation}
\bar{h}^{\alpha\beta} = \mathrm{A}^{\alpha\beta}exp\left( i k_\gamma x^\gamma\right)
\label{eq:gr11}
\end{equation}

\vspace{0.1cm}

\noindent where $\left\{ \mathrm{A}^{\alpha\beta} \right\}$ are constant components of
some tensor and $\left\{ k_\alpha \right\}$ are the constant components of some
one form satisfying the equation: 
\begin{equation}
\eta^{\mu\nu}k_\mu k_\nu = k^\nu k_\nu = 0
\label{eq:gr12}
\end{equation}

\vspace{0.1cm}

\noindent This means that~(\ref{eq:gr11}) represents a solution of~(\ref{eq:gr10}) provided
$k_\gamma$ is null one form and the associated four vector $k^\alpha$ is null. 
The value of $\bar{h}^{\alpha\beta}$
is constant on a hyper surface on which $k_\alpha x^\alpha$ is constant i.e.
\begin{equation}
k_\alpha x^\alpha = k_ot + {\bf k.x} = const. \, ; \qquad {\bf k } = \{  k^i \}
\label{eq:gr13}
\end{equation}

\noindent It is conventional to refer $k^o$ as $w$, which is called the frequency of
wave:
\begin{equation}
k^\alpha : \,\left( w , {\bf k } \right)
\label{eq:gr14}
\end{equation}

\noindent The gauge condition~(\ref{eq:gr5}) now requires
\begin{equation}
\mathrm{A}^{\alpha\beta}k_\beta = 0
\label{eq:gr15}
\end{equation}

\noindent which means that $\mathrm{A}^{\alpha\beta}$ must be orthogonal to $ k_\beta$.

\par The solution~(\ref{eq:gr11}) represents a plane wave propagating with velocity
of light. In physical applications one has to consider the real part of the
solution. Having obtained the
solution $\bar{h}^{\alpha\beta}$ one still has the freedom of choosing specific
$\xi^\alpha$ with the requirement that it represents some solution of the
Eq.~(\ref{eq:gr7}); which in view of~(\ref{eq:gr10}) becomes
\begin{equation}
\left( - \frac{\partial^2}{\partial t^2} + \bigtriangledown^2 \right)\xi^\alpha = 0
\end{equation}

\noindent Let us take its solution as
\begin{equation}
\xi_\alpha = \mathrm{B}_\alpha\exp (i k_\mu x^\mu )
\end{equation}

\noindent where $\mathrm{B}_\alpha$ is constant. It can be shown that the freedom available
in choosing the values of $\mathrm{B}_\alpha$ may be employed such that new $\mathrm{A}_{\alpha\beta}$
satisfy the conditions
\begin{eqnarray}
\label{eq:gr17}
\mathrm{A}^\alpha_{\ \alpha} & = & 0  \\
\mathrm{A}_{\alpha\beta}\mathrm{U}^\beta & = & 0
\label{eq:gr18}
\end{eqnarray}

\noindent where $\mathrm{U}^\alpha$ is some fixed four velocity.
Eqs.~(\ref{eq:gr15}),~(\ref{eq:gr17}) and~(\ref{eq:gr18}) are called the transverse
traceless (TT) gauge conditions. We choose $\mathrm{U}^\alpha$ as time basis vector
$\mathrm{U}^\alpha = \delta^\alpha_{\ 0}$. Let us take the direction of propagation of the
wave as the $z$ axis of the coordinate system. Now using~(\ref{eq:gr12}) and~(\ref{eq:gr14})
we have $k^\mu :\, (w,\, 0, \, 0, \, w)$. Now Eq.~(\ref{eq:gr18}) in view
of~(\ref{eq:gr15}) implies (i) $\mathrm{A}_{\alpha 0} = 0$ and (ii) $\mathrm{A}_{\alpha z} = 0$ for
all $\alpha$. This is the reason the gauge is called `transverse'
gauge. Further the trace free condition~(\ref{eq:gr17}) require $\mathrm{A}_{xx} = - \mathrm{A}_{yy}$.
Now the non vanishing component of $\mathrm{A}_{\alpha\beta}$ may be expressed as

\begin{eqnarray}
\mathrm{A}^{TT}_{\mu\nu} &=& 
\left(
\begin{array}{cccc}
0 & 0 & 0 & 0 \\
0 & \mathrm{A}_+ & \mathrm{A}_{\times} & 0 \\
0 & \mathrm{A}_{\times} & - \mathrm{A}_+ & 0  \\
0 & 0 & 0 & 0 
\end{array}\right)\, ; \quad \mathrm{A}_{xx} = \mathrm{A}_+\, , \ \mathrm{A}_{xy} = \mathrm{A}_\times
\label{eq:gr20}
\end{eqnarray}

\vspace{0.1cm}

\noindent Thus there are only two independent components of $\mathrm{A}_{\alpha\beta}; \, \mathrm{A}_+$ and
$\mathrm{A}_\times$. Let us note that traceless condition~(\ref{eq:gr17}) results into
\begin{equation}
\bar{h}^{TT}_{\alpha\beta} = h^{TT}_{\alpha\beta}
\label{eq:gr21}
\end{equation}

We have considered the plane wave solution of Eq~(\ref{eq:gr10}).
We know that any solution of Eqs.~(\ref{eq:gr15})
and~(\ref{eq:gr10}) may be expressed, because of theorems on Fourier analysis,
as a superposition of plane waves. Hence
if we consider the waves propagating along $z$ axis we can put all such planes
waves in the form~(\ref{eq:gr21}). Thus any wave has only two independent
components $h^{TT}_{xx}$ and $h^{TT}_{xy}$ represented respectively as $h_+$
and $h_\times$ corresponding to $\mathrm{A}_+$ and $\mathrm{A}_\times$.

\vspace{0.65cm}

\subsection{The effect of waves on free particles and its polarisation}
\label{sec:gweffect}
The independent components $h_+$ and $h_\times$ represent the polarisation
states of the wave. To understand their nature in little detail it is instructive 
to discuss the effect of GW as it hits a free particle. Let us choose a
background Lorentz frame in which the particle is initially at rest. We may
employ the initial four velocity of the particle $\left( \mathrm{U}^\alpha = \delta^\alpha_{\ 0}\right)$
to define the TT gauge of the wave. A free particle obeys the geodesic equation
\begin{equation}
\frac{d\mathrm{U}^\alpha}{d\tau} + \Gamma^\alpha_{\ \mu\nu} \mathrm{U}^\mu \mathrm{U}^\nu = 0
\end{equation}

\noindent This equation may be used to obtain the initial acceleration of the
particle:
\begin{equation}
\left( \frac{d\mathrm{U}^\alpha}{d\tau}\right)_{t = 0} = - \Gamma^\alpha_{ \ 00} = -
\frac{1}{2}\eta^{\alpha\beta}\left( h_{\beta 0 , \, 0} + h_{0 \beta, \, 0} -
h_{ 0 0 , \, \beta}\right)
\end{equation}

In view of Eqs.~(\ref{eq:gr11}),~(\ref{eq:gr20}) and~(\ref{eq:gr21}). the initial
acceleration of the particle vanishes. This means that the particle will be at
rest a moment later and consequently will be there forever. What does it mean ? Is
there no effect of the GW on a free particle ? No, the interpretation presented
at its face value is quite misleading. The result only means that the choice
of the TT gauge employed resulted into a coordinate system for the wave which
stays attached to the individual particles.

\par To get a better measure of the effect of the wave let us consider two
nearby particles situated at the origin $(0 , 0 , 0)$ and on the x axis, $(\bar{\epsilon} ,
0 , 0)$ separated by a distance $\bar{\epsilon}$. In view of our above discussion the
particles remain at their initial coordinate positions. The proper distance between
them is
\begin{eqnarray}
\bigtriangleup l & = & \int \left|ds^2\right|^{1/2} \; = \; \int \left|g_{\alpha\beta}dx^\alpha dx^\beta \right|^{1/2} \nonumber\\
& = & \int^\epsilon_0 \left|g_{xx}\right|^{1/2} dx \; \approx \; \left|g_{xx} (x = 0)\right|^{1/2}\bar{\epsilon}  \nonumber \\
\bigtriangleup l &=& \left\{ 1 + \frac{1}{2} h^{TT}_{xx} (x = 0)\right\}\bar{\epsilon}
\label{eq:gr25}
\end{eqnarray}

\vspace{0.1cm}

\noindent Thus the proper distance between two particles (as opposed to their
coordinate distance) does change with time. The effects
of the wave may also be described in terms of the geodesic deviation of the
separation vector, $\eta^\alpha$, connecting the two particles. It obeys the
equation
\begin{equation}
\frac{d^2}{d\tau^2}\eta^\alpha = \mathrm{R}^\alpha_{\ \mu\nu\beta}\mathrm{U}^\mu \mathrm{U}^\nu \eta^\beta
\label{eq:gr31}
\end{equation}

\noindent It can be shown that for the particles initially having the separation
vetor, $\eta^\alpha \longrightarrow (0 , \bar{\epsilon} , 0 , 0)$ one would get
\begin{equation}
\frac{\partial^2}{\partial t^2}\eta^x = \frac{1}{2}\bar{\epsilon}\frac{\partial^2}{
\partial t^2}h^{TT}_{xx}, \quad \frac{\partial^2}{\partial t^2}\eta^y \; = \;
\frac{1}{2}\bar{\epsilon}\frac{\partial^2}{\partial t^2}h^{TT}_{xy}
\end{equation}

\vspace{0.1cm}

\noindent Similarly, for initial separation vector, $\eta^\alpha
\longrightarrow (0 , 0 , \bar{\epsilon} , 0)$ we would get
\begin{eqnarray}
\frac{\partial^2}{\partial t^2}\eta^y& =& \frac{1}{2}\bar{\epsilon}\frac{\partial^2}{ 
\partial t^2}h^{TT}_{yy}\; = \; - \frac{1}{2}\bar{\epsilon}\frac{\partial^2}{
\partial t^2}h^{TT}_{xx}  \nonumber\\  
\frac{\partial^2}{\partial t^2}\eta^x & = & \frac{1}{2}\bar{\epsilon}\frac{\partial^2}{
\partial t^2}h^{TT}_{xy}
\end{eqnarray}

\vspace{0.1cm}

\noindent Let us note that in view of the results of the previous section we
may write~(\ref{eq:gr11}) as 
\begin{equation}
h_{\alpha\beta} = \mathrm{A}_{\alpha\beta}\exp (wt - kz)
\end{equation}

\noindent Thus the separation vector $\eta^\alpha$ of the
particles oscillates.

\begin{figure}[hbt]
\centering
\epsfig{file=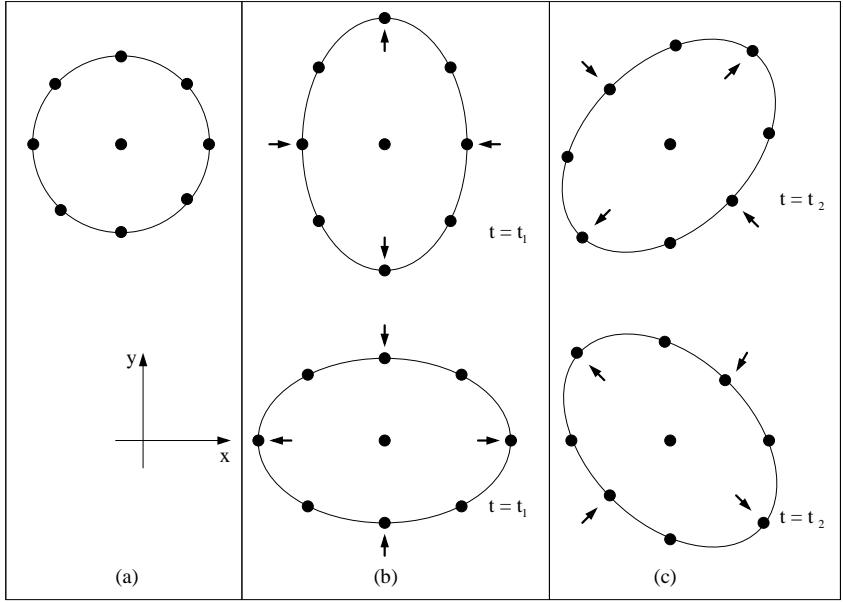,height=8.0cm}


\caption{(a) A circle of free particles before a wave travelling in the $z$
direction reaches them. (b) and (c) Distortions of the circle by a wave with
the `$+$' and `$\times$' polarisation. The two pictures represent the same wave
at phases separated by $180^o$. Particles are positioned according to their
proper distances from one another.}
\label{fig:pol}
\end{figure}

\par Let us consider a ring of particles initially at rest in the x-y plane
as shown in Fig.~(\ref{fig:pol}-a). Suppose a wave having $h^{TT}_{xx} \neq 0, \, h^{TT}_{xy} = 0$
hits the system of particles. The particles will be moved (in terms of proper
distance relative to the one in the centre) in the way shown in
Fig.~(\ref{fig:pol}-b).
Similarly, for a wave with $h^{TT}_{xx} = 0 = h^{TT}_{yy}, \, h^{TT}_{xy} \neq 0$
the picture would distort as in Fig.~(\ref{fig:pol}-c). Since $h^{TT}_{xx}$ and $h^{TT}_{xy}$
are independent, Fig.~(\ref{fig:pol}-b) and~(\ref{fig:pol}-c) provide the pictorial representation of the
polarisation states of the wave. Let us note that the two polarisation states are
simply rotated by $45^o$ relative to each other. This is in contrast to
electromagnetic waves where the two polarisation states are at $90^o$ to each
other.

\vspace{0.65cm}

\subsection{Generation of gravitational wave}

To understand the generation of GW it is sufficient to discuss the weak field
limit equation~(\ref{eq:10gr}), rewritten as
\begin{equation}
\left(-\frac{\partial^2}{\partial t^2} + \bigtriangledown^2\right) \bar{h}_{\mu\nu}
= - 16\pi \mathrm{T}_{\mu\nu}
\label{eq:gn1}
\end{equation}

\noindent Let us assume, for sake of simplicity, the time dependence of $\mathrm{T}_{\mu\nu}$
as sinusoidal oscillation of angular frequency $\omega$:
\begin{equation}
\mathrm{T}_{\mu\nu} = \mathrm{S}_{\mu\nu}(x^j)e^{-i\omega t}
\label{eq:gn2}
\end{equation}

\noindent and look for a solution for $\bar{h}_{\mu\nu}$ of the form
\begin{equation}
\bar{h}_{\mu\nu} = \mathrm{B}_{\mu\nu}(x^j)e^{- i \omega t}
\end{equation}

\noindent This, in view of Eqs.~(\ref{eq:gn1}) and~(\ref{eq:gn2}), would require
$B_{\mu\nu}$ to satisfy
\begin{equation}
\left( \bigtriangledown^2 + \omega^2 \right) \mathrm{B}_{\mu\nu} = - 16\pi \mathrm{S}_{\mu\nu}
\end{equation}

\noindent Outside the source i.e where $\mathrm{S}_{\mu\nu} = 0$ we want a solution which
would represent outgoing radiation far away. Hence we may take the solution as
\begin{equation}
\mathrm{B}_{\mu\nu} = \left( \frac{\mathbb{A}_{\mu\nu}}{r}\right) e^{i \omega r}
\end{equation}

\noindent Obviously, $\mathbb{A}_{\mu\nu}$ is to be related to $\mathrm{S}_{\mu\nu}$. Under
the assumption that the region of the space in which $\mathrm{S}_{\mu\nu} \ne 0$ is small
compared to $2\pi /\omega $ it can be deduced that
\begin{equation}
\mathbb{A}_{\mu\nu} = 4 \mathrm{J}_{\mu\nu} \, ; \quad \mathrm{J}_{\mu\nu} = \int \mathrm{S}_{\mu\nu}d^3x
\end{equation}

\noindent This assumption is referred to as slow motion approximation since it
implies that typical velocity inside the source, which is $\omega$ times the
size of that region, should be much less than 1. All, except the most powerful
sources, satisfy this assumption. Thus we get
\begin{equation}
\bar{h}_{\mu\nu} = \left(\frac{4}{r}\right) \mathrm{J}_{\mu\nu} e^{i\omega (r - t)}
\label{eq:gn41}
\end{equation}

\noindent This means that the generated GW has frequency $\omega$. This relation
may be expressed in terms of useful quantities with the help of the
following results.

\begin{enumerate}[(i)]
\item The energy momentum satisfies the conservation equation
\begin{equation}
\mathrm{T}^{\mu\nu}_{\ \ ,\, \nu} = 0
\end{equation}
and obeys the identity:
\item \begin{equation}
\frac{d^2}{dt^2}\int \mathrm{T}^{oo}x^lx^md^3x = 2\int \mathrm{T}^{lm}d^3x
 \end{equation}
\item The quadrupole moment tensor $\mathbf{I}^{lm}$ is defined as 
\begin{equation}
\mathbf{I}^{lm} = \int \mathrm{T}^{oo}x^lx^md^3x
\end{equation}
and which, in view of~(\ref{eq:gn2}), may be expressed as  
\begin{equation}
\mathbf{I}^{lm} = \mathrm{D}^{lm}e^{- i\omega t}
\end{equation}
where $\mathrm{D}^{lm}$ represents the time independent factor of $\mathbf{I}^{lm}$ (Misner et al., 1973). 

\par The first result gives
\begin{equation}
\mathrm{J}^{\mu o} = 0 \ \Rightarrow \ \bar{h}_{\mu o}
\end{equation}
\end{enumerate}

\noindent whereas the others let us write~(\ref{eq:gn41}) as
\begin{equation}
\bar{h}_{jk} = \left( \frac{- 2}{r}\right)\omega^2 \mathrm{D}_{jk}e^{i\omega (r - t)}
\end{equation}

\noindent We have still freedom of adopting TT gauge and may use it to achieve further
simplification. Let us choose $z$ axis along the direction of propagation of
the wave. We will then have 
\begin{eqnarray}
\bar{h}^{TT}_{zi}& = &0                            \\
\bar{h}^{TT}_{xx} & = & - \bar{h}^{TT}_{yy}\; = \; - \omega^2 \left( \mathbb{I}_{xx} - \mathbb{I}_{yy}
\right)\left(\frac{e^{i\omega r}}{r}\right)          \\
\bar{h}^{TT}_{xy}& =& - \left(\frac{2}{r}\right)\omega^2 \mathbb{I}_{xy}e^{i\omega r}
\end{eqnarray}

\noindent where $\mathbb{I}_{jk}$ represent the trace free part of quadrupole moment
tensor i.e.
\begin{equation}
\mathbb{I}_{jk} = \mathbf{I}_{jk} - \frac{1}{3}\delta_{jk}\mathbf{I}^i_{\ i}
\end{equation}

\noindent As an application of our results let us determine the amplitude of GW generated
by a laboratory source.

\subsubsection{Laboratory generator (Bar)}
\begin{figure}[hbt]
\centering
\epsfig{file=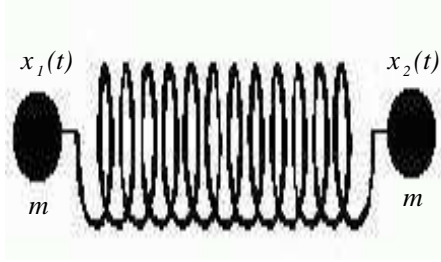,height=3.50cm}
\vspace{1.0cm}
\caption{An idealised gravitational wave detector with two identical masses connected
with a spring.}
\label{fig:idealbar}
\end{figure}

Let us take a system of two equal mass points capable of oscillating about their
mean position, such as shown in Figure~(\ref{fig:idealbar}). Let the system
oscillates longitudinally with angular frequency $\mathrm{w}$ about its mean position i.e.
\begin{equation}
\left.\begin{array}{ccl}
\vspace{0.2cm}
x_1 & = & - \frac{1}{2}l_o - \mathrm{A}\cos  \mathrm{w}t  \\
x_2 & = &  \frac{1}{2}l_o + \mathrm{A}\cos \mathrm{w}t
\end{array}\right\}
\end{equation}

\noindent where $l_o$ is the normal separation between the mass points and $\mathrm{A}$
represents the amplitude of oscillation. Now it is straight-forward to compute
$\mathbf{I}_{jk}$. The only non-zero component is
\begin{equation}
\left.\begin{array}{ccl}
\vspace{0.2cm}
\mathbf{I}_{xx}& = & \mathrm{m}\left[ (x_1)^2 + (x_2)^2\right]\\
     & = & const. + \mathrm{m}\mathrm{A}^2\cos 2\mathrm{w}t + 2\mathrm{m}l_o\mathrm{A}\cos \mathrm{w}t
\end{array}\right\}
\end{equation}

\noindent For purpose of wave generation the constant term is irrelevant. We
may obtain the non-vanishing components of $\mathbb{I}_{jk}$ as

\begin{equation}
\left.\begin{array}{ccl}
\vspace{0.2cm}
\mathbb{I}_{xx}& = & \frac{4}{3}\mathrm{m}l_o\mathrm{A}e^{-i\mathrm{w}t} + \frac{2}{3}\mathrm{m}l_o\mathrm{A}^2e^{-2i\mathrm{w}t}\\
\mathbb{I}_{yy}& = & \mathbb{I}_{zz} \; = \; -\frac{2}{3}\mathrm{m}l_o\mathrm{A}e^{-i\mathrm{w}t} - \frac{1}{3}\mathrm{m}l_o\mathrm{A}^2e^{-2i\mathrm{w}t}
\end{array}\right\}
\end{equation}

\noindent Finally one obtains after taking the real part
\begin{equation}
\left.\begin{array}{ccccl}
\vspace{0.2cm}
\bar{h}^{TT}_{xx}& = & -\bar{h}^{TT}_{xy} & = & -\left[ 2\mathrm{mw}^2l_o\mathrm{A}\cos (\mathrm{w}(r - t))
\, + \right. \\
&&&& \left. 4\mathrm{m}\mathrm{w}^2\mathrm{A}^2\cos (2\mathrm{w}(r - t))\right]/r\\
\bar{h}^{TT}_{xy}& = & 0 &&
\end{array}\right\}
\end{equation}

\indent For a laboratory generator, let us take
\begin{equation}
\mathrm{m} = 10^3\;kg\, , \quad l_o = 1\; m\, , \quad \mathrm{A} = 10^{-4}\; m\, , \quad
\mathrm{w} = 10^{-4}\, s^{-1}
\end{equation}

\noindent The data chosen represents a typical bar detector. Substituting the values
after converting them into geometrised units $(\mathrm{G} = 1 = c)$ the amplitude of
generated wave is about $10^{-34}/r$
\begin{equation}
\left|h\right| \simeq 10^{-34}/r  \, ; \qquad Laboratory\; source
\end{equation}

This demonstrates that laboratory generators are unlikely to produce useful
GW for its demonstration. For sake of comparison, let us conclude this subsection
by making an estimate of the strength of waves from strongest astrophysical sources.

\subsubsection{Astrophysical sources}

For a strong GW we should have $h_{\mu\nu} = O(1)$. This would occur near the
source where the Newtonian potential would be of the order 1. For a source of
mass $\mathrm{M}$ this should be at a distance of order $\mathrm{M}$. As we have seen the amplitude
of GW falls off as $r^{-1}$ far from the source. This means that the largest
amplitude expected to be incident on earth would be $\sim \mathrm{M}/\mathrm{R}$ where
$\mathrm{R}$ is the distance of the source from earth. For the formation of $10$ $\mathrm{M}_{\odot}$
black hole in a supernova explosion in a nearby galaxy $10^{23}$ away this is
about $10^{-17}$. Thus
\begin{equation}
\left|h\right|_{max} \simeq 10^{-17}  \, ; \qquad Astrophysical \; sources
\label{eq:maxh}
\end{equation}

\noindent Thus is in fact an upper limit and less violent events will lead to very
much smaller amplitudes.

\vspace{0.65cm}

\section{Gravitational wave detectors}

Equation~(\ref{eq:gr25}) may be recast as 
\begin{equation}
\frac{\bigtriangleup l - \bar{\epsilon}}{\bar{\epsilon}} = \frac{1}{2}h^{TT}_{xx}
\end{equation}

\noindent This means that the strain produced due to the hitting of a GW
in two mass points separated by $\bar{\epsilon}$ is of the order of $h$. In
view of Eq.~(\ref{eq:maxh})
let us note that even a strong GW signal would produce a space strain of the
order of $10^{-17}$, an unbelievably small effect which would jerk masses
spaced at one kilometer by a mere $10^{-20}$-one thousand of the diameter of
a proton!

\par Josheph Weber (1960) who pioneered the direct detection of GW
constructed an instrument consisting of massive cylinders of aluminium 
so-called ``bar'' detectors. These detectors exploit the sharp resonance of
the cylinder to get their sensitivity which is normally confined to a
narrow bandwidth (one or a few Hz) around the resonant
frequency. 

\par Despite their great potential sensitivity the primary drawback of
resonant bars is that they are by definition resonant. They are
sensitive mainly to the signal with a frequency corresponding to the
bar mechanical ringing frequency of the order of 1 kHz. A bar would
respond to the hammer blow of the asymmetrical supernova explosion by
simply ringing at its own bell tones and would be excited by a twin
neutron star inspiral only in that brief instant when the two stars
crib up through the bell tone frequency. 

\par Bar detectors continue to be developed, and have until very recently had 
a sensitivity to broadband bursts. However, the best hope for the first
detection of GW lies with large-scale interferometers. Within ten
years from now, we may see the launch of space-based interferometer,
LISA, to search for signals at frequencies lower than those that are
accessible from the ground. To measure the strain produced by GW to bar or
interferometric detector, one must fight against the different sources of noises.

\par An interesting additional twist is given by the fact that the 
gravitational wave  may 
be accompanied by gamma ray bursts. The GW
detectors will then work in coincidence not only with themselves and
GW bar antennas, but also with conventional
high-energy physics detectors like the underground neutrino
experiments and the orbital gamma rays burst monitors.

\par In this section we review, in brief, these developments.
In subsection~(\ref{sec:bar}) the underlying physical principles of the
bar detectors have been described.
Subsection~(\ref{sec:laser}) deals with large-scale
interferometers. Finally
subsection ~(\ref{sec:lisa}) describe the space-based project LISA, the exciting
prospects for the long-term future of GW detection in space. For 
details, one may refer to Thorne (1987, 1995), Saulson (1994).\\

\subsection{Bar detectors}
\label{sec:bar}
A bar detector, in its simplest form, may be idealised as a system of two
mass points coupled to a spring and may be depicted as shown in
Fig.~(\ref{fig:idealbar}). Let the system lies on the $x$ axis of our $TT$ coordinate
system with the masses at coordinate positions $x_1$ and $x_2$. The force free
oscillation of the system, in flat space time, may be expressed via
\begin{equation}
\left.\begin{array}{ccl}
\vspace{0.2cm}
\mathrm{m}x_{1,\, 00}& = &  -\kappa ( x_1 - x_2 + l_o ) - \nu ( x_1 - x_2 )_{\,,\, 0}\\
\mathrm{m}x_{2,\, 00}& = &  -\kappa ( x_2 - x_1 + l_o ) - \nu ( x_2 - x_1 )_{\,,\, 0}
\end{array}\right\}
\end{equation}

\noindent where $l_o$, $\kappa$ and $\nu$ represent respectively unstretched
length of the spring, spring constant and damping constant. We can combine
these equations to obtain the usual damped harmonic oscillator equation
\begin{equation}
\xi_{,\, 00} + 2\gamma\xi_{,\, 0} + w_o^2\xi = 0
\end{equation}

\noindent by introducing
\begin{equation}
\xi = x_2 - x_1 - l_o\, , \quad w_o^2 = 2\kappa /\mathrm{m} \, , \quad \gamma = \nu /m
\end{equation}

Let us recall that the $TT$ coordinate system is not convenient for
discussion of the dynamics of such system because in this frame a free particle
\underline{always} (before the arrival and after the passage of the wave)
remains at rest. However, this fact is useful in assigning a local inertial
frame $\left\{x^{\alpha '}\right\}$ at some $TT$ coordinate. Suppose that the
only motions in the system are those produced by the wave then masses velocities
will be very small and we may apply Newton's equations of motion for the masses:
\begin{equation}
\mathrm{m}x^{j '}_{,\, 0'0'} = \mathrm{F}^{j '}
\end{equation}

\noindent where $\left\{ \mathrm{F}^{j '} \right\}$ are the components of any non
gravitational forces on the mass. Further as the coordinates $\left\{x^{j '}\right\}$
differ negligibly to the order of $h_{\mu\nu}$ from that of its value
$\left\{x^j\right\}$ in $TT$ coordinate system, we may write this equation with
negligible error:
\begin{equation}
\mathrm{m}x^j_{,\, 0 0} = \mathrm{F}^j
\end{equation}

The only non gravitational force on each mass is due to the spring. The
spring will exert a force proportional to its instantaneous proper extensions.
If the proper length of the spring is $l$ and the direction of propagation of
the wave, for simplicity, is assumed to be along $z$ axis then
\begin{equation}
l(t) = \int_{x_1(t)}^{x_2(t)} \left[ 1 + h^{TT}_{xx} (t)\right]^{1/2} dx \approx
\left[ 1 + \frac{1}{2}h^{TT}_{xx}(t)\right]\left( x_2 - x_1 \right)
\end{equation}

\noindent (refer to~(\ref{eq:gr25})). Hence, equation of motion of the system
after hitting of the wave is given via
\begin{equation}
\left.\begin{array}{ccl}
\vspace{0.2cm}
\mathrm{m}x_{1,\, 00}& = &  -\kappa ( l_o - l ) - \nu ( l_o - l )_{\,,\, 0}\\
\mathrm{m}x_{2,\, 00}& = &  -\kappa ( l - l_o ) - \nu ( l - l_o )_{\,,\, 0}
\end{array}\right\}
\label{eq:31gr}
\end{equation}

\noindent Let us define
\begin{equation}
\xi = l - l_o =  \left[ 1 + \frac{1}{2}h^{TT}_{xx}\right]\left( x_2 - x_1 \right) - l_o
\end{equation}

\noindent leading to
\begin{equation}
x_2 - x_1 \simeq \left( \xi + l_o \right) \left( 1 - \frac{1}{2}h^{TT}_{xx}\right)
= \xi + l_o - \frac{1}{2}h^{TT}_{xx}l_o + O\left(\left|h_{\mu\nu}\right|^2\right)
\end{equation}

\noindent Using this equation, we may obtain from Eq.~(\ref{eq:31gr})
\begin{equation}
\xi_{,\, 00} + 2\gamma \xi_{,\, 0} + w_o^2\xi = \frac{1}{2}l_oh^{TT}_{xx, 00}
\end{equation}
\noindent correct to first order in $h^{TT}_{xx}$. This is the fundamental
equation governing the response of the detector to the GW. It has the simple
form of a forced, damped harmonic oscillator . 

\par Let a GW of frequency $\omega$ described via
\begin{equation}
h^{TT}_{xx} = \mathrm{A}\cos\omega t
\end{equation}

\noindent hits the detector then the steady solution for $\xi$ may be taken as
\begin{equation}
\left.\begin{array}{ccl}
\vspace{0.2cm}
\xi & = & \mathrm{R}\cos (\omega t + \varphi )\, ; \\
\vspace{0.2cm}
\mathrm{R} & = & \frac{1}{2}l_o\omega^2\mathrm{A}/\left[(w_o - \omega )^2 + 4\omega^2\nu^2\right]^{1/2}\, , \\
\tan \varphi & = & 2\nu\omega / \left(w_o^2 - \omega^2\right)
\end{array}\right\}
\end{equation}

\noindent The average energy of oscillation of the detector over one period,
$2\pi /\omega$:
\begin{equation}
\langle \mathrm{E} \rangle = \frac{1}{8} \mathrm{mR}^2(w_o^2 + \omega^2 )
\end{equation}

If we wish to detect a specific source whose frequency $\omega$ is known, then
we should adjust $w_o$ equal to $\omega$ for maximum response (resonance). The
resonance amplitude and energy of the detector will be
\begin{eqnarray}
\label{eq:37gr}
\mathrm{R}_{resonant} & = & \frac{1}{4}l_o \mathrm{A}\omega /\gamma\\
\mathrm{E}_{resonant} & = & \frac{1}{64}\mathrm{m}l_o^2\omega^2 \mathrm{A}^2(\omega /\gamma)^2
\end{eqnarray}

\noindent The ratio $(\omega /\gamma)$ is related to what is called the quality
factor $\mathrm{Q}$.
\begin{eqnarray}
\mathrm{Q}& = &\omega /2\gamma\\
\mathrm{E}_{resonant} & = & \frac{1}{16}\mathrm{m}l_o^2\omega^2 \mathrm{A}^2\mathrm{Q}^2
\end{eqnarray}

\par The bar detectors are massive cylindrical bars; its elasticity provides the
function of the spring. When waves hit the bar broadside, they excite its
longitudinal modes of vibration. The first detectors built by Weber were
aluminium bars of mass $1.4 \times 10^3$ kg, length $l_o = 1.5$ m, resonant
frequency $w_o = 10^4\, s^{-1}$ and $\mathrm{Q}$ about $10^5$. This means that a strong
resonant GW of $\mathrm{A} = 10^{-20}$ will excite the bar to an energy of the order of
$10^{-20}$ J. The resonant amplitude given by~(\ref{eq:37gr}) is only about
$10^{-15}$ m, roughly the diameter of an atomic nucleus.

\begin{figure}[hbt]
\centering
\epsfig{file=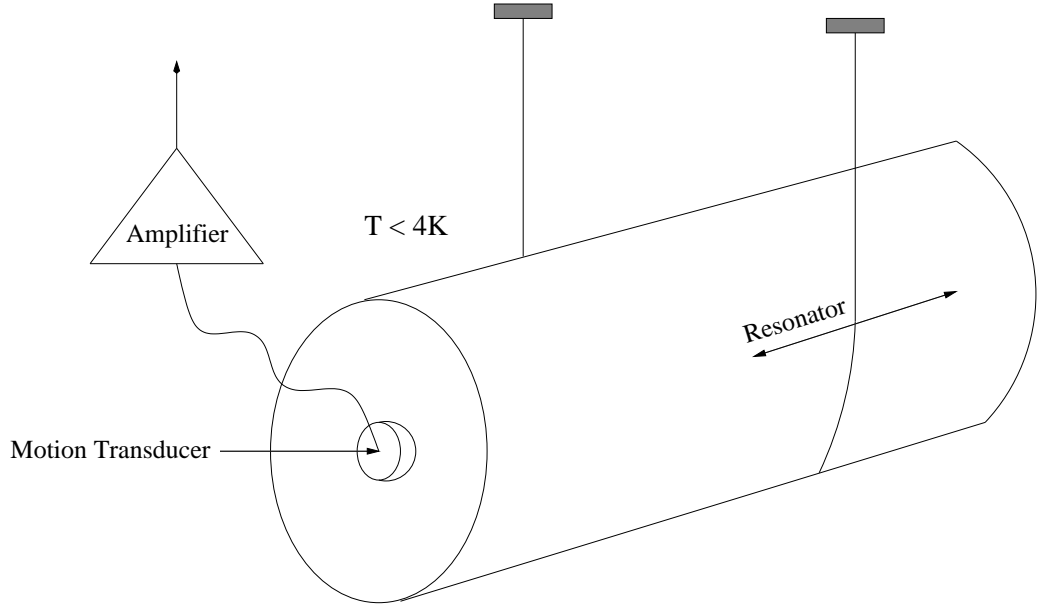,height=8.0cm}
\vspace{1.0cm}
\caption{A schematic of resonant bar detector.}
\label{fig:barfig}
\end{figure}

\par Clearly, the detections of such small levels of excitations will be
hampered by random noise in the oscillator. For example, thermal noise in any
oscillator induces random vibration with a mean energy of $\mathrm{kT}$; $\mathrm{k}$ the
Boltzmann constant having value $1.38 \times 10^{-23}$ $\mathrm{J}/\mathrm{K}$. At room temperature
$(\mathrm{T} \sim 300 \mathrm{K})$ the thermal noise amounts to the energy $\sim 4 \times 10^{-21}$ J.
Other sources of noise such as vibrations from passing vehicles and every day
seismic disturbances could be considerably larger than this, so the apparatus
has to be carefully isolated.

\begin{table}
\centering
\begin{tabular}{|l|c|c|}
\hline
&&\\
Detector & Location & Taking data since\\
&&\\  \hline
&&\\
NAUTILUS & Frascati, Rome    & 1993\\ 
&&\\ \hline
&&\\
EXPLORER & Cern (Rome group) & 1990\\
&&\\ \hline
&&\\
ALLEGRO  & Louisiana, USA    & 1991\\
&&\\ \hline
&&\\
AURIGA   & Padua, Italy      & 1997\\ 
&&\\ \hline
&&\\
NIOBE    & Perth, Australia  & 1993\\ 
&& \\
\hline
\end{tabular}
\vspace{1.0cm}
\caption{The existing resonant bar detectors.}
\label{table:barsite}
\end{table}

\par A typical ``bar'' detector consists of a cylinder of aluminium with a
length $l_o \sim 3$ m, a resonant frequency of order $w_o \sim 500$ Hz to 1.5 
kHz, and a mass $\sim 1000$ kg whose mechanical oscillations are driven by 
GW [Fig.~(\ref{fig:barfig})]. A transducer converts mechanical vibrations of the bar into an 
electrical signal, which is then amplified by an amplifier
and recorded. 

\par Currently there are a number of bar detectors in operation [see Table~(\ref{table:barsite})]. Some of 
these operate at room temperature and some others at cryogenic temperature. 
Some detectors (NAUTILUS \& EXPLORER) may be cooled down to ultra cryogenic 
temperature. They can detect signal amplitudes $h \sim 10^{-20}$ in a band 
width of 10-20 Hz around a central frequency of 1 kHz. Asymmetric supernovae 
in our Galaxy are the best candidates  for these detectors. For example,
a supernova collapse in our galaxy at a distance of 10 kpc emits GW of the 
amplitude $h \sim 10^{-17}$. At present this sensitivity has been achieved by 
some of the bar detector. They may also see continuous 
radiation emitted by a neutron star if the frequency happens to lie in their 
sensitivity band.

\vspace{0.65cm}

\subsection{Ground based laser interferometric detectors}
\label{sec:laser}
The effect of GW is to produce a transverse shear strain and this fact
makes the Michelson interferometer an obvious candidate for a
detector. The Michelson interferometers must have kilometric arms, constituted by 
``high fineness'' Fabry Perot cavities to trap the light for long period and to increase the
sensitivity. Laser standing power measured in KW will be stored
within the cavities. Beam losses at the level of $10^{-6}$ per
passages
 are required. Mirrors must be 20 or 30 cm in diameter just
to hold diffraction losses, and coating with parts per
million reflection losses have been developed. Vacuum pipe with 
diameters of the order of a meter are necessary just to contain the
diffraction limited laser beams and vacuum exceeding 10 torr is
required to prevent the introduction of spurious signal by refraction
index fluctuations. The list of ground based laser interferometric detector
site and orientation are shown in Table~(\ref{table:intsite}).

\par Interferometric detectors currently under construction will 
increase our ability to directly observe GW. 
The Japanese TAMA-300 meter interferometric detector having sensitivity
$10^{-20}$ is already
operational and some initial hours of data has been analysed. Several other
projects are now nearing to completion:
The British-German collaboration is 
constructing a 600 m interferometer (GEO) in Hannover, Germany, 
the French-Italian collaboration is building a 3 km detector 
(VIRGO) near Pisa, Italy and the Americans are
building two 4 km antennas (LIGO), one in Livingston and the other 
in Hanford in the USA which will have the sensitivity $10^{-21}$. These detectors 
will start taking data between 2002 and 2003. The larger of these
detectors, LIGO and VIRGO, are likely to  be upgraded in sensitivity
by an order of magnitude with a better low-frequency performance
in 2005. These ground based interferometers will eventually be
sensitive to sources in the frequency range from 10 Hz to several kHz.

\begin{table}
\centering
{\footnotesize
\begin{tabular}{|l|c|c|c|r|r|} \hline
{\sl Detector}&{\sl Location}& {\sl Length} (m)&{\sl Corner
Location}&{\sl Arm 1}&{\sl Arm 2}\\ \hline
Glasgow & Glasgow, GBR &10 & $55.87^\circ$N $\quad
 - \quad 4.28^\circ$W  & $77.00^\circ$ & $167.00^\circ$ 
\\ \hline
CIT & Pasadena, CA, USA & 40 & $34.17^\circ$N $\quad
 -  \quad 118.13^\circ$W  & $180.00^\circ$ & $270.00^\circ$  \\ \hline
MPQ & Garching, GER &  30 & $48.24^\circ$N $\quad - \quad 11.68^\circ$W
& $329.00^\circ$ & $239.00^\circ$  \\ \hline
ISAS-100 & Tokyo, JPN & 100 & $35.57^\circ$N $\quad
- \quad 139.47^\circ$W  & $42.00^\circ$ & $135.00^\circ$ 
\\ \hline
TAMA-20 & Tokyo, JPN & 20 & $35.68^\circ$N $\quad
- \quad 139.54^\circ$W  & $45.00^\circ$ & $315.00^\circ$ 
\\ \hline
Glasgow & Glasgow, GBR &10 & $55.87^\circ$N $\quad
 - \quad 4.28^\circ$W  & $62.00^\circ$ & $152.00^\circ$ 
\\ \hline
TAMA-300 & Tokyo, JPN & 300 & $35.68^\circ$N $\quad
- \quad 139.54^\circ$W  & $90.00^\circ$ & $180.00^\circ$ 
\\ \hline
GEO-600 & Hannover, GER & 600 & $52.25^\circ$N $\quad
- \quad 9.81^\circ$W  & $25.94^\circ$ & $291.61^\circ$ 
\\ \hline
VIRGO   & Pisa, ITA & 3000        & $43.63^\circ$N $\quad
- \quad 10.5^\circ$W  & $71.50^\circ$      & $341.50^\circ$  \\ \hline
LIGO    & Hanford, WA, USA & 4000        & $46.45^\circ$N
$\quad - \quad 119.41^\circ$W & $36.80^\circ$      & $126.80^\circ$  \\ \hline
LIGO    & Livingston, LA, USA  & 4000        & $30.56^\circ$N
$\quad - \quad 90.77^\circ$W  & $108.00^\circ$     & $198.00^\circ$
\\ \hline
\end{tabular}}
\vspace{0.30cm}
\caption{ Site and orientation of Earth based interferometric
gravitational wave detectors.}
\label{table:intsite}
\end{table}

\par When a GW, with frequencies high compared to the masses $\sim 1$ Hz pendulum
frequency, passes through the detector of arm length  $l_o$ then one arm of the 
detector get stretched in one direction whereas the other arm gets compressed.
If the total change in the the arm-length is
$\delta l$,  then the response $R(t)$ of the detector which is defined as
$\frac{\delta l}{l_o}$ may be given via

\begin{equation}
R(t) = \frac{\delta l}{l_o} = F_+h_+(t) + F_\times h_\times (t) 
\end{equation}

\noindent where, $h_+$, $h_\times$  are two polarisations of the signal and  
$F_+$, $F_\times$ are beam pattern functions arising because of transformation
equations relating to the wave frame and the detector frame [refer to
Eq.~(\ref{eq:gr31})]. Let the detector frame $(x, y, z)$ and the wave frame
$(X, Y, Z)$ are related via transformation matrix ${\bf A\/}$ i.e
\begin{equation}
\xi^i = {\bf A\/}^i_R\xi^R\, , \quad \xi^k = {\bf A\/}^k_j\xi^j
\end{equation}

\noindent where capital latin letters refers to wave frame quantities and lower
case for the detector frame quantities. It can be shown that the response of the
detector may be expressed as
\begin{equation}
\frac{\delta l}{l_o} = - \sin 2\Omega \left[ \left({\bf A\/}^x_X{\bf A\/}^y_Y
- {\bf A\/}^x_Y{\bf A\/}^y_Y\right) h_+ + \left( {\bf A\/}^x_X{\bf A\/}^y_Y +
{\bf A\/}^x_Y{\bf A\/}^y_X\right) h_\times\right]
\end{equation}

\noindent where $2\Omega$ is the angle between the arms of the detector, normally
kept equal to $90^o$ [Schutz and Tinto (1987)].
\begin{figure}[hbt]
\centering
\epsfig{file=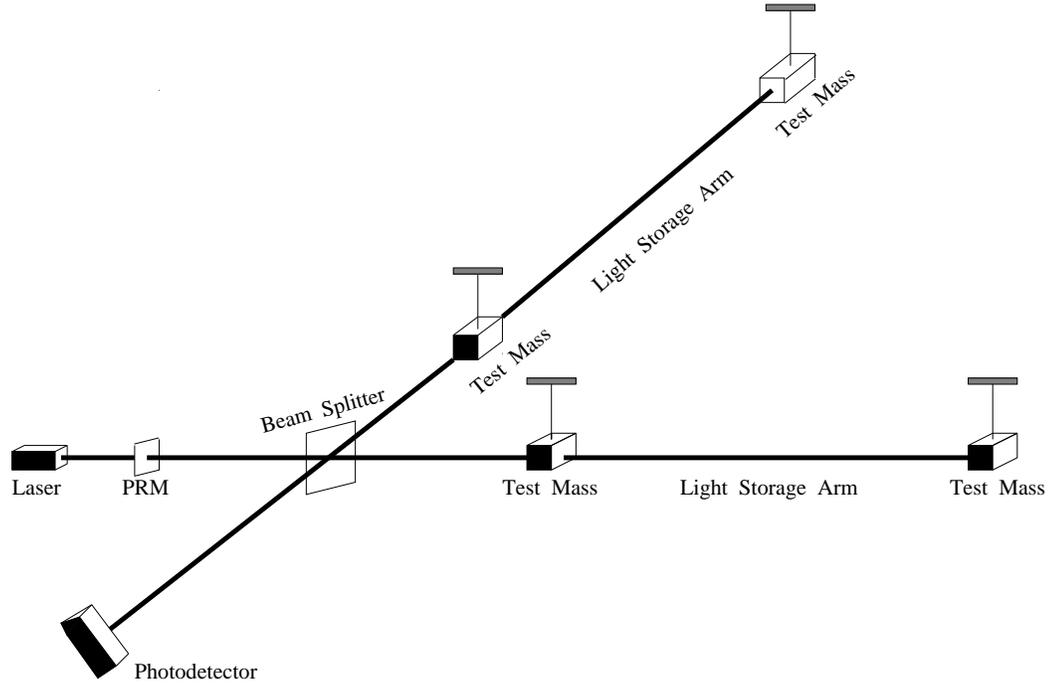,height=9.0cm}
\vspace{1.0cm}
\caption{A schematic of Earth based laser interferometric gravitational wave 
detector.}
\label{fig:intfig}
\end{figure}

\par A laser interferometric detector consists of four test masses/mirrors
that hang from vibration isolated supports as shown in
Fig.~(\ref{fig:intfig}). The mirrors facing each other on each arm form
a Fabry-Perot cavity. A beam splitter splits a carefully prepared
laser beam in two, and directs the resulting beams down the two
arms. Each beams penetrates through the mass, and through the
dielectric coating (the mirror). The length of the
arm's of the Fabry-Perot cavity is adjusted to be nearly an integral number of
half wave-length of light. The
cavity's end mirror has much higher reflectivity than its corner
mirror, so the trapped light leaks back out through the corner mirror,
and then hits the beam splitter where it recombines with light from
the other arm. Most of the recombined light goes back toward the laser
(where it can be returned to the interferometer by a ``power recycled
mirror'' labeled PRM, but a tiny portion goes toward the
photo-diode.

\par When a GW hits the detector and moves the masses, thereby changing the 
lengths of the arms of the two cavities, it shifts each cavity's
resonant frequency slightly relative to the laser frequency, and the
phase of the light that exists from the cavity toward the beam
splitter. Correspondingly, the relative phase of the two beams
returning to the splitter is altered by an amount $\bigtriangleup \Phi
\propto \delta l$, and this relative phase shift causes a
change in the intensity of the recombined light at photo-diode,
$\bigtriangleup I_{pd} \propto \bigtriangleup \Phi \propto
\delta l \propto R(t)$. Thus, the change of photo-diode output
current is directly proportional to the GW strain.

\par Scientists are now studying the technologies that may be needed for a 
further improvement of the sensitivity of the next-generation
detectors. This may involve cooling mirrors, using ultra-massive
mirrors of special materials, using purely non-transmissive optics,
and even circumventing the quantum limit in interferometers, as has
been studied for bars. The goal of next-generation detectors would be
limited just by gravity-gradient noise and quantum effects.

\vspace{0.65cm}

\subsection{Laser interferometric space antenna}
\label{sec:lisa}
\par Gravity-gradient noise on Earth is much larger than the amplitude of 
any expected waves from astronomical sources at frequencies below about 1
Hz, but this noise falls off a $1/r^3$ as one moves away from the
Earth. A detector in space would not notice the Earth's noisy
environment. The Laser interferometric space antenna (LISA) project,
efforts underway by both European Space Agency (ESA) and NASA with a
view toward a collaborative mission around 2010, would open up the
frequency window between 0.1 mHz and 0.1 Hz.

\begin{figure}[hbt]
\centering
\epsfig{file=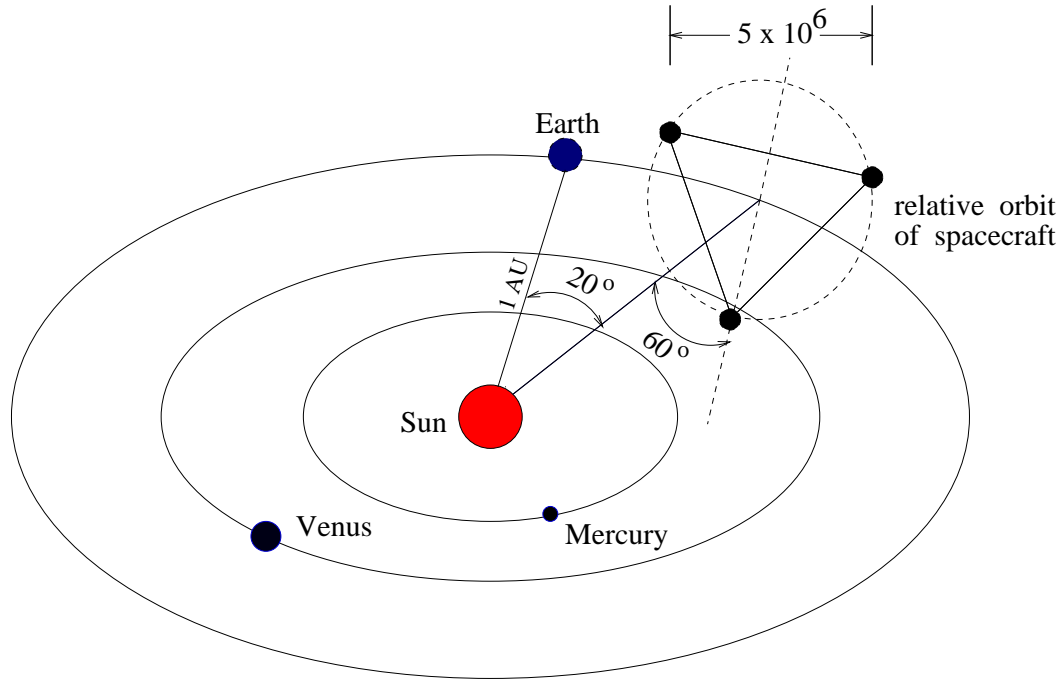,height=9.0cm}
\vspace{1.0cm}
\caption{A schematic diagram of the laser interferometric space antenna.}
\label{fig:lisafig}
\end{figure}

\par A concept of the project is shown in Fig.~(\ref{fig:lisafig}). Three 
spacecraft are placed in solar orbit at 1 A.U, about 20 degrees behind
the Earth in it orbit. The spacecraft are located at the corners of an
equilateral triangle with $5 \times 10^6$ km-long sides. Two arms of
the triangle comprise a Michelson interferometer with vertices at the
corners. The third arm permits another interferometric observable to
be measured, which can determine a second polarization. The
interferometers use one micron light as the terrestrial detectors but
need only a single pass in the arms to gain the desired sensitivity.
The end points of the interferometers are referenced to proof masses
free-floating within and shielded by the spacecraft. The spacecraft is
incorporated in a feedback loop with precision thrust control to
follow the proof masses. 

\par The main environmental disturbances to LISA are the forces from the Sun.: 
fluctuations in solar radiation pressure and pressure from the solar wind. To 
minimize these, LISA incorporates drag-free technology. Interferometry is 
referenced to an internal proof mass that falls freely, not attached to the 
spacecraft. The job of the spacecraft is to shield this mass from external 
disturbances. It does this by sensing the position of the mass and firing its 
own jets to keep itself (the space craft) stationary relative to the proof 
mass. To do this, it needs thrusters of very small thrust that have accurate 
control. The availability of such thrusters, of the accelerometers needed to 
sense disturbances to the spacecraft, and of lasers capable of continuously 
emitting 1 W infrared light for years, have enabled the LISA mission.

\par LISA is supposed to see 
many exciting sources for example the coalescences of giant black holes in 
the centre of galaxies. LISA will see such events with extraordinary 
sensitivity, recording typical signal-to-noise-ratios of 1000 or more for 
events at redshift 1.

\vspace{0.65cm}

\section{Sources of gravitational waves}
Astronomical observations have led to the belief that 
luminous matter constitutes
a small fraction of the total matter content of the Universe. More
than 90\% of the mass in the Universe is electro-magnetically silent. The 
presence of dark matter is inferred from the gravitational influence it
causes on luminous matter. It is possible that some fraction
of this dark matter is a strong emitter of GW. There are many reviews on GW
sources (Thorne, 1987; Blair, 1991; Schutz, 1989, 1993, 1999, Sathyaprakash,
1999 and Grishchuk, et. al., 2000). The discussion 
on the GW sources in this section is introductory and for details one may
refer to these reviews. The anticipated
GW sources can be classified into (i) Transients (ii)
Continuous and (iii) Stochastic.

\vspace{0.65cm}

\subsection{Transient sources}
Supernovae and inspiralling compact stars in binary systems are prime examples of
sources of this type. These sources last for a
duration of about $10^{-3} - 10^4$ seconds.

\vspace{0.65cm}

\subsubsection{Supernovae explosions}
The type II Supernovae explosions, which are believed to occur as a result of the core collapse of an 
evolved massive ($> 9M_\odot$) star and which are associated with 
violent mass ejection with velocities of order 0.03 with formation of a 
compact remnant---neutron star or a black hole--may emit significant amount of
GW depending upon how asymmetric the collapse is. The emission occurs
essentially during the rotational core 
collapse, bounce and oscillations, rotation-induced bars and convective 
instabilities set up in the core of the new born neutron star.

\par Rapid rotation flattens the collapsing core inducing large quadrupole
moment; thus generating GW. Study of a wide range of 
rotational core collapse models suggests that the largest signals are 
produced by models which are (i) initially slowly rotating and have a stiff 
equation of state or (ii) initially rapidly rotating and have a soft equation 
of state. In the first case bounce occurs at densities above nuclear matter 
density, with a fast deceleration of the collapsing core resulting in the 
emission of GW signals. In the second case, the quadrupole moment is large 
due to rapid rotation which facilitates emission of GW. However, in either
case the signals are not strong enough to be interesting sources for 
the first generation detectors. 

\par When the core's rotation is {\it high enough \/} it may cause the core to flatten 
before it reaches nuclear density leading to an instability which transforms 
the flattened core into a bar-like configuration spinning about its 
transverse axis. Some of these instabilities could also fragment the core into 
two or more pieces which then rotate about each other. Both are efficient 
ways of losing energy in the form of GW.

\par Instabilities in the core of the newly born neutron star, which last for 
about a second after the collapse, are likely to produce GW  
due to anisotropic mass distribution and motion.

\vspace{0.65cm}

\subsubsection{Inspiraling compact binaries}

The binary systems whose either member is a compact star e.g. 
neutron star (NS) and black hole (BH) are the most promising transient sources 
of GW during the phase of their coalescence. The well known binary pulsar
PSR 1913 + 16 is such a system but it will coalesce in a time scale of $10^9$
years from now --- not a right candidate. However, there are binaries in our
galaxy with coalescence time scale much shorter than this. Further, statistical
analysis of binary pulsars estimates three NS-NS coalescence per year out to a
distance of 200 Mpc (Phinney, 1991; Narayanan et. al 1991). The initial
LIGO/VIRGO interferometers have a fair chances to see the inspiral events.
Compact inspiraling binaries emit quasi-periodic GW with a frequency that
sweeps upward toward a maximum frequency. The maximum frequency may be of the
order of 1 KHz for neutron stars. In the lower frequency regime the wave form
is easily computed from the quadrupole formalism. At higher frequencies
Post-Newtonian corrections would be required (Thorne, 1987; Krolak, 1989). In
view of the strong potentialities of such sources the various aspects related
to the emission of GW has been dealt extensively in full details and one may
refer to Sathyprakash, 1999 and Grischuk, et al. 2000.

\vspace{0.65cm}

\subsection{Continuous gravitational wave}

The continuous gravitational wave (CGW) sources are of prime interest because such sources can be
observed again and again and hence single interferometer is 
sufficient to confirm its detection. However, one can't expect a source to be
continuous as well as strong. For emission of GW from pulsars, there should be
some asymmetry in it. There are several mechanisms which may lead to deformations
of the star, or to precession of its rotation axis. The
characteristic amplitude of GW from pulsars scales as
\begin{equation}
h \sim \frac{\mathrm{I} f^2_o\varepsilon }{r}
\end{equation}
where $\mathrm{I}$ is the moment of inertia of the pulsar, $f_o$ is the GW frequency,
$\varepsilon $ is a measure of deviation from axissymmetry and $r$ is the distance
of the pulsar.

\par As remarked earlier, pulsars are born in supernovae explosions. The outer layers
of the star crystallizes as the newborn star pulsar cools by neutrino emission.
Estimates, based on the expected breaking strain of the crystal lattice, suggest
that anisotropic stresses, which build up as the pulsar looses rotational energy,
could lead to $\varepsilon \le 10^{-5}$; the exact value depends on the breaking
strain of the neutron star crust as well as the neutron star's ``geological history'', 
and could be several orders of magnitude smaller. Nonetheless, this upper
limit makes pulsars a potentially interesting source for kilometer scale
interferometers.

\par Large magnetic fields trapped inside the super fluid interior of a pulsar may
also induce deformation of the star. This mechanism has been explored recently,
indicating that the effect is extremely small for standard neutron star models
( $\varepsilon \le 10^{-9}$).

\par Another plausible mechanism for the CGW is the Chandrasekhar-Friedman-Schutz
(CFS) instability, which is driven by GW back reaction. It
is possible that newly-formed neutron stars may undergo this instability
spontaneously as they cool soon after formation. The frequency of the emitted wave 
is determined by the frequency of the unstable normal mode, which may
be less than the spin frequency.

\par Accretion is another way to excite neutron stars. There
is also the Zimmermann - Szedinits mechanism where the principal axes of
moment inertia are driven away from the rotational axes by accretion from a
companion star. Accretion can in principle produce relatively strong wave 
since the amplitude is related to the accretion rate rather than to structural
effects in the star.

\vspace{0.65cm}

\subsection{Stochastic waves}

Catastrophic processes in the early history of Universe, as well
as the astrophysical sources distributed all over the cosmos,
generate stochastic background of GW. A given stochastic background
will be characteristic of the sources that are responsible for it and it may
be possible to discriminate different backgrounds provided their
spectral characteristics are sufficiently different.\\

\par {\it Primordial background:} It is believed that similar to  
cosmic microwave background (CMBR) the GW background was also produced at the same 
time as a result of quantum fluctuations in the early Universe (Grischuk, 1997).
Primordial background radiation would freely travel to us from almost the
very moment of creation because GW couples very weakly with matter. Hence,
its detection would help us to get a picture of the first moments after the
big bang. The COBE data have set limits on the strength of the GW background.
The strength are far too weak 
to be detected by any planned ground-based detectors. However, advanced LIGO
detectors may observe the background generated by the collisions of a cosmic
string network (for details one may refer to Allen, 1997).

\par Phase transitions in the early Universe, inspired by fundamental
particle physics theories, and cosmological strings and domain walls
are also sources of a stochastic background. These processes are expected
to generate a background which has a different spectrum and strength than
the primordial one. Future ground-based detectors will achieve good enough
sensitivity to measure this background GW and such measurements will prove to
be a good test bed for these cosmological models.\\

\par {\it Supernovae background:} Even though an individual supernova may
not be detectable out to a great distance, the background produced by all 
supernovae within the Hubble radius might indeed be detected (Blair \& Ju, 
1996). The coincidence between an advanced interferometer and a resonant bar
within 50 km of the interferometer will enable the detection of this background.
These studies may shed light on the history of star formation rate, a subject
of vigorous debate amongst astrophysicists.\\

\par {\it Galactic binary background:} Binaries with orbital periods $P
\sim 10^{-4}-10^{-2} s $ will be observable in space-based detectors. 
A large number of them are present in our galaxy but they will not be identifiable
separately because they are at a large distances and consequently have a feeble
amplitude.  However they would contribute to background GW. In addition to binaries of compact stars,
there are also other binaries consisting of white dwarfs, cataclysmic 
variables, etc., which will also contribute to the background radiation 
produced by compact binaries. The net effect is that these sources will 
appear as a background noise in space interferometers. By studying the nature 
of this background one can learn a lot about binary population in our 
Galaxy. \\

\par {\it Galactic pulsar background} The CGW from pulsars could also produce 
background of radiation which will limit the sensitivity of the ground-based 
laser interferometers (Giazotto, Gourgoulhon and Bonazzola, 1997). There are 
about $10^9$ neutron stars in our Galaxy of which about $2 \times 10^5$ will 
contribute to the background radiation. This background radiation will be 
prominent, and observable in the LIGO/VIRGO detectors, in the frequency range 
of 5-10 Hz, at an rms amplitude of $h_{rms}\sim 2 \times 10^{-26}$, where the 
rms is computed over $10^5$ sources. At frequency of 10 Hz, wavelength of GW 
will be around 30,000 Km. Hence, it would be possible to cross-correlate data 
from two distant detectors, such as two LIGOs, or LIGO and VIRGO, and 
discriminate the background against other sources of noise.\\

\chapter{Data analysis concept}
\section{Introduction}
The GW data analysis strategy is different in many ways
from conventional astronomical data analysis.
This is due to the following:
\begin {itemize}
\item GW antennas are essentially omni-directional with
their response better than 50\% of the average over 75\% of the sky.
Hence our data analysis systems will have to carry out all-sky searches
for the sources. 

\item Interferometers are typically broad-band covering 3 to 4
orders of magnitude in frequency. While this is obviously to our
advantage, as it helps to track sources whose frequency 
may change rapidly, it calls for searches to be carried over 
a wide-band of frequencies. 

\item In contrast to electromagnetic (EM) radiation, most astrophysical GWs
are tracked in phase and the signal-to-noise ratio (SNR) 
is built up by coherent superposition of many wave cycles emitted 
by a source. Consequently, the SNR is proportional to the 
amplitude and only falls off, with the distance to
the source $r,$ as $1/r.$ Therefore,
the number of sources of a limiting SNR increases as $r^{3}$ 
for a homogeneous distribution of sources in a flat Universe, as opposed
to EM sources that increase only as $r^{3/2}.$ 

\item The GW antennas
acquire data continuously for many years at the rate of several
mega-bytes per second. It is expected that about a hundredth of
this data will have to pass through our search analysis systems. Unless
on-line processing can be done we cannot hope to make our searches.
This places huge demands on the speed of our data analysis hardware. 
A careful study of our search algorithms with a view to making
them as optimal (maximum SNR) and efficient (least search times) 
as one possibly can is required.

\end {itemize}

\vspace{0.65cm}

\section{Gravitational wave antenna sensitivity}
The performance of a GW detector is characterised by the one sided
{\it power spectral density} (PSD) of its noise background.
The analytical fits to noise power spectral densities $S_n(f)$ of ground
based interferometers are given in Table~(\ref{table:noise})
where $S_o$ and $f_{n_o}$ represent respectively the value of minimum noise and the
corresponding frequency. At the lower-frequency cutoff $f_l$ and the
high-frequency cutoff $f_u$,
$S_n(f)$ can be treated as infinite. One can construct
the noise PSD as follows: 

\begin{table}
{\footnotesize
\centering
\begin{tabular}{|l|c|c|r|c|r|}
\hline
&&&&&\\
Detector & Fit to noise power spectral density & $S_o$ &
$f_{n_o}$ & $f_l$  & $f_u$  \\ 
&& $(Hz)^{-1}$& (Hz) & (Hz) & (Hz) \\
&&&&&\\
\hline
&&&&&\\
LIGO I & $S_o/3\,\left[(f_{n_o}/f)^4+2(f/f_{n_o})^2\right]$ &
$8.0\times10^{-46}$ & 175 & 40 & 1300 \\
&&&&&\\
\hline
&&&&&\\
LIGO II & $S_o/11\,\left\{2(f_{n_o}/f)^{9/2}+9/2[1+(f/f_{n_o})^2]\right\}$ &
$7.9\times10^{-48}$ & 110 & 25 & 900 \\
&&&&&\\
\hline
&&&&&\\
LIGO III & $S_o/5\,\left\{(f_{n_o}/f)^4+2[1+(f/f{n_o})^2]\right\}$ &
$2.3\times10^{-48}$ & 75 & 12 & 625 \\
&&&&&\\
\hline
&&&&&\\
VIRGO & $S_o/4\,\left[290(f_s/f)^5+2(f_{n_o}/f)+1+(f/f_{n_o})^2\right]$ &
$1.1\times10^{-45}$ & 475 & 16 & 2750 \\
&&&&&\\
\hline
&&&&&\\
GEO600 & $S_o0/5\,\left[4(f_{n_o}/f)^{3/2}-2+3(f/f_{n_o})^2\right]$ &
$6.6\times10^{-45}$ & 210 & 40 & 1450 \\
&&&&&\\
\hline
&&&&&\\
TAMA & $S_o/32\,\left\{(f_{n_o}/f)^5+13(f_{n_o}/f)+9[1+(f/f_{n_o})^2]\right\}$ &
$2.4\times10^{-44}$ & 400 & 75 & 3400\\
&&&&&\\
\hline
\end{tabular} }
\vspace{0.30cm}
\caption{Analytical fits to noise power spectral densities $S_n(f)$ of ground 
based interferometers.}
\label{table:noise}
\end{table}

\par A GW detector output represents a dimensionless data train,
say $x(t)$. In the absence of any GW signal. The detector 
output is just an instance of noise $n(t),$ that is, $x(t)=n(t).$ 
The noise auto-correlation function $\mathrm{c}$ is defined as
\begin{equation}
\mathrm{c}(t_1,t_2) \equiv \langle n(t_1) n(t_2) \rangle ,
\label{eq:noise correlation}
\end{equation}
where $\langle \; \; \rangle $ represents the average over an ensemble of noise realisations.
In general, $\mathrm{c}$ depends both on $t_1$ and $t_2.$ However, if the detector
output is a stationary noise process, i.e. its performance is, statistically
speaking, independent of time, then $\mathrm{c}$ depends only on 
$\tau\equiv t_2 - t_1.$ We shall, furthermore, assume that $\mathrm{c}(\tau)
= \mathrm{c}(-\tau).$ 
For data from real detectors the above average can be replaced by a time
average under the assumption of ergodicity: 
\begin{equation}
\mathrm{c}(\tau) = \frac{1}{T} \int_{-T/2}^{T/2} n(t) n(t-\tau) dt.
\end{equation}

The assumption of stationarity is not strictly valid in the case of real GW
detectors; however, if their performance does not vary greatly over 
time scales much larger than typical observation time scales, stationarity
could be used as a working rule.  While this may be good enough in the case
of binary inspiral and coalescence searches, it is a matter of concern for
the observation of continuous and stochastic GW. Under such an assumption the 
one-sided noise PSD, defined only at
positive frequencies, is the Fourier transform (FT) of the noise 
auto-correlation function:
\begin{eqnarray}
S_n(f) & \equiv & \frac{1}{2} \int_{-\infty}^{\infty} \mathrm{c}(\tau) 
e^{- 2\pi i f \tau} d\tau,\ \ f\ge 0, \nonumber\\
       & \equiv & 0, \ \ f<0,
\label{eq:psd1}
\end{eqnarray}
where a factor of 1/2 is included by convention because it has been 
assumed that $\mathrm{c}(\tau)$ is an even function. This equation
implies that $S_n(f)$ is real. It is straightforward to show that
\begin {equation}
\langle\tilde n(f)\tilde n^*(f')\rangle = S_n(f) \delta (f-f'),
\label{eq:psdinfourier}
\end {equation}
where $\tilde n(f)$ represents the Fourier transform of $n(t)$
and $^*$ denotes complex conjugation.   
The above identity implies that $S_n(f)$ is positive definite.

\par The autocorrelation function $\mathrm{c}(\tau)$ at $\tau=0$ can be expressed
as an integral over $S_n(f).$ Indeed, it is easy to see that
\begin{equation}
\langle n^2(t) \rangle = 2 \int_0^{\infty} S_n(f) df.
\label{eq:psd3}
\end{equation}
The above equation justifies the name {\it power spectral density} given to
$S_n(f).$ 
It is obvious that $S_n(f)$ has dimensions of time but it is conventional
to use, instead, the dimensions of Hz$^{-1}$ since it is a quantity defined in the
frequency domain. The square-root of $S_n(f)$ is 
the noise amplitude, $\sqrt{S_n(f)},$ and has dimensions of {\it per root Hz}.
It is often useful to define the dimensionless quantity 
$h_n^2(f) \equiv fS_n(f),$ called the {\it effective noise.} 
In GW interferometer literature one also comes 
across the {\it displacement noise} or {\it strain noise} defined 
as $h_{l_o} (f) \equiv l_o h_n(f),$ and the corresponding noise spectrum
$S_{l_o}(f) \equiv l_o^2 S_n(f),$ where $l_o$ is the arm length of the 
interferometer. The displacement noise gives the smallest strain 
$\delta l/l_o$ in the arms of an interferometer which can be measured at a 
given frequency.

\subsubsection{Sensitivity vs source amplitudes}
One compares the GW amplitudes of astronomical 
sources with the instrumental sensitivity and assesses
what sort of sources will be observable in the following way.
Firstly, as comparisons are almost always made in the 
frequency-domain it is important to note that the Fourier component 
$\tilde h(f)$
of a deterministic signal $h(t)$ has dimensions of Hz$^{-1}$ and 
the quantity $f |\tilde h(f)|,$ is dimensionless. It is this last
quantity that should be compared with $h_n(f)$ to deduce the strength of a
source relative to detector noise.  Secondly, it is quite common also to 
compare the amplitude spectrum per logarithmic bin of a source, 
$\sqrt {f} |\tilde h(f)|,$ with the amplitude spectrum of noise, 
$\sqrt {S_n(f)},$ 
both of which have dimensions of per root Hz. 
For monochromatic sources, one compares the effective noise in a long 
integration period with the expected ``instantaneous'' amplitudes 
in the following way: A monotonic wave of frequency $f_o$ observed 
for a time $T_o$ is simply a narrow line in a frequency bin
of width $\Delta f\equiv 1/T_o$ around $f_o.$ The noise in this
bin is $S_n(f) \Delta f=S_n(f)/T_o.$ 
Thus the SNR after a period of 
observation $T_o$ is 
\begin {equation}
\frac{S}{N} = \frac {h_o} {\sqrt {S_n(f_o)/T_o}}.
\label{eq:snrcw1}
\end {equation}
One, therefore, computes this dimensionless noise spectrum for
a given duration of observation, $S_n(f)/T_o,$  to assess the 
detectability of  a continuous GW. 
Surely, if the observation time is $T_o$ then
the total {\it energy} (that is, the integrated 
power spectrum) of both the signal and noise must increase in 
proportion to $T_o^2.$ Then how does the SNR for a continuous 
wave improve with the duration of observation? 
The point is that while the signal energy is all concentrated in
one bin, the noise is distributed over the entire frequency band. 
As $T_o$ increases, the frequency resolution improves as $1/T_o$ and
the number of frequency bins increase in proportion to $T_o.$
Consequently, the noise intensity {\it per frequency bin} increases only as 
$T_o.$ Now, the signal intensity is concentrated in just one bin since the 
signal is assumed to be monochromatic. Therefore, the power SNR
increases as $T_o,$ or, the amplitude SNR increases as $\sqrt{T_o}.$

\vspace{0.65cm}

\section{Noises in the earth based interferometric detectors}

As mentioned in the earlier section the performance of GW detector
is characterised by the one-sided PSD. The sensitivity of ground based
detectors is limited at
frequencies less than a Hertz by the time-varying local gravitational 
field caused by a variety of different noise sources, e.g. low frequency 
seismic vibrations, density variation in the atmosphere due to winds, etc.
Thus, for data analysis purposes, the noise PSD is assumed to be essentially
infinite below a certain lower cutoff $f_s.$ The effective noise
$h_n(f)$ expected in these detectors is equal to $\sqrt{fS_n(f)}$. Type of
noises that one have to fight to detect the signals are:\\

\par {\bf Ground vibration:}  External mechanical vibrations  
are a problem for bar detectors, too, but are more serious for 
interferometers, not least because interferometers bounce light back
and forth between the mirrors, and so each reflection introduces
further vibrational noise.  Suspension/isolation systems are based on
pendula. A pendulum is a good mechanical filter for frequencies above
its natural frequency.  By hanging the mirrors on pendula of perhaps
0.5~m length, one achieves filtering below a few Hz.  Since the
spectrum of ground noise falls at higher frequencies, this provides
suitable isolation. But these systems can be very sophisticated; the
GEO600 detector has a three-stage pendulum and other vibration
isolation components. The most ambitious isolation system is being
developed for the VIRGO detector.\\

\par {\bf Thermal noise:} Vibrations of the mirrors and of the 
suspending pendulum can mask gravitational waves. As with vibrational
noise, this is increased by the bouncing of the light between the
mirrors.  Unlike bars, interferometers measure only at frequencies far
from the resonant frequency, where the amplitude of vibration is
smaller. Thus, the pendulum suspensions have thermal noise at a few
Hz, but measurements will be made above 20 or 30~Hz in the first
detectors. Internal vibrations of the mirrors have natural frequencies
of several kilohertz. By ensuring that both kinds of oscillations have
very high $\mathrm{Q}$, one can confine most of the vibration energy to a small
bandwidth around the resonant frequency, so that at the measurement
frequencies the vibration amplitudes are small. This allows
interferometers to operate at room temperature.  But mechanical $\mathrm{Q}$s
of $10^7$ or higher are required, and this is technically demanding.\\

\par {\bf Shot noise:}  The photons that are used to do interferometry 
are quantized, and so they arrive at random and make random
fluctuations in the light intensity that can look like a gravitational
wave signal.  The more photons one uses, the smoother will be the
interference signal.  As a random process, the error improves with the
square-root of the number $\mathrm{N}$ of photons.  Using infrared light with a
wavelength $\lambda\sim 1\;\mu\rm m$, one can expect to measure to an
accuracy of
\[\bigtriangleup \mathrm{L}_{shot} \sim \lambda/(2\pi\sqrt{\mathrm{N}})\]
To measure at a frequency $f$, one has to make at least $2f$ measurements 
per second i.e. one can accumulate photons for a time $1/2f$.  
With light power $\mathrm{P}$ and GW amplitude $h$ one gets $\mathrm{N} =
\mathrm{P}/(hc/\lambda)/(2f)$
photons.  In order that $\bigtriangleup 
\mathrm{L}_{shot}$ should be below $10^{-16}$~m one needs large light power, 
far beyond the output of any continuous laser.

\par Light-recycling techniques overcome this problem by using light
efficiently.  An interferometer actually has two places where light
leaves.  One is where the interference is measured. The other goes
back towards the input laser.  Normally one arranges that no light
goes to the interference sensor, so that only when a gravitational
wave passes does a signal register there. This means that all the
light normally returns to the mirror, apart from small losses at the
mirrors.  Since mirrors are of good quality, only one part in $10^3$
or less of the light is lost during a 1~ms storage time. By placing a
power-recycling mirror in front of the laser, one can reflect this
wasted light back in, allowing power to build up in the arms until the
laser merely resupplies the mirror losses. This can dramatically
reduce the power requirement for the laser.  The first interferometers
will work with laser powers of 5-10~W.  This is attainable with modern
laser technology.  \\

\par {\bf Quantum effects:} Shot noise is a quantum noise, but in addition 
there are effects like bar detectors face: zero-point vibrations of
mirror surfaces and so on.  These are small compared to present
operating limits of detectors, but they may become important in 5
years or so.  Practical schemes to reduce this noise have already been
demonstrated in principle, but they need to be improved considerably.
They can be reduced by making the mirror masses large, since the
amplitude of vibration scales inversely as the square-root of the
mass. \\

\par {\bf Gravity gradient noise:} One noise which cannot be screened out is 
that arises due to changes in the local Newtonian gravitational field on the
time scale of the measurements. A gravitational wave detector will
respond to tidal forces from local sources just as well as to
gravitational waves.  Environmental noise comes not only from man-made
sources, but even more importantly from natural ones: seismic waves
are accompanied by changes in the gravitational field, and changes in
air pressure are accompanied by changes in air density.  The spectrum
falls steeply with increasing frequency, so for first-generation
interferometers this will not be a problem, but it may limit the
performance of detectors a decade from now. And it is the primary
reason that detecting gravitational waves in the low-frequency band
around 1~mHz must be done in space.

\vspace{0.65cm}

\section{Matched filtering and optimal signal-to-noise ratio}
\label{sec:matched filtering}
Matched filtering is a data analysis
technique that efficiently searches for a signal of known shape 
buried in noisy data (Helstrom, 1968). The technique
consists in correlating the noisy output of each interferometer data with a
set of theoretical waveform templates. If the maximum
of the correlation exceeds a preset threshold, a detection is
announced. However, even if a signal is present in the data, the
accompanying noise can mask its presence by preventing such crossing
of the threshold in some cases. Thus a signal can be detected with
only a certain probability, called its detection probability. Also,
there will be a non-zero probability, called the false alarm
probability, of a false detection due to noise alone. A formalism to choose
the optimal set of templates using a criterion was first given by
Sathyaprakash and Dhurandhar (1991) and Dhurandhar and Sathyaprakash (1994) 
known as the S-D formalism. 

\par These search templates will be
discrete subset of a continuous, multi parameter family, each of
which approximate a possible signal. Given a signal $h(t)$ buried in noise
$n(t),$ the task is to find an `optimal' template $q(t)$ that 
would produce, on the average, the best possible SNR. The enhancement in the
signal-to-noise ratio not only increases the number of detectable
events but, more importantly, it also allows a more accurate
determination of signal parameters---the error in the estimation of
parameter being inversely proportional to the signal-to-noise ratio.

\par If the functional form of the template is identical to that
of the signal, the mean signal-to-noise ratio in the presence of a
signal is highest possible for any linear data processing technique,
which is why matched filtering is also known as optimal filtering
(Helstrom, 1968). A {\it matched filter\/}, in the frequency domain, is a
best-guess template of the expected signal waveform divided by the
interferometer's spectral noise density. 

\par In order to take full advantage of matched filtering it is essential
that the phase of the waveform be known to a very high degree of accuracy
to accumulate the SNR (Cutler et al. 1993). This is referred to as
coherent integration of the signal, where the SNR grows $\sqrt{T_o}$. 
This is in contrast with incoherent methods, such as {\it stacking\/} and
{\it tracking\/} (Brady, et. al. 1998, 2000), where one disregards the
phase information and the SNR does not accumulate as quickly. {\it Tracking \/}
involves tracking of lines in the time-frequency plane built from the FT
of short (around 40 minutes long) stretches of data. The other
{\it stacking \/} involves dividing the data into shorter (around a day long) 
stretches, searching each stretch for signals, and enhancing the detectability
by incoherently summing the FT of data stretches.

\par There are several questions that must be answered in order to determine
the feasibility of a matched filtering search strategy and, if
feasible, to implement it. Which parameters significantly affect
the wave form ? How should the spacing of the template parameters 
(lattice points) be chosen? Is there a parameterization that is in
some sense preferred by the template waveforms? How many templates
are needed to cover a given region of interest in the parameter
space and how much computing power and memory will it cost to process
the data through them? In the case of modulated Pulsar signal we must
also ask what approximation to the true wave form is good enough?

\par Ambiguity function, well known in
statistical theory of signal detection (Helstrom, 1968), is a very
powerful tool in signal analysis. It helps to access the number of
templates required to span the parameter space of the signal. However,
Apostolatos (1995) defined a term called {\it Fitting Factor \/} $(FF)$ which
quantitatively 
describes the closeness of the true signals to the template manifold in terms
of the reduction in SNR due to cross correlation of a signal outside the
manifold with all the templates lying inside the manifold.  

\vspace{0.65cm}

\subsubsection{Fitting factor}
\label{sec:ff}
One of the standard measure for deciding what class of wave form is good enough
is the {\it Fitting Factor \/}  
and has already  been used by various earlier authors
(Sathyaprakash and Dhurandhar, 1991 and Dhurandhar and Sathyaprakash, 1994) as a measure  of
adequateness of a template family, but previous discussions have
not shown explicitly how this intuitively well-formed function is
connected with detectability. Previously $FF$ has sometimes been called
the ``correlation", a number that is often used in so many different
ways. The name comes from the fact that, by maximising the quantity
FF, the parameters of the templates
 can be adjusted to best fit the
true waveform.

\par Let us consider a detector receiving almost monochromatic 
GW signals from a pulsar. The signal will be Doppler
modulated due to the motion of Earth and stars, represented by
$S(t)$. The detector's output $X(t)$ of the detector will be 
\begin{equation}
X(t) = S(t) + n(t)
\end{equation}

\noindent If the search template is the exact waveform $h(t)$, then we would
achieve
 the highest possible signal-to-noise ratio given as 
\begin{equation}
\left(\frac{S}{N}\right)_{max} = \langle S(f)|S(f)\rangle^2
\end{equation}

\noindent where the inner product of two waveforms $\langle h_1|h_2\rangle $ is
defined as
\begin{eqnarray}
\langle h_1|h_2\rangle & =& 2\int_0^\infty \frac{\tilde{h}_1^*(f)\tilde{h}_2(f)
+ \tilde{h}_1(f)\tilde{h}_2^*(f)}{S_n(f)}df \nonumber \\
 & = &
4\int_0^\infty \frac{\tilde{h}_1^*(f)\tilde{h}_2(f)}{S_n(f)}df
\label{eq:ip}
\end{eqnarray}

\par The SNR ratio will be reduced below $(S/N)_{max}$ whenever the
template is not the exact waveform but some other approximate one. If
search is performed with some family of templates $S_T(f;\vec{\xi})$
for a signal vector $\vec{\xi}$ (The signal is considered a vector in parameter
space). Then the SNR will be given by
\begin{eqnarray}
\left(\frac{S}{N}\right) & = & \stackrel{max}{\vec{\xi}}\,\frac{\langle X(f)|
S_T(f;\vec{\xi})\rangle}{rms\langle n|S_T(f;\vec{\xi})\rangle}
\end{eqnarray}

For an ensemble of realizations of the detector noise, the expectation
values
 $\langle n|S_T(f;\vec{\xi})\rangle$ and $\langle
n|S_T(f;\vec{\xi})\rangle
\langle n|S_T(f;\vec{\xi})\rangle$ are zero and $\langle S_T(f;\vec{\xi})| 
S_T(f;\vec{\xi})\rangle$, respectively. Thus the ensemble-averaged 
signal-to-noise ratio turns out to be
\begin{eqnarray}
\left(\frac{S}{N}\right) & = &
\stackrel{max}{\vec{\xi}}\,\frac{\langle S(f)|
S_T(f;\vec{\xi})\rangle}{\sqrt{\langle
S_T(f;\vec{\xi})|S_T(f;\vec{\xi})\rangle}}\nonumber \\
 & = &\left[
\stackrel{max}{\vec{\xi}}\,\frac{\langle S(f)| S_T(f;\vec{\xi})
\rangle }{\sqrt{\langle S_T(f;\vec{\xi})|S_T(f;\vec{\xi})\rangle
\langle S(f)|S(f)\rangle}}
 \right]\left(\frac{S}{N}\right)_{max}
\end{eqnarray}
The name {\it Fitting Factor\/} is given to the reduction in
signal-to-noise ratio that results because of using the chosen template family,
rather than the true signal $h$, in the search:
\begin{eqnarray}
FF(\vec{\xi}) & = &\stackrel{max}{\vec{\xi}}\,\frac{\langle S(f)|
S_T(f;\vec{\xi})\rangle}{\sqrt{\langle
S_T(f;\vec{\xi})|S_T(f;\vec{\xi})\rangle}}\nonumber \\
 & =&
\stackrel{max}{\vec{\xi}}\,\frac{\langle S(f)| S_T(f;\vec{\xi})
\rangle}{\sqrt{\langle S_T(f;\vec{\xi})|S_T(f;\vec{\xi})\rangle\langle
S(f)|S(f)\rangle}}
\label{eq:ff}
\end{eqnarray}
Therefore, if the family of search template used at the detection
stage does not
 contain the true signal waveform, then the reduction in the 
signal-to-noise ratio will be given by
\begin{equation}
\left(\frac{S}{N}\right) = FF(\vec{\xi}) \times \left(\frac{S}{N}\right)_{max}
\end{equation}

\par The spacing of the grid of filters is decided by the fractional loss
due to to
 the imperfect match that can be tolerated. Given the
parameters space that one
 needs to scan, it is then easy to estimate
the total number of filters required
 to carry out the search for the
signal. The signal strength scales inversely with
 distance, the
fraction of event rate retained is approximately $FF^3$. Therefore
 it
has become conventional to regard $FF = 0.9$ as the lowest acceptable
$FF$
 value for some template family to be considered adequate. Since
a reduction in
 SNR by $10\%$ means a $27\%$ loss in the event
rate. On the other hand a $10\%$
 reduction in SNR is equivalent to
roughly $10\%$ shortening of the detectors arms.

\vspace{0.65cm}

\section{Computational costs}
\label{sec:compute costs}
Matched filtering places stringent demands on
 the knowledge of the
signal's phase evolution which depends on two things:
 (1) our
modelling of the signal and (2) the parameters characterising
 the
signal. If our signal model is inaccurate or if the signal's
parameters
 are unknown, there could be a loss in the SNR extracted.
For instance, in the case of inspiral signals, a mismatch of one cycle
in $10^4$ cycles leads to a drop in the SNR by more than a factor
of two, losing a factor of eight in the number of potentially
detectable events. Recall that the SNR is inversely proportional to
the distance to a source; thus an SNR loss by a factor $a$ will
reduce the
 span of a detector by the same factor, resulting in a
decrease in the volume of observation, and hence the number of
events, by a factor $a^3.$ Moreover, since the parameters of a
signal will not be known in advance,
 it is necessary to filter the
data with a family of templates located
 at various points in the
parameter space e.g., placed on a lattice such that
 any signal
will lie close enough to at least one of the templates
 to have a good
cross-correlation with that template.
 The number of such templates is
typically very large. This places
 a great demand on the computational
resources needed
 to make an on-line search.

\par The search problem for continuous waves from spinning neutron stars
is the most computer-intensive job in gravitational wave data
analysis. Today, there is little hope that all-sky searches lasting
for a year or more, can be made. It is easy to see why this is
 such
an intensive job: Firstly, the data has to be collected continuously
for months together and at a good sensitivity. 
Secondly, though a neutron star emits a
periodic signal
 in its rest frame, save for the neutron star
spin-down which indeed
 induces some modulation in the waveform,
because of Earth's acceleration relative to the source, the detector
does not see a periodic wave. The wave is both frequency and
amplitude modulated. One can, fortunately, de-modulate these effects
since Earth's motion is known quite accurately, and hence recover the
original periodic signal. But de-modulation requires a
knowledge of the source's direction
 and its frequency, which are
unknown in a blind search. The angular resolution one obtains in a
year's integration is $\Delta \theta = \lambda/D,$ where $\lambda$ is
the wave length of radiation and $D$ is the baseline of the detector
in a year's integration, namely 1 A.U. Thus, for $f=1$~kHz we have
$\Delta\theta=10^{-6}$~rad or about two arcsec. Now, assuming that
the source may be in any one of the 4 arcsec$^2$ patches on the sky
we get the number of patches in the sky for which we will have to
try out a de-modulation correction
 to be $4 \pi/(\Delta \theta)^2 = 4
\pi 10^{12}.$ It is quite an impossible task to apply Doppler
de-modulation to the detector output for each of these $\sim 10^{13}$
patches and compute as many Fourier transforms. 

\par One, therefore,
asks the question given a compute power what is the
 best possible
search one can do? Is there any advantage in going from
 a one-step
search to a two or multi-step hierarchical search? What
 about
directional searches? These are some of the problems for which
 we
have some answer; but a great deal of work is needed and
 is currently
under progress, to improve and optimise search algorithms.

\par The differential geometric studies shows that with a TFLOPS-class computer, 
the number of days of data can be analysed on-line and we can carry out a blind search. 
Unfortunately, the longest
 data we can
integrate on-line, for neutron stars with spin frequencies
 $f\le
100$~Hz and spin-down rates less than 1000 years, is about 18 days.
This yields a SNR lower by a factor of 5 as compared to a year's
worth of observing. On-line searches for neutron stars with $f\le
500$~Hz (largest observed
 frequencies of millisecond pulsars) and
spin-down rates
 of 40 years (shortest observed spin-down rates), can
only be made for a data set lasting for a duration of 20 hours or
less. If source's position is
 known in advance, but not its
frequency, then one can carry out an on-line
 search, again with a
TFLOPS-class computer, for the frequency of the source
 in a data set
that is worth 3 months long. This is good news since there
 are many
known pulsars and X-ray binary systems that are potential sources
 of
radiation. In addition, the obvious targeted search locations are
 the
centre of the Galaxy and globular clusters.

\par There have been efforts to study the effectualness of a two-step
hierarchical method for a blind search. Here the basic
idea is to construct Fourier transforms of data sets of duration
smaller than the period in which Doppler modulations will be
important and to stack spectral densities obtained in this way and to
add them all up. This is an incoherent way of building the signal
since one adds spectral
 densities that have no phase
information. Therefore, 
 one gains in SNR less than what an optimal
matched filtering method
 is able to achieve.  However, this does not
matter since
 (i) the targeted SNR's are quite high $\sim 10$ and (ii)
candidate
 events can always be followed-up using coherent
integration
 methods. These methods afford an on-line all-sky blind
search for continuous gravitational waves for a
period of 4 months or less using
 a 20 GFLOPS computer. Detector
groups are planning to build computers
 of this kind to aid in their
search for continuous GW.

\vspace{0.65cm}

\section{Detection criteria}
A signal needs to be above the noise experienced in the instrument and 
environment. However, this alone is insufficient to establish it as a 
gravitational wave in the terrestrial detectors. The most satisfying 
circumstance is that a gravitational wave observation be made in a set of 
widely distributed detectors [the Gravitational Wave Network (GWN)] and the 
resorted wave forms allow the solutions for the polarization of the wave and 
the position of the source. Armed with this information an electromagnetic 
(or neutrino) search could be attempted in the error circle of the
gravitational wave detection: a time honored approach bringing 
gravitational wave of observation into the main stream of Astrophysics. The 
strategy would apply to all classes of sources: impulsive chirps, 
quasiperiodic and periodic.

\par The confident detection of impulsive sources is most difficult, while 
the periodic and quasiperiodic detection will have confidence limits based on 
quasi-stationary system noise (the signals last long enough to take a 
meaningful sample of the noise spectrum), the impulsive signals
especially if rare, will be particularly dependent of the non-Gaussian 
component of the noise; the noise most difficult to reduce and control in a 
signal detectors. The technique of multiple coincidence of several detectors 
is one of the best means to gain confidence. The coincidence 
must occur within a time window to permit a consistent solution for a location in 
the sky. If the general character of the source can be guessed in advance 
(for example, a binary coalescence chirp, or a black-hole normal mode 
oscillation), the signal is filtered prior to the coincidence 
measurement to improve the sensitivity. The more detectors involved, the 
greater the confidence assigned to the detection.

\par There is still a possibility of coincidence due to environmental or 
anthropogenic causes. The various sites throughout the world are far enough 
apart that most environmental perturbations should not correlate between them. 
The acoustic noise, the seismic noise, and the power line 
(especially if the network includes detectors in a different power grids and 
significantly different time zones) will be uncorrelated. There are 
correlations in the magnetic field fluctuations (thunderstorms) and in radio 
frequency emittions. As part of the detection strategy a large number of 
environmental parameters will be measured along with the gravitational wave 
signals at each site. One of the requirements for the authenticity of 
impulsive sources will be the lack of correlation with environmental 
perturbations and other ancillary internal signals developed to 
monitor the performance of the instruments.

\chapter{Data analysis of continuous gravitational wave: Fourier transform-I}
\section{Introduction}
\indent The first generation of long-baseline laser interferometers 
and ultra cryogenic bar detectors will start collecting data very soon.
The network of detectors will not only confirm the existence of GWs but will
also yield information about its important parameters viz; amplitude,
frequency and phase. In addition, a careful monitoring of the signal will also
provide
the information about the structure and the dynamics of its source. At the
present stage, the data analysis depends largely upon the study of the expected
characteristic of its potential sources and the waveforms. The majority of the
experimental searches are focussed on the detection of burst 
and {\it chirp\/} signals. However, the interest in the data analysis for 
CGW signals is growing. A prime example of sources of this type is
a spinning neutron star. Many research groups around the globe are working
extensively on the data analysis for spinning neutron stars
[Jaranwoski, et.al (1998, 1999, 2000), Brady et. al. (1998, 2000),
Kr\'olak (1999)]. 

\par Our Galaxy is expected to have, on logistic grounds, at least $10^8$
spinning neutron stars; with a birth rate of one in every 30 years.
However, a very insignificant fraction of them are observed---the
number of known
pulsars has gone up to $1100.$ There are compelling arguments that
nearby millisecond pulsars can provide a \underline{detectable} source of CGW.
Hence, there is a strong case for searching all sky for the presence of neutron
stars with the philosophy that they are emitting GWs but otherwise unknown 
optically. This study will help us to understand 
the pulsar population in our Galaxy. 

\par The detection of GW signals in the output has its own 
problems, not the least of which is the sheer volume of data analysis. Bar 
detectors have essentially the same problems as interferometers 
in reference to CGW sources. The possibility that these 
detectors are capable to search for different 
kinds of signals makes them very versatile, but also complicates the data
analysis. Each detector produces a single data stream that may contain
many kinds of signals. Detectors don't point, but rather sweep their broad 
quadrupolar beam pattern across the sky as the earth moves. So possible 
sources could be anywhere on the sky and accordingly the data analysis
algorithms need to accommodate signals from any arbitrary location of its
source.

\par In this and the next Chapter we present analysis of FT of the output
data of a ground based laser interferometer. The output data has prominently
dominant broad band noise and the signal is to be extracted out of it. For
this, one has to enhance SNR. This is achieved by analyzing long
observation time data as SNR is directly proportional to the square root of
observation time $\sqrt {T_o}$.
However, in a data for long duration, the monochromatic signal gets Doppler
modulated due to (i) orbital motions of Earth around Sun and (ii) spin of 
Earth. The Frequency modulation (FM) will spread the signal in a very large number
of bins depending on the source location and the frequency. In addition 
there is Amplitude modulation (AM). As we will see in the sequel the
amplitude of the detector output consists of simple harmonic terms with
frequencies $w_{rot}$ and $2 w_{rot}$ where, $w_{rot}$ stands for angular
rotational frequency of Earth. Accordingly, the
AM results in splitting of FT into frequencies $\pm w_{rot}$
and $\pm 2 w_{rot}$.

\par In the next section, we present the noise free response of the laser
interferometric detector and obtain the explicit beam pattern functions. In
section 3, we discuss the Doppler effect and obtain the FT of the FM signal
for arbitrary source and detector locations taking into account Earth's
rotational motion about its axis and its revolution around Sun. In
section 4, the FT of the Doppler modulated complete response
of the detector has been obtained. In the last section we discuss the results
of the Chapter.

\vspace{0.65cm}

\section{The noise free response of detector: Beam pattern and amplitude
modulation}
\label{sec:nfr}
Let a plane GW falls on a laser interferometer and produces changes in the
arms of the detector. In order to express these changes quantitatively we would require
to specify the wave and the detector. Let $X Y Z$ and $x y z$ represent
respective frames characterising wave and detector. We assume the direction
of propagation of the wave to be the $Z$ axis and the vertical at place of
detector to be the $z$ axis [see Fig.~(\ref{fig:od})]. In view of our discussion in subsection~(\ref{sec:laser}),
the difference of the changes $\delta l$ in the arm lengths of the detector is
given via

\begin{equation}
\label{eq:deltal}
R(t) = \frac{\delta l}{l_o} = - \sin 2\Omega\left[\left( {\bf A}^x_X{\bf A}^y_X -
{\bf A}^x_Y{\bf A}^y_Y\right)h_+ +
\left( {\bf A}^x_X{\bf A}^y_Y + {\bf A}^x_Y{\bf A}^y_X\right) h_\times\right]
\end{equation}

\begin{figure}
\centering

\vspace{6.1cm}

\caption{Orientation of the detector.}
\label{fig:od}
\end{figure}

\begin{figure}
\centering

\vspace{6.1cm}

\caption{Earth frame $(x' , y' , z').$}
\label{fig:earth}
\end{figure}

\begin{figure}
\centering

\vspace{6.1cm}

\caption{Solar System Barycentre frame $(X' , Y' , Z').$}
\label{fig:ssb}
\end{figure}

\vspace{0.5cm}

\noindent where $l_o$ is the normal length of the arms of the detector and $2\Omega$
express the angle between them (Schutz and Tinto, 1987). The matrix $\left( {\bf A}^j_K \right)$
represents the transformation expressing the rotations to bring the wave
frame $(X , Y , Z)$ to the detector frame $(x , y , z)$. The direction of the source may be expressed in any of the coordinates
employed in Spherical Astronomy. However, we find it convenient to define it
in Solar System Barycentre (SSB) frame $(X' , Y' , Z')$ [see Fig.~(\ref{fig:ssb})]. This SSB frame is
nothing but
astronomer's ecliptic coordinate system. Let $\theta$ and $\phi$ denote the celestial
colatitude and celestial longitude of the source. These coordinates are related
to right ascension $\bar{\alpha}$  and declination $\bar{\delta}$ of
the source via

\begin{equation}
\left.\begin{array} {rcl}
\vspace{0.2cm}
\cos\theta & = & \sin\bar{\delta}\cos\epsilon - \cos\bar{\delta}\sin
\epsilon\sin\bar{\alpha}\\
\vspace{0.2cm}
\sin\theta\cos\phi & = & \cos\bar{\delta}\cos\bar{\alpha} \\
\sin\theta\sin\phi & = & \sin\bar{\delta}\sin\epsilon + \cos\bar{\delta}
\cos\epsilon\sin\bar{\alpha}
\end{array}\right\}
\vspace{0.3cm}
\end{equation}

\vspace{0.5cm}

\noindent where $\epsilon$ represents obliquity of the ecliptic. We choose $x$ axis as the bisector of the angle between the arms of the
detector. At this stage the orientation of the detector in the horizontal plane is arbitrary.
It is assigned with the help of the angle $\gamma$ which $x$ axis makes with
the local meridian. The location of the detector on earth is characterised by
the angle $\alpha$, colatitude and $\beta$, the local sidereal time expressed
in radians.
The transformation matrix $\left( {\bf A}^j_K\right)$ may be expressed as
\begin{equation}
{\bf A\/} = {\bf DCB \/}
\end{equation}

\noindent where

\noindent ${\bf B\/}:$ rotation required to bring $X Y Z$ to $X' Y' Z'$\\
${\bf C\/}:$ rotation required to bring $X' Y' Z'$ to $x' y' z'$\\
${\bf D\/}:$ rotation required to bring $x' y' z'$ to $x y z$\\

\noindent Here $x' y' z'$ represents the frame associated
with earth [see Figs.~(\ref{fig:od}-\ref{fig:earth})]. The Euler angles  
defining the corresponding rotation matrices (Goldstein, 1980) are given via

\begin{equation}
\left.\begin{array}{ccl}
\vspace{0.2cm}
{\bf B \/}& : & (\theta , \phi , \psi )\\
\vspace{0.2cm}
{\bf C \/}& : & (0 , \epsilon , 0 )\\
{\bf D \/}& : & (\alpha , \beta + \pi /2 , \gamma - \pi /2 )
\end{array}\right\}
\end{equation}

\vspace{0.5cm}

\noindent where $\psi$ is a measure of the polarisation of the wave. Let us
express Eq.~(\ref{eq:deltal}) as
\begin{equation}
\label{eq:rt}
R(t) = \frac{\delta l}{l_o} = - \sin 2\Omega \left[ F_+h_+ + F_\times h_\times \right]
\end{equation}

\vspace{0.1cm}

The functions $F_+$ and $F_\times$ involve the angles $\theta , \phi , \psi ,
\epsilon  , \alpha , \beta , \gamma$ and express the effect of the interaction of
the wave and detector. These are called antenna or beam patterns. After a
straight forward calculation one obtains (Jotania and Dhurandhar, 1994)

\begin{eqnarray}
\label{eq:fpk}
F_+(t)& =& \frac{1}{2}\left[ \left\{ (\cos\psi\cos\phi -
\cos\theta\sin\phi\sin\psi )^2
- (\cos\psi\sin\phi + \cos\theta\cos\phi\sin\psi)^2\right\}\right. \nonumber\\
&&\times\,\left\{2(\cos\alpha\cos\beta\cos\gamma - \sin\beta\sin\gamma)(-\cos
\alpha \cos\beta\sin\gamma - \sin\beta\cos\gamma)\right\}\nonumber \\
&&+\, \left\{(-\sin\psi\cos\phi - \cos\theta\sin\phi\cos\psi)^2 -
(\cos\theta\cos
\phi\cos\psi - \sin\psi\sin\phi)^2\right\}\nonumber \\
&&\times\left[\cos^2\epsilon\left\{2(\cos\alpha\sin\beta\cos\gamma +
\cos\beta\sin
\gamma)(\cos\beta\cos\gamma - \cos\alpha\sin\beta\sin\gamma)\right\}
\right.\nonumber \\
& & -\,\sin^2\epsilon\sin^2\alpha\sin 2\gamma + \sin
2\epsilon\left\{(\cos\alpha\sin
\beta\cos\gamma + \cos\beta\sin\gamma )\sin\alpha\sin\gamma \right.\nonumber\\
&&-\, \left.\left.(\cos\beta\cos\gamma - \cos\alpha\sin\beta\sin\gamma
)\sin
\alpha\cos\gamma\right\}\right]\nonumber \\
&& -\,(\sin^2\theta\cos 2\phi)\nonumber \\
&&\times\,\left[\sin^2\epsilon\left\{2(\cos\alpha\sin\beta\cos\gamma +
\cos\beta
\sin\gamma )(\cos\beta\cos\gamma - \cos\alpha\sin\beta\sin\gamma )\right\}
\right.\nonumber \\
&& -\, \cos^2\epsilon\sin^2\alpha\sin 2\gamma - \sin 2\epsilon\left\{
(\cos\alpha
\sin\beta\cos\gamma + \cos\beta\sin\gamma )\sin\alpha\sin\gamma\right. 
\nonumber \\
&&-\, \left.\left.\left.(\cos\beta\cos\gamma -
\cos\alpha\sin\beta\sin\gamma )
\sin\alpha\cos\gamma\right\}\right]\right]\nonumber \\
&&+\,\left\{ (\cos\psi\cos\phi - \cos\theta\sin\phi\sin\psi
)(-\sin\psi\cos\phi
 - \cos\theta\sin\phi\cos\psi)\right. \nonumber
\\
 &&-\,\left.(\cos\psi\sin\phi + \cos\theta\cos\phi\sin\psi
)(\cos\theta\cos\phi\cos\psi
 - \sin\psi\sin\phi )\right\} \nonumber
\\
 &&\times\,\left[\cos\epsilon\left\{ (\cos\beta\cos\gamma -
\cos\alpha\sin\beta
\sin\gamma )(\cos\alpha\cos\beta\cos\gamma - \sin\beta\sin\gamma )
\right.\right.\nonumber \\
&&-\,\left.(\cos \alpha\sin\beta\cos\gamma + \cos\beta\sin\gamma
)(\cos\alpha
\cos\beta\sin\gamma + \sin\beta\cos\gamma )\right\}\nonumber \\
&&+\,\sin\epsilon\left\{ (\cos\alpha\cos\beta\cos\gamma -
\sin\beta\sin\gamma )
 (\sin\alpha\sin\gamma)\right.\nonumber \\
&&+\, \left.\left.(\cos\alpha\cos\beta\sin\gamma + \sin\beta\cos\gamma
)
 (\sin\alpha\cos\gamma )\right\}\right]\nonumber \\
 &&+\,\left\{
(\cos\psi\cos\phi - \cos\theta\sin\phi\sin\psi )(\sin\theta\sin\phi
)\right.\nonumber \\
 &&+\,\left. (\cos\psi\sin\phi +
\cos\theta\cos\phi\sin\psi )(\sin\theta\cos\phi )\right\}\nonumber \\
&&\times\,\left[-\sin\epsilon\left\{ (\cos\beta\cos\gamma -
\cos\alpha\sin\beta
\sin\gamma )(\cos\alpha\cos\beta\cos\gamma - \sin\beta\sin\gamma )\right.
\right.\nonumber \\
&&-\,\left.(\cos \alpha\sin\beta\cos\gamma + \cos\beta\sin\gamma
)(\cos\alpha
\cos\beta\sin\gamma + \sin\beta\cos\gamma )\right\}\nonumber \\
&&+\,\cos\epsilon\left\{ (\cos\alpha\cos\beta\cos\gamma -
\sin\beta\sin\gamma )
 (\sin\alpha\sin\gamma)\right.\nonumber \\
&&+\, \left.\left.(\cos\alpha\cos\beta\sin\gamma + \sin\beta\cos\gamma
)
 (\sin\alpha\cos\gamma\right\}\right]\nonumber \\
 &&+\,\left\{
(-\sin\psi\cos\phi - \cos\theta\sin\phi\cos\psi )(\sin\theta\sin\phi
)\right.\nonumber \\
 &&+\,\left. ( \cos\theta\cos\phi\cos\psi -
\sin\psi\sin\phi )(\sin\theta\cos\phi )\right\}\nonumber \\
&&\times\,\left[ -(\cos\epsilon\sin\epsilon)\left\{
2(\cos\alpha\sin\beta\cos
\gamma + \cos\beta\sin\gamma)\right.\right.\nonumber \\
&&\;\;\left.(\cos\beta\cos\gamma - \cos\alpha\sin\beta\sin\gamma ) +
\sin^2\alpha\sin
 2\gamma \right\}\nonumber \\
 && +\,\cos
2\epsilon\left\{ (\cos\alpha\sin\beta\cos\gamma + \cos\beta\sin\gamma
)
 (\sin\alpha\sin\gamma )\right.\nonumber \\
 &&-\,
\left.\left.(\cos\beta\cos\gamma - \cos\alpha\sin\beta\sin\gamma
)(\sin\alpha\cos\gamma )\right\}\right]\; ;
\end{eqnarray}

\begin{eqnarray}
\label{eq:fck}
F_\times (t)& =& 2(\cos\psi\cos\phi - \cos\theta\sin\phi\sin\psi )
(\cos\psi\sin\phi + \cos\theta\cos\phi\sin\psi ) \nonumber \\
&&\times\,(\cos\alpha\cos\beta\cos\gamma - \sin\beta\sin\gamma )(-\cos
\alpha \cos\beta\sin\gamma - \sin\beta\cos\gamma)\nonumber \\
&&+\, (-\sin\psi\cos\phi - \cos\theta\sin\phi\cos\psi)(\cos\theta\cos
\phi\cos\psi - \sin\psi\sin\phi)\nonumber \\
&&\times\left[\cos^2\epsilon\left\{2(\cos\alpha\sin\beta\cos\gamma +
\cos\beta\sin
\gamma)(\cos\beta\cos\gamma - \cos\alpha\sin\beta\sin\gamma)\right\}\right.
\nonumber \\
& & -\,\sin^2\epsilon\sin^2\alpha\sin 2\gamma + \sin
2\epsilon\left\{(\cos\alpha\sin
\beta\cos\gamma + \cos\beta\sin\gamma )\sin\alpha\sin\gamma \right.\nonumber\\
&&-\, \left.\left.(\cos\beta\cos\gamma - \cos\alpha\sin\beta\sin\gamma
)\sin
\alpha\cos\gamma\right\}\right]\nonumber \\
&& -\,\frac{1}{2}(\sin^2\theta\sin 2\phi)\nonumber \\
&&\times\,\left[\sin^2\epsilon\left\{2(\cos\alpha\sin\beta\cos\gamma +
\cos\beta
\sin\gamma )(\cos\beta\cos\gamma - \cos\alpha\sin\beta\sin\gamma )\right\}
\right.\nonumber \\
&& -\, \cos^2\epsilon\sin^2\alpha\sin 2\gamma - \sin 2\epsilon\left\{
(\cos\alpha
\sin\beta\cos\gamma + \cos\beta\sin\gamma )\sin\alpha\sin\gamma\right. 
\nonumber \\
&&-\, \left.\left.(\cos\beta\cos\gamma - \cos\alpha\sin\beta\sin\gamma
)
\sin\alpha\cos\gamma\right\}\right]\nonumber \\
&&+\,\left\{ (\cos\psi\cos\phi - \cos\theta\sin\phi\sin\psi
)(-\sin\psi\sin\phi
 + \cos\theta\cos\phi\cos\psi)\right. \nonumber
\\
 &&-\,\left.(\cos\psi\sin\phi + \cos\theta\cos\phi\sin\psi
)(\cos\theta\sin\phi\cos\psi
 + \sin\psi\cos\phi )\right\} \nonumber
\\
 &&\times\,\left[\cos\epsilon\left\{ (\cos\beta\cos\gamma -
\cos\alpha\sin\beta
\sin\gamma )(\cos\alpha\cos\beta\cos\gamma - \sin\beta\sin\gamma )\right.
\right.\nonumber \\
&&-\,\left.(\cos \alpha\sin\beta\cos\gamma + \cos\beta\sin\gamma
)(\cos\alpha
\cos\beta\sin\gamma + \sin\beta\cos\gamma )\right\}\nonumber \\
&&+\,\sin\epsilon\left\{ (\cos\alpha\cos\beta\cos\gamma -
\sin\beta\sin\gamma )
 (\sin\alpha\sin\gamma)\right.\nonumber \\
&&+\, \left.\left.(\cos\alpha\cos\beta\sin\gamma + \sin\beta\cos\gamma
)
 (\sin\alpha\cos\gamma )\right\}\right]\nonumber \\
 &&+\,\left\{-
(\cos\psi\cos\phi - \cos\theta\sin\phi\sin\psi )(\sin\theta\cos\phi
)\right.\nonumber \\
 &&+\,\left. (\cos\psi\cos\phi +
\cos\theta\cos\phi\sin\psi )(\sin\theta\sin\phi )\right\}\nonumber \\
&&\times\,\left[-\sin\epsilon\left\{ (\cos\beta\cos\gamma -
\cos\alpha\sin\beta
\sin\gamma )(\cos\alpha\cos\beta\cos\gamma - \sin\beta\sin\gamma )\right.
\right.\nonumber \\
&&-\,\left.(\cos \alpha\sin\beta\cos\gamma + \cos\beta\sin\gamma
)(\cos\alpha
\cos\beta\sin\gamma + \sin\beta\cos\gamma )\right\}\nonumber \\
&&+\,\cos\epsilon\left\{ (\cos\alpha\cos\beta\cos\gamma -
\sin\beta\sin\gamma )
 (\sin\alpha\sin\gamma)\right.\nonumber \\
&&+\, \left.\left.(\cos\alpha\cos\beta\sin\gamma + \sin\beta\cos\gamma
)
 (\sin\alpha\cos\gamma\right\}\right]\nonumber \\
 &&+\,\left\{
(-\sin\psi\sin\phi + \cos\theta\cos\phi\cos\psi )(\sin\theta\sin\phi
)\right.\nonumber \\
 &&+\,\left. ( \cos\theta\sin\phi\cos\psi +
\sin\psi\cos\phi +)(\sin\theta\cos\phi )\right\}\nonumber \\
&&\times\,\left[ -(\cos\epsilon\sin\epsilon)\left\{
2(\cos\alpha\sin\beta\cos
\gamma + \cos\beta\sin\gamma)\right.\right.\nonumber \\
&&\;\;\left.(\cos\beta\cos\gamma - \cos\alpha\sin\beta\sin\gamma ) +
\sin^2\alpha\sin
 2\gamma \right\}\nonumber \\
 && +\,\cos
2\epsilon\left\{ (\cos\alpha\sin\beta\cos\gamma + \cos\beta\sin\gamma
)
 (\sin\alpha\sin\gamma )\right.\nonumber \\
 &&-\,
\left.\left.(\cos\beta\cos\gamma - \cos\alpha\sin\beta\sin\gamma
)(\sin\alpha\cos\gamma )\right\}\right]\; ;
\end{eqnarray}

\vspace{0.5cm}

\noindent It is easy to understand the structure of the above complicated looking
expressions for $F_+$ and $F_{\times}$ by introducing following abbreviations 

\begin{equation}
\left.\begin{array}{lcl}
\vspace{0.2cm}
U &=& \cos \alpha \cos \beta \cos \gamma - \sin \beta \sin \gamma\, , \\
\vspace{0.2cm}
V &= & - \cos \alpha \cos \beta \sin \gamma - \sin \beta \cos \gamma \, ,\\
\vspace{0.2cm}
X &= &\cos \alpha \sin \beta \cos \gamma + \cos \beta \sin\gamma \, ,\\
\vspace{0.2cm}
Y &=& - \cos \alpha \sin \beta \sin \gamma + \cos \beta\cos \gamma \, 
\end{array} \right\}\hspace{4.05cm}
\end{equation}

\begin{equation}
\left.\begin{array}{lcl}
\vspace{0.2cm}
L &= &\cos\psi\cos\phi - \cos\theta\sin\phi\sin\psi\, ,\\
\vspace{0.2cm}
M &=& \cos\psi\sin\phi + cos\theta\cos\phi\sin\psi\, , \\
\vspace{0.2cm}
N &=& - \sin\psi\cos\phi - cos\theta\sin\phi\cos\psi\, ,\\
\vspace{0.2cm}
P& = &-\sin\psi\sin\phi + cos\theta\cos\phi\cos\psi \, ,\\
Q &=&\sin\theta\sin\phi\, ,\qquad R\; =\; \sin\theta\cos\phi \, ,
\end{array} \right\} \hspace{4.0cm}
\end{equation}

\begin{equation}
\left.\begin{array}{lcl}
\vspace{0.2cm}
A &= &2 X Y\cos^2\epsilon - \sin^2\epsilon \sin^2\alpha \sin 2\gamma +
 \sin 2\epsilon ( X \sin \alpha \sin \gamma \\
\vspace{0.2cm}
 && -\, Y \sin \alpha \cos\gamma ) \, ,\\
\vspace{0.2cm}
B& = & 2 X Y\sin^2\epsilon - \cos^2\epsilon \sin^2\alpha \sin 2\gamma -
\sin 2\epsilon ( X \sin \alpha \sin \gamma \\
\vspace{0.2cm}
&&-\, Y \sin \alpha \cos \gamma ) \, ,\\
\vspace{0.2cm}
C& =& \cos\epsilon ( Y U + X V ) + \sin\epsilon ( U \sin \alpha \sin
\gamma  - V\sin\alpha \cos \gamma )\, ,\\
\vspace{0.2cm}
D &= &- \sin\epsilon ( Y U + X V ) + \cos\epsilon ( U \sin \alpha
\sin\gamma - V \sin\alpha \cos \gamma ]\, ,\\
\vspace{0.2cm}
E& = &- 2 X Y \cos\epsilon \sin\epsilon - \cos\epsilon\sin\epsilon
\sin^2\alpha\sin 2\gamma   \\
&& +\,\cos 2\epsilon ( X \sin\alpha\sin\gamma - Y \sin \alpha \cos\gamma )
\end{array} \right\}
\end{equation}

\vspace{0.5cm}

\noindent One obtains:
\begin{eqnarray} F_+(t) & = &{1\over 2}\left[ 2 ( L^2 - M^2 ) U V + (
N^2 - P^2 ) A \: +  ( Q^2 - R^2) B \right] + ( L N - M P ) C
\nonumber\\
\label{eq:fplust}
 &&+\, ( L Q + M R ) D + ( N Q + P R ) E\, ,\\
F_\times (t)&
= & 2 L M U V + N P A - {1\over 2} B \sin^2\theta\sin 2\phi\: +
 ( L P
+ M N ) C \nonumber\\
 &&+\, ( M Q - L R ) D + ( P Q - N R ) E 
\label{eq:fcrosst}
\end{eqnarray}

\vspace{0.1cm}

\noindent The compactification achieved here is because of the fact that these 
abbreviations find places in the transformation matrices as:

\begin{equation}
{\bf B \/}= \left( \begin{array}{ccc}
L &   N & Q \\
M & P & - R\\
\sin\theta\sin\psi & \sin\theta\cos\psi & \cos\theta \end{array}\right) 
\end{equation}

\begin{equation}
{\bf C \/}= \left( \begin{array}{ccc}
1 &   0 & 0 \\
0 & \cos\epsilon & \sin\epsilon\\
0 & - \sin\epsilon & \cos\epsilon\end{array}\right) 
\end{equation}

\begin{equation}
{\bf D \/}= \left( \begin{array}{ccc}
U &   V & \sin\alpha\cos\beta \\
X & Y & \sin\alpha\sin\beta\\
- \sin\alpha\cos\gamma & \sin\alpha\sin\gamma & \cos\alpha \end{array}\right)\; ;
\end{equation}

\vspace{0.5cm}

\noindent After algebraic manipulation Eqs.~(\ref{eq:fplust})
and~(\ref{eq:fcrosst}) may be expressed as
\begin{eqnarray}
\label{eq:fps}
F_+(t) &= &F_{1_+}\cos 2\beta + F_{2_+}\sin 2\beta + F_{3_+}\cos \beta + 
 F_{4_+}\sin \beta + F_{5_+}\; ;
\end{eqnarray}
\begin{eqnarray}
\label{eq:fcs}
F_\times (t)&= &F_{1_\times}\cos 2\beta + F_{2_\times}\sin 2\beta +
F_{3_\times}\cos \beta + 
 F_{4_\times}\sin \beta +
F_{5_\times}\end{eqnarray}

\vspace{0.1cm}

\noindent where $F_{i _+}$ and $F_{i_\times}$ $( i = 1,2,3,4,5 )$ are time
independent expressions given via

\begin{equation}
\left.\begin{array}{lcl}
\vspace{0.2cm}
F_{1_+} &= & - 2 G \cos\alpha\cos 2\gamma + {H \sin 2\gamma\over 2}(
\cos^2\alpha + 1 ) \, ,\\
\vspace{0.2cm}
F_{2_+} &= &H \cos\alpha\cos 2\gamma + G\sin 2 \gamma (\cos^2\alpha + 1 )\, ,\\
\vspace{0.2cm}
F_{3_+} &= &I\sin\alpha\cos 2\gamma + J \sin 2\alpha\sin 2\gamma\, ,\\
\vspace{0.2cm}
F_{4_+} &= &2 J \sin\alpha\cos 2\gamma - {I\over 2}\sin 2\alpha\sin 2\gamma\, ,\\
F_{5_+} &=& {3\sin^2\alpha\sin 2\gamma\over 2}[ H + L^2 - M^2 ]\, ,
\end{array} \right\} \; ;
\end{equation}

\begin{equation}
\left.\begin{array}{lcl}
\vspace{0.2cm}
G& =& {1\over 2}[ ( L Q + M R )\sin\epsilon - ( L N - M P )\cos\epsilon ]\, ,\\ 
\vspace{0.2cm}
H &=& {1\over 2}[ ( N^2 - P^2 ) \cos^2\epsilon
- ( L^2 - M^2 ) + 
 ( Q^2 - R^2 )\sin^2\epsilon \\ 
\vspace{0.2cm}
&& -\, ( N Q + P R )\sin 2\epsilon ]\, ,\\
\vspace{0.2cm}
I& = &{1\over 2}[ (Q^2 - R^2 ) \sin 2\epsilon - (
N^2 - P^2 )\sin 2\epsilon - \\
&& 2 ( N Q + P R )\cos 2\epsilon ]\, ,\\
J& =& {1\over 2}[ ( L N - M P )\sin\epsilon + ( L Q + M R
)\cos\epsilon ] \\
\end{array} \right\}
\end{equation}

\vspace{0.5cm}

\noindent Let us note that $ F_{i_\times} $ is related to $ F_{i_+} $ via
\begin{eqnarray}
F_{i_\times}( \theta , \phi , \psi , \alpha , \beta , \gamma ,
\epsilon )& = & F_{i_+}( \theta , \phi - {\pi\over 4}, \psi , \alpha ,
\beta , \gamma , \epsilon ) \; ;\\
&& i\; = \;1,2,3,4,5  \nonumber
\end{eqnarray}

\vspace{0.1cm}

\noindent This symmetry is representative of the quadrupolar nature of the
detector and the wave. A detector in different orientations will record
different amplitudes in the response. The explicit beam pattern functions may
be computed easily for any instant of time. Due to
symmetries involved in $F_+$ and $F_\times$ it is sufficient to
discuss either of the beam pattern.

\par  The amplitude modulation of the received signal is a
direct consequence of the non-uniformity of the sensitivity pattern. The
antenna response is not peaked in any direction
but is anisotropic. As remarked earlier they are fairly complicated function
of their arguments. Equations.~(\ref{eq:fps}) and~(\ref{eq:fcs})
reveal that the monochromatic signal
frequency will split, due to AM, into five lines. This results in the
distribution of energy in various
frequencies and consequent reduction of the amplitude of the signal. The periodicity of the beam
patterns $F_+$ and $F_\times$ with a period equal to one sidereal day
 is due
to the diurnal motion of Earth.

\vspace{0.65cm}

\section{Doppler shift and Frequency modulation}
\label{sec:fmd}
\indent The frequency of a monochromatic signal will be Doppler shifted
due to the translatory motion of the detector, acquired from the motions of
Earth. Let us  consider a CGW signal of constant frequency $f_o$.
The frequency $f'$ received at the instant $t$ by the detector is given by

\begin{equation}f'(t) = f_o\gamma_o\left( 1 + {{\bf v . n} \over c}(t)\right)\quad ; \quad
\gamma_o = \left(1 - {v^2\over c^2}\right)^{- 1/2}
\end{equation}

\vspace{0.5cm}

\noindent  where ${ \bf\textstyle n}$ is the unit vector from the antenna to source,
${\displaystyle {\bf v}}$ is the relative velocity of the source and the antenna,
and $c$ is the velocity of light. The unit vector ${ \bf\textstyle n}$ from the
antenna to the source, because of the fact that the distance of the source is
very large compared to the average distance of the centre of the SSB frame and the detector, may
be taken parallel to the unit vector drawn from the centre of the SSB frame to
the source. Hence,

\begin{equation}
{ \bf\textstyle n}\quad = \quad \left(\sin\theta\cos\phi \quad , \quad \sin\theta\sin\phi
\quad , \quad \cos\theta\right)
\end{equation}

\vspace{0.5cm}

\noindent As {\bf v\/} keeps on changing continuously
both in its amplitude and direction $f'$ is a continuous function of t. Further,
since $v \ll c $ we take $\gamma_o = 1$. \\
\noindent The radius vector ${\displaystyle\bf r\/}(t)$ in
the SSB frame is given by
\begin{eqnarray}
{\bf r\/}(t) & = & \left[ R_{se} \cos (w_{orb} t) + R_e \sin\alpha \cos\beta\right. \, , \nonumber \\
&& R_{se} \sin (w_{orb} t) + R_e \sin\alpha\sin\beta\cos\epsilon - R_e \cos
\alpha \sin\epsilon \, ,  \nonumber \\
&& \left. R_e \sin\alpha \sin\beta\sin\epsilon + R_e \cos\alpha \cos\epsilon\right]\; ;
\end{eqnarray}
\begin{equation}
\label{eq:beta}
\beta = \beta_o + w_{rot}t
\end{equation}

\vspace{0.1cm}

\noindent  where $R_{e}$, $R_{se}$ and $w_{orb}$ represent respectively 
Earth's radius, average distance between Earth's centre from the centre of
SSB frame and the orbital angular velocity of Earth. Here, $t$ represents the
time in seconds elapsed from the instant Sun is at the Vernal Equinox and $\beta_o$
is local sidereal time at that instant. The Doppler shift is now given via 
\begin{eqnarray}
\frac{f' - f_o}{f_o}&=&{{\bf v . n}\over c}(t)\; = \; {\Huge{{\bf\dot{r}_t . n}\over c}} \; = \;
 { R_{se} w_{orb}\over c}\sin\theta\sin ( \phi - w_{orb} t ) +  \nonumber\\
&& {R_e w_{rot} \over c} \sin\alpha \left[ \sin\theta\{ \cos\beta\cos\epsilon
\sin\phi  -  \right. \nonumber\\
&& \left. \cos\phi\sin\beta \}  + \cos\beta\sin\epsilon\cos\theta \right]
\label{eq:ds}
\end{eqnarray}

\vspace{0.1cm}

\noindent The phase $\Phi (t)$ of the received signal is given by
\begin{eqnarray}
 \Phi (t) & = & 2 \pi \int_0^t f' (t') dt'  \nonumber\\
& = & 2 \pi f_o \int_0^t  \left[ 1 + {{\bf v . n}\over c}
(t') \right]dt' 
\end{eqnarray}

\vspace{0.1cm}

\noindent Here, we assume the initial phase of the wave to be zero. After a
straight forward calculation we obtain
\begin{eqnarray}
\Phi (t) & = &  2\pi f_o \left[ t + {R_{se}\over c} \sin
\theta\cos\phi' + \right.\nonumber \\
&& {R_e\over c}\sin\alpha \{\sin\theta (\sin\beta\cos\epsilon\sin\phi + \cos\phi\cos\beta) + \nonumber \\
&&\sin\beta\sin\epsilon\cos\theta\} - {R_{se}\over c}
\sin\theta\cos\phi - \nonumber \\
&& {R_e\over c}\sin\alpha\{\sin\theta (\sin\beta_o\cos\epsilon\sin\phi 
 +  \cos\phi\cos\beta_o) +\nonumber\ \\
&&\left. \sin\beta_o\sin\epsilon\cos\theta\} \right]\nonumber \\
& =& 2\pi f_o t + {\cal Z}\cos (w_{orb} t - 
\phi ) + {\cal P}\sin (w_{rot}t) + {\cal Q}\cos (w_{rot}t) - {\cal R} - {\cal Q} \nonumber\\
& = & 2\pi f_o t + {\cal Z}\cos (a\xi_{rot} - \phi ) + 
{\cal N}\cos (\xi_{rot} - \delta ) - {\cal R} - {\cal Q}
\label{eq:phase}
\end{eqnarray}

\vspace{0.1cm}

\noindent where \\
\begin{equation}
\label{eq:daypq}
\left.\begin{array}{lcl}
\vspace{0.2cm}
{\cal P}& = & 2\pi f_o {R_e\over c} \sin\alpha (\cos\beta _o(\sin
\theta \cos\epsilon \sin\phi + \cos\theta \sin\epsilon )\\
\vspace{0.2cm}
&& -\, \sin\beta _o \sin\theta \cos\phi )\, ,\\
\vspace{0.2cm}
{\cal Q}& = & 2\pi f_o {R_e\over c}\sin\alpha (\sin\beta _o (\sin\theta \cos
\epsilon \sin\phi + \cos\theta \sin\epsilon ) \\
\vspace{0.2cm}
&&+\, \cos\beta _o \sin\theta\cos\phi ) \, ,\\
\vspace{0.2cm}
{\cal N}& = & \sqrt{ {\cal P}^2 + {\cal Q}^2 }\, ,\\
\vspace{0.2cm}
{\cal Z}& = & 2\pi f_o {R_{se}\over c}\sin\theta \, , \\
{\cal R}& = & {\cal Z}\cos\phi\, ,\\
\end{array} \right\}
\end{equation}

\begin{equation}
\label{eq:daydelta}
\left.\begin{array}{lcl}
\vspace{0.2cm}
\delta & = &  \tan^{- 1}\frac{{\cal P}}{{\cal Q}}\, ,\\
\vspace{0.2cm}
\phi' & = & w_{orb}t - \phi\, ,\\
\vspace{0.2cm}
\xi_{orb} & = & w_{orb}t\; = \; a\xi_{rot};\quad  a \;= \; w_{orb}/w_{rot}\; \approx \; 1/365.26\, , \\
\xi_{rot} & = & w_{rot}t 
\end{array} \right\}
\end{equation}

\vspace{0.5cm}

\noindent The two polarisation states of the signal can be taken as
\begin{equation}
h_+(t) = h_{o_+}\cos [\Phi (t)] 
\label{eq:hpt}
\end{equation}
\begin{equation}
h_\times (t) = h_{o_\times}\sin [\Phi (t)]\, ;
\label{eq:hct}
\end{equation}

\noindent $h_{o_+}$, $h_{o_\times}$ are the time independent amplitude of
$h_+(t)$, and  $ h_\times (t)$ respectively.

\indent To understand the nature of the FM, let us consider the FT of $h_+(t)$
and $h_\times (t).$ We may begin by considering the function 
\begin{equation}
h(t) = \cos[\Phi (t)]
\label{eq:cosphit}
\end{equation}

\noindent and analysing the data for one day observation time. The FT is given via

\begin{equation}
\left[\tilde{h}(f)\right]_d = \int_0^T \cos[\Phi (t)]e^{-i2\pi ft}dt\; ;
\label{eq:hf1}
\end{equation}

\vspace{0.5cm}

\indent\indent\indent\indent\indent $T =$ one sideral day $= 86164$ sec.

\noindent This may be splitted into two terms as 
 \begin{equation} \left[\tilde{h}(f)\right]_d = I_{\nu_-} + I_{\nu_+} \; ;
 \label{eq:inu}
 \end{equation}
 \begin{eqnarray}
\label{eq:dayinu}
 I_{\nu_-}& =& {1\over 2 w_{rot}}\int_0^{2\pi} e^{i \left[
 \xi\nu_- +
{\cal Z}\cos (a\xi - \phi ) + {\cal N}\cos (\xi - \delta ) -{\cal R}  - {\cal 
Q} \right] } d\xi\, ,\\
I_{\nu_+}& = & {1\over 2 w_{rot}}\int_0^{2\pi} e^{- i \left[
\xi\nu_+
+ {\cal Z} \cos (a\xi - \phi ) + {\cal N}\cos (\xi - \delta ) - {\cal R} - {
\cal Q} \right] } d\xi\, , \\
\nu_{\pm}& = & \frac{f_o \pm f}{f_{rot}} ; \quad \xi \; = \; \xi_{rot} \; = \;
w_{rot}t 
\end{eqnarray}

\vspace{0.1cm}

\noindent Numerical result shows that $I_{\nu_+}$ oscillates very fast and
contributes very little to $\left[\tilde{h}(f)\right]_d$. Hence, hereafter, we drop $I_{\nu_+}$ 
from Eq.~(\ref{eq:inu}) and write $\nu$ in place of $\nu_-$. Using 
the identity

\begin{equation}
e^{\pm i\kappa\cos\vartheta} = J_o (\pm\kappa) + 2 \sum_{l = 1}^{l = 
\infty} i^l J_l (\pm\kappa)\cos l\vartheta
\label{eq:bessel}
\end{equation}

\vspace{0.1cm}

\noindent we obtain
\begin{eqnarray}
\left[\tilde{h}(f)\right]_d &\simeq & {1  \over 2 w_{rot}} e^{i ( 
- {\cal R} - {\cal Q})} \int_0^{2\pi} e^{i\nu\xi} \left[ J_o( {\cal Z} ) + 2 
\sum_{k = 1}^{k =  \infty} J_k ({\cal Z}) i^k \cos k (a\xi - \phi )\right] 
 \nonumber \\
&& \times\,\left[J_o( {\cal N} ) + 2 \sum_{m = 1}^{m =  \infty} J_m ({\cal N}) i^m 
\cos m (\xi - \delta )\right] d\xi
\end{eqnarray}

\vspace{0.5cm}

\noindent where $J$ stands for the Bessel function of first kind of integral
order. After performing the integration we get

\begin{equation}
\label{eq:hfd}
\left[\tilde{h}(f)\right]_d  \simeq  {\nu \over 2 w_{rot}} \sum_{k  =  - 
\infty}^{k = \infty} \sum_{m = - \infty}^{m =  \infty} e^{ i {\cal A}}{\cal 
B}[ {\cal C} - i{\cal D} ] \; ; \;
\end{equation}  

\begin{equation}
\left.\begin{array}{lcl}
\vspace{0.2cm}
{\cal A}&  = &{(k + m)\pi\over 2} - {\cal R} - {\cal Q} \nonumber \\
\vspace{0.2cm}
{\cal B} & = & {J_k({\cal Z}) J_m({\cal N})\over {\nu^2 - (a k + m)^2}} \nonumber \\
\vspace{0.2cm}
{\cal C} &= & \sin 2\nu\pi \cos ( 2 a k \pi - k \phi - m \delta ) -  \nonumber\\
\vspace{0.2cm}
&&{ a k + m \over \nu}\{\cos 2 \nu \pi \sin ( 2 a k \pi - k \phi - m \delta )
+ \sin ( k \phi + m \delta )\}\nonumber \\
\vspace{0.2cm}
 {\cal D} & = & \cos 2\nu\pi \cos ( 2 a k \pi - k \phi - m \delta ) + \nonumber\\
&& {k a +m \over \nu}\sin 2 \nu \pi \sin ( 2 a k \pi - k \phi - m \delta )
 - \cos ( k \phi + m \delta )  \nonumber 
\end{array} \right\}
\end{equation}

\vspace{0.5cm}

\noindent The FT of the two polarisation states of the wave can now be written as
\begin{eqnarray}
\label{eq:hfpd}
\left[\tilde{h}_+(f)\right]_d&=&h_{o_+}\left[\tilde{h}(f)\right]_d \nonumber \\
& \simeq & {\nu h_{o_+}\over 2 w_{rot}} \sum_{k  =  - \infty}^{k =  
\infty} \sum_{m = - \infty}^{m =  \infty} e^{ i {\cal A}}{\cal B}[ {\cal C} - i{\cal D} ] \; ;
\end{eqnarray}
\begin{eqnarray}
\label{eq:hfcd}
\left[\tilde{h}_\times (f)\right]_d &=&- i h_{o_\times}\left[\tilde{h}(f)\right]_d \nonumber \\
& \simeq & {\nu h_{o_\times}\over 2 w_{rot}} \sum_{k  =  - \infty}^{k =  
\infty} \sum_{m = - \infty}^{m =  \infty} e^{ i {\cal A}}{\cal B}[ {\cal D} - i{\cal C} ] 
\end{eqnarray}

\vspace{0.5cm}

\begin{figure}[b]
\epsfig{file=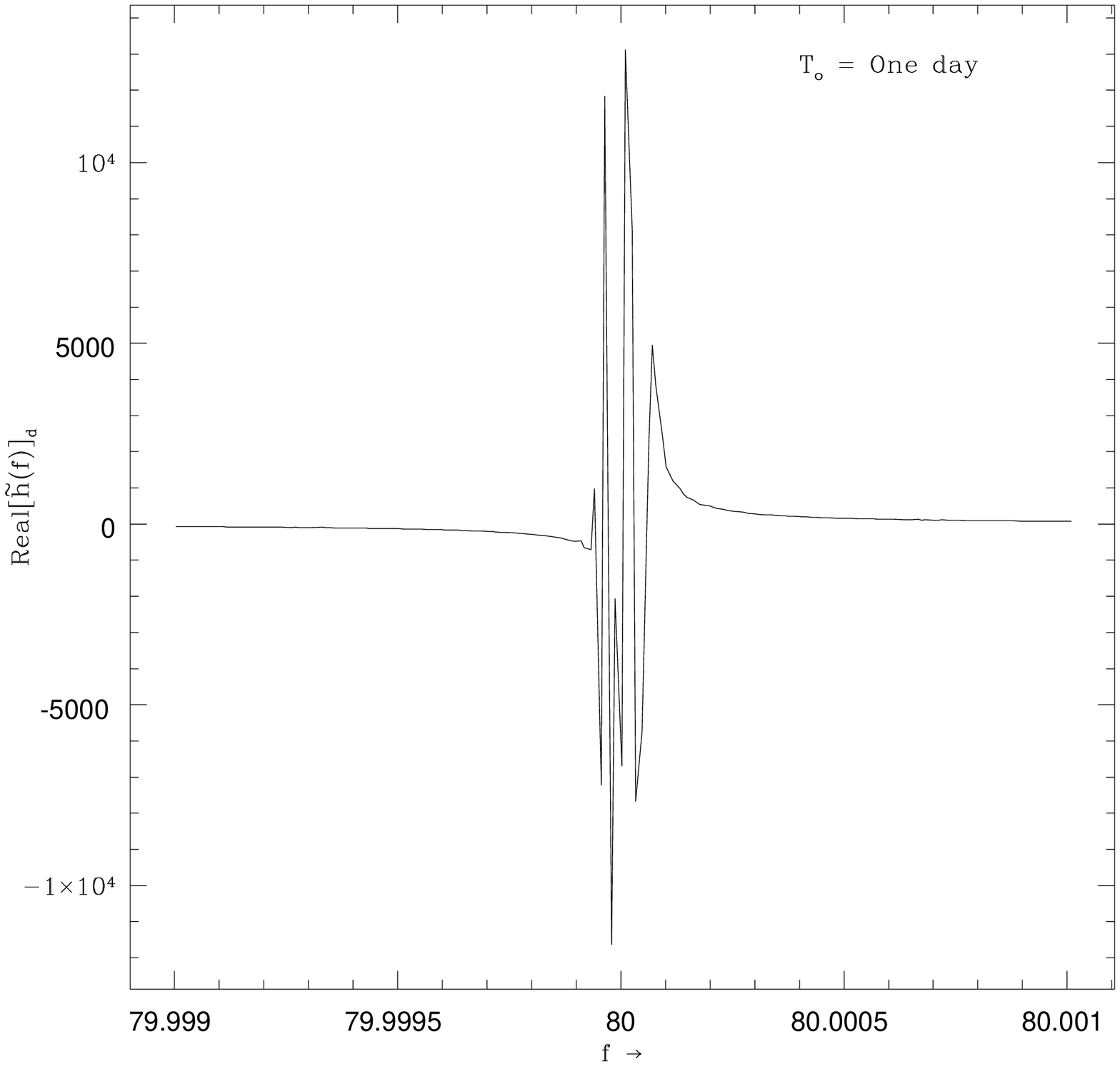,height=14.6cm}
\caption{FT of a FM signal of a source located at 
$(\pi /36 , \pi )$ with a resolution of $1.16 \times 10^{-5}.$}
\label{fig:drealfmrt}
\end{figure}

\noindent The FT of the FM signal contains the
double series Bessel functions. The Bessel functions has contributions
due to the rotational as well as the orbital motion of the earth.
It is remarked that
Jotania et al. (1996) have analysed FT of FM signal for one day observation
time. They have taken specific detector as well as source location. They have
also neglected the orbital motion. Our analysis generalizes their results.
We may now compute
$\left[\tilde{h} (f)\right]_d$ and may plot its behaviour. To achieve this we
have made use of {\it Mathematica\/} software. Fig.~(\ref{fig:drealfmrt}) represents
such a plot for

\begin{equation}
\label{eq:dayloc}
\left.\begin{array}{lll}
\vspace{0.2cm}
f_o = 80\; Hz\, , & h_o = h_\times = 1 & \\
\vspace{0.2cm}
\alpha = \pi /3\, , & \beta_o = \pi /4\, , & \gamma =  2\pi/5 \, , \\
\theta = \pi /36\, , & \phi = \pi\, , &  \psi = \pi /6. \\
\end{array} \right\}
\end{equation}

\begin{figure}
\centering
\epsfig{file=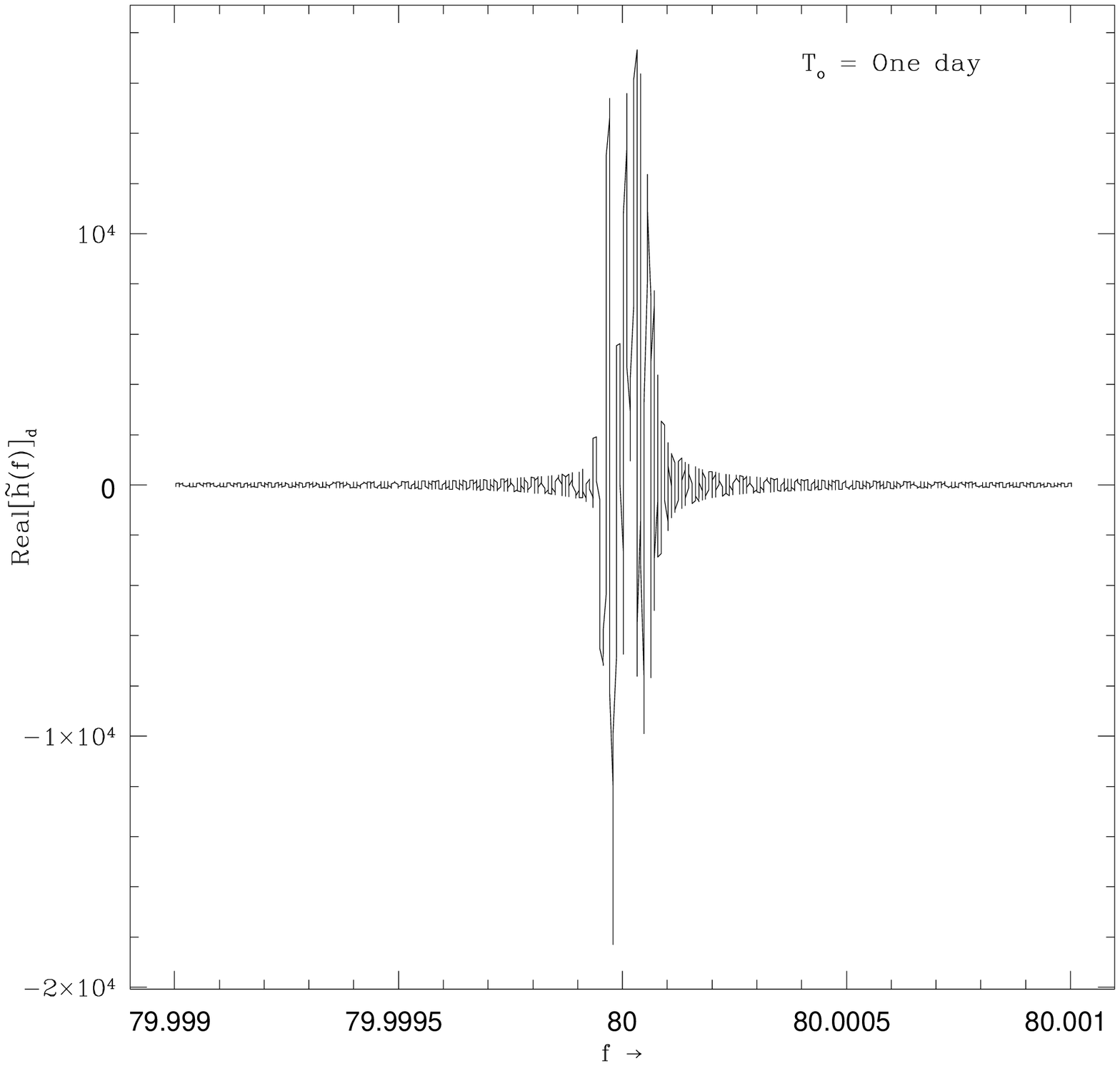,height=14.6cm}
\caption{FT of a FM signal of a source located at 
$(\pi /36 , \pi )$ with a resolution of $10^{-6}.$}
\label{fig:drealfmr6}
\end{figure}

\begin{figure}[b]
\epsfig{file=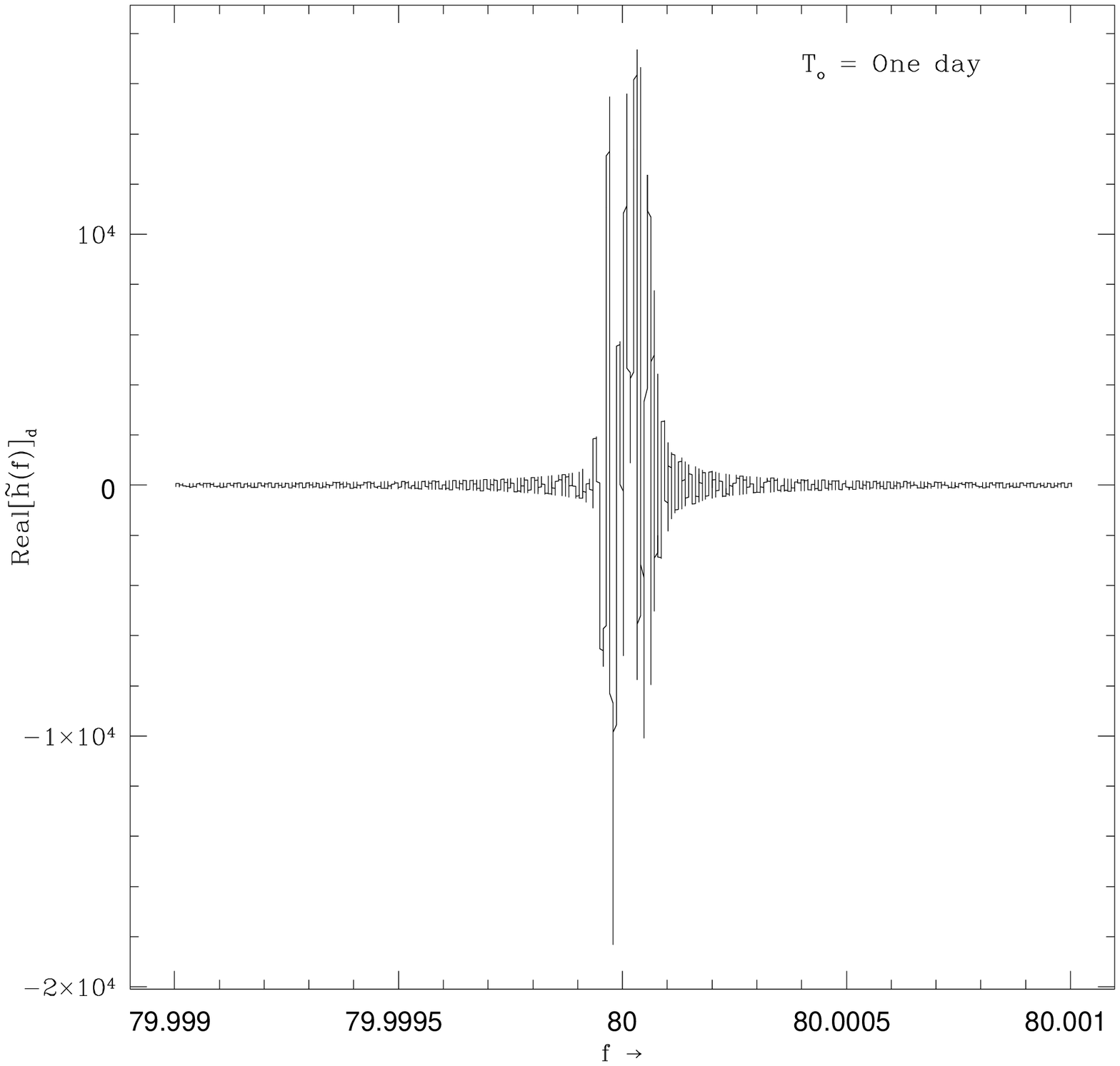,height=14.6cm}
\caption{FT of a FM signal of a source located at 
$(\pi /36 , \pi )$ with a resolution of $10^{-7}.$}
\label{fig:drealfmr7}
\end{figure}

\begin{figure}[b]
\epsfig{file=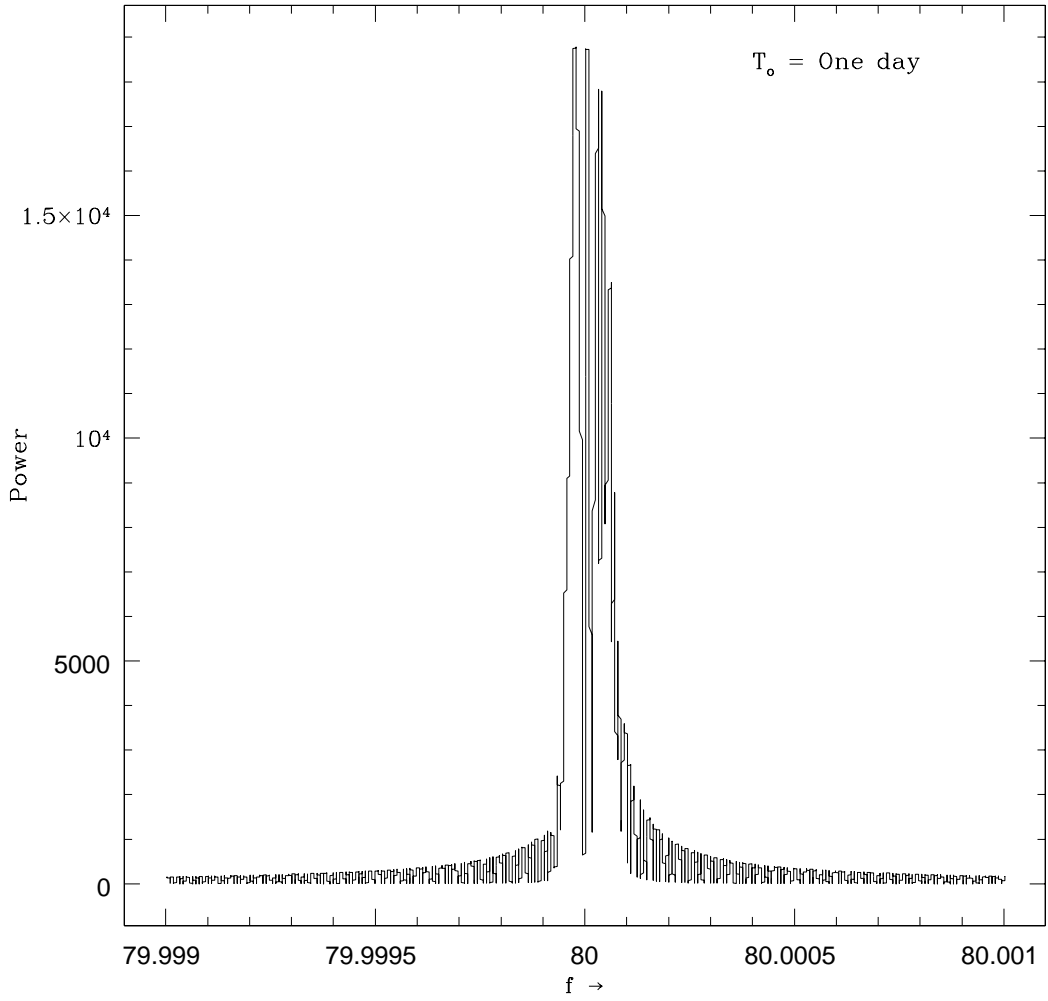,height=14.6cm}
\caption{Power spectrum of a FM signal of a source located at $(\pi /36 , \pi )$
with a resolution of $10^{-7}$.}
\label{fig:fmpowerd}
\end{figure}

\vspace{0.5cm}

\noindent with the resolution equal to $1/T_o = 1.16 \times 10^{-5}$ Hz.
Figures~(\ref{fig:drealfmr6}) and~(\ref{fig:drealfmr7}) represent the plot
$\;$of$\,$ the$\;$ FT$\;$ at resolution $10^{-6}$ Hz and $10^{-7}$ Hz. A careful
look at these plots reveals that the resolution of Fig.~(\ref{fig:drealfmrt})
$\;$ does not represent the details of dominant peaks around $f_o$,
whereas, Fig.~(\ref{fig:drealfmr7}) does not
give any new behaviour as compared to Fig.~(\ref{fig:drealfmr6}). Hence,
we may say that a resolution of about $10^{-6}$ Hz is required to understand the correct
behaviour of the FT for one day observation time data analysis. In this reference
let us recall that the data analysis for Fast Fourier Transform (FFT) limits
the resolution to $1/T_o$. However, detector output may provide
us higher
resolution. Thus the semi-analytical analysis presented here may provide
more information as compared to FFT. Further, in order to see the distribution
of amplitudes at various frequencies we plot power spectrum in
Fig.~(\ref{fig:fmpowerd}).

\clearpage
\clearpage

\vspace{0.65cm}

\section{Fourier transform of the complete response}
\label{sec:crd}
\indent The complete response $R(t)$, in view of
Eqs.~(\ref{eq:rt}),~(\ref{eq:fps}),~(\ref{eq:fcs}),~(\ref{eq:hpt})
and~(\ref{eq:hct}) may be written as
\begin{equation}
R(t) = R_+(t) + R_\times (t)\; ;
\end{equation}
\begin{eqnarray}
\label{eq:rpt}
R_+(t)& =& h_{o_+}\left[ F_{1_+}\cos 2\beta + F_{2_+}\sin 2\beta + F_{3_+}\cos \beta +\right.  \nonumber \\
&&\left. F_{4_+}\sin \beta + F_{5_+} \right]\cos [\Phi (t)]\, , \\
 R_\times (t)& = & h_{o_\times}\left[ F_{1_\times}\cos 2\beta + F_{2_\times}\sin 2\beta +
F_{3_\times}\cos \beta + \right.\nonumber \\
&&\left. F_{4_\times}\sin \beta + F_{5_\times}\right]\sin [\Phi (t)]
\label{eq:rct}
\end{eqnarray}

\vspace{0.1cm}

\noindent Here we have taken for simplicity the angles between the arms of
the detector to be $\pi /2$ i.e. $\Omega = \pi /4$. Now the FT of the
complete response may be expressed as
\begin{eqnarray} \tilde {R}(f) &= &\tilde{R}_+(f) + \tilde{R}_\times (f)
\end{eqnarray}

\vspace{0.1cm}

\noindent To evaluate these let us substitute $\beta$ as given
by~(\ref{eq:beta}). One obtains
\begin{eqnarray}
R_+(t)& =& h_{o_+}\left[\frac{1}{2}\left\{e^{-i 2 \beta_o}( F_{1_+} +
i F_{2_+} ) e^{-i 2 w_{rot}t} + e^{i 2 \beta_o}( F_{1_+} - i F_{2_+} ) e^{i 2 w_{rot}t}
+\right.\right. \nonumber \\
&& \left. e^{-i \beta_o}( F_{3_+} + i F_{4_+} )e^{-i w_{rot}t} + 
 e^{i \beta_o} ( F_{3_+} - i F_{4_+}) e^{i w_{rot}t}\frac{}{}\right\}\nonumber \\
&& + \, \left.F_{5_+}\right]\cos [\Phi (t)]
\end{eqnarray}

\vspace{0.1cm}

\noindent and similar expression for $R_\times (t)$. Now it is straight forward
to obtain the expression for $\tilde{R}_+(f)$ and $\tilde{R}_\times (f)$.
We get
 \begin{eqnarray}
\label{eq:rfpd}
\left[\tilde{R}_+(f)\right]_d & = & h_{o_+}\left[e^{-i 2 \beta_o}( F_{1_+} +
i F_{2_+} ) \left[\tilde{h} ( f + 2 f_{rot})/2\right]_d + \right. \nonumber \\
&& e^{i 2 \beta_o}( F_{1_+} - i F_{2_+} )\left[\tilde{h} ( f - 2 f_{rot})/2\right]_d
+ \nonumber \\
&& e^{-i \beta_o}( F_{3_+} + i F_{4_+} )\left[\tilde{h} ( f +  f_{rot})/2\right]_d + \nonumber \\
&& \left. e^{i \beta_o} ( F_{3_+} - i F_{4_+}) \left[\tilde{h} ( f -  f_{rot})/2\right]_d
+ F_{5_+}\left[\tilde{h}(f)\right]_d\right]
\end{eqnarray}

\vspace{0.1cm}

\noindent Similarly we have
 \begin{eqnarray}
\label{eq:rfcd}
\left[\tilde{R}_\times(f)\right]_d & = & h_{o_\times}\left[e^{-i 2 \beta_o}(
F_{2_\times} -i F_{1_\times} )\left[\tilde{h}( f + 2 f_{rot})/2\right]_d + \right. \nonumber \\
&&- e^{i 2 \beta_o}( F_{2_\times} + i F_{1_\times} )\left[\tilde{h}( f -
2 f_{rot})/2\right]_d + \nonumber \\
&& e^{-i \beta_o}( F_{4_\times} - i F_{3_\times} )\left[\tilde{h}( f +  f_{rot})/2 \right]_d + \nonumber \\
&& \left.- e^{i \beta_o} ( F_{4_\times} + i F_{3_\times}) \left[\tilde{h}
( f -  f_{rot})/2\right]_d - i F_{5_\times}\left[\tilde{h}(f)\right]_d\right]
\end{eqnarray}

\vspace{0.1cm}

\noindent Collecting  our results the FT of the complete response
of the detector for one day integration will be 
\begin{eqnarray}
\label{eq:rfd}
\left[\tilde{R}(f)\right]_d & = & e^{-i 2 \beta_o}\left[\tilde{h}( f
+ 2 f_{rot})/2\right]_d\left[ h_{o_+}( F_{1_+} + i F_{2_+} )
+ h_{o_\times} ( F_{2_\times} - i F_{1_\times} )\right] +\nonumber \\
&& e^{i2\beta_o}\left[\tilde{h}( f - 2 f_{rot})/2\right]_d\left[ h_{o_+}
( F_{1_+} - i F_{2_+} )
- h_{o_\times} ( F_{2_\times} + i F_{1_\times} )\right] +\nonumber \\
& & e^{-i\beta_o}\left[\tilde{h}( f + f_{rot})/2\right]_d\left[ h_{o_+}( F_{3_+} + i F_{4_+} )
+ h_{o_\times} ( F_{4_\times} - i F_{3_\times} )\right] +\nonumber \\
&& e^{i\beta_o}\left[\tilde{h}( f - f_{rot})/2\right]_d\left[ h_{o_+}( F_{3_+} - i F_{4_+} )
- h_{o_\times} ( F_{4_\times} + i F_{3_\times} )\right] +\nonumber \\
&&\left[\tilde{h}(f)\right]_d\left[ h_{o_+}F_{5_+} - i h_{o_\times}F_{5_\times}\right]
\end{eqnarray} 

\vspace{0.1cm}

\noindent This shows that due to AM every Doppler modulated FM signal
will split in four additional lines at $f \pm 2 f_{rot}$ and $f \pm f_{rot}$, 
where $f_{rot}$ is the rotational
frequency of Earth ($f_{rot} \approx 1.16 \times 10^{-5}$ Hz). In terms of
dimensionless scaled frequency $\nu$ the power of the signal will be at $\nu 
\pm 2 a$, $\nu \pm a$ \& $\nu$.

\par
\begin{figure}[b]
\epsfig{file=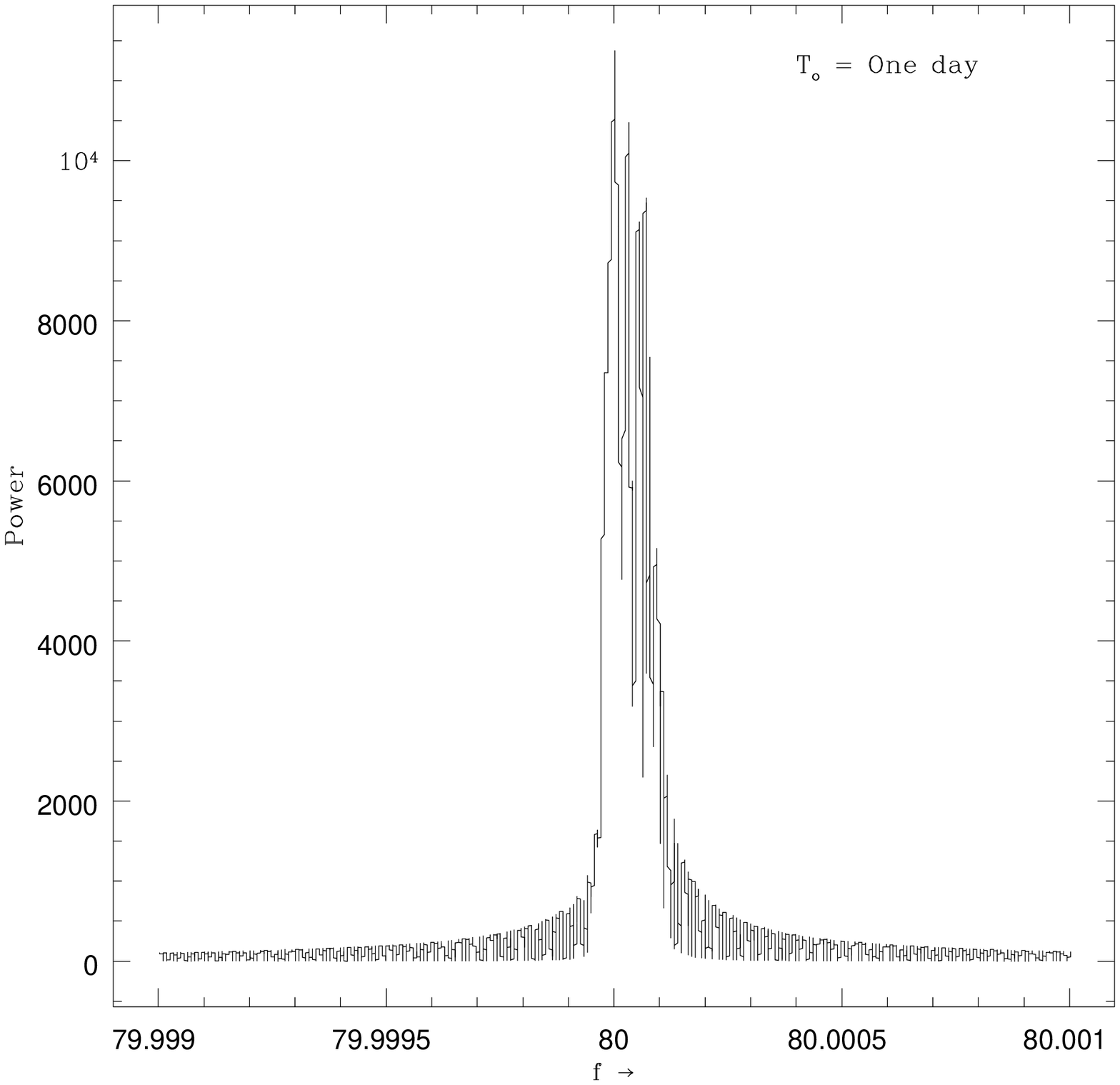,height=14.6cm}
\caption{Power spectrum of the complete response of a Doppler modulated signal of
a source located at $(\pi /36 , \pi )$ with a resolution of $10^{-7}$.}
\label{fig:crd}
\end{figure}

\begin{figure}[b]
\epsfig{file=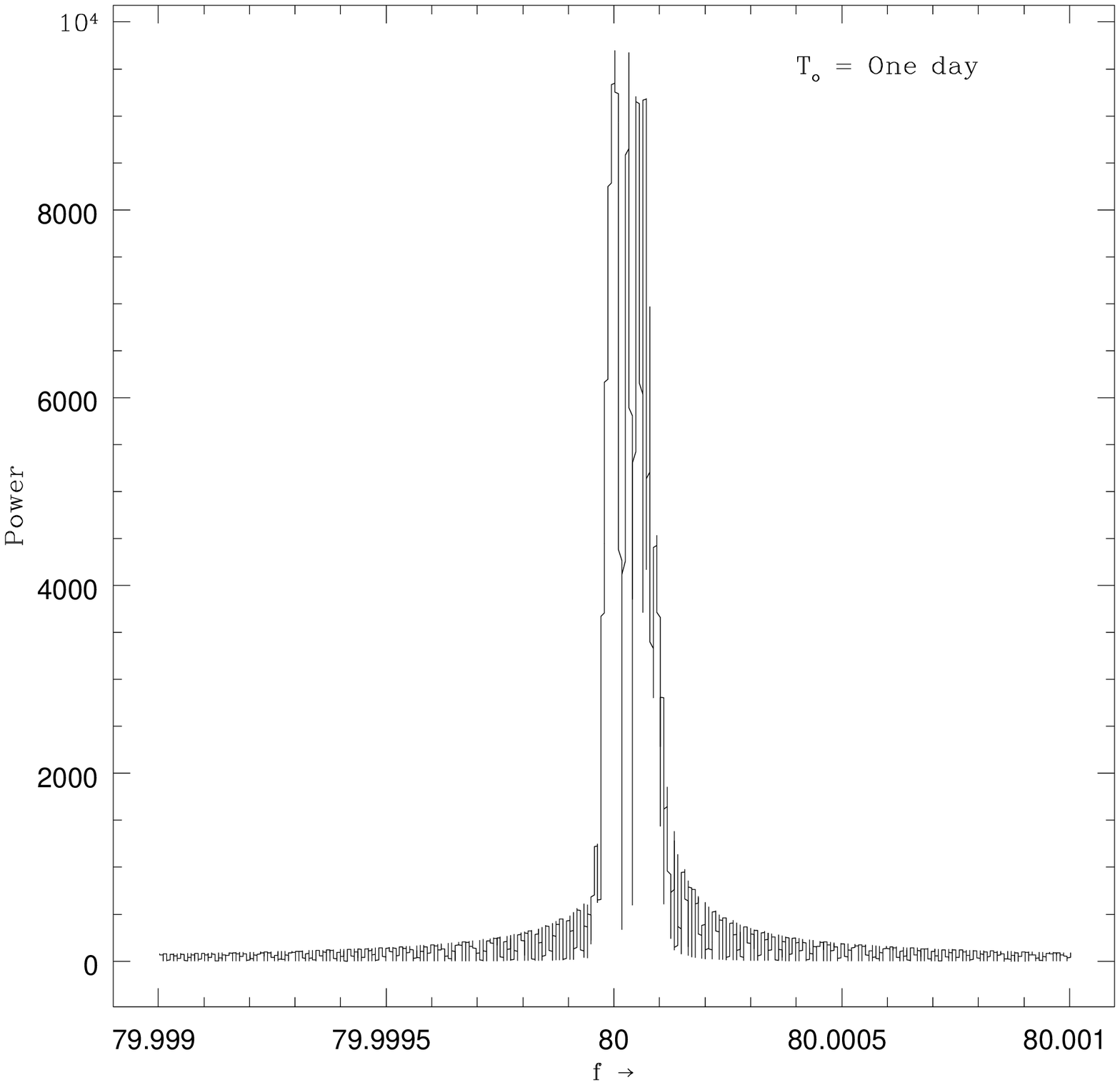,height=14.6cm}
\caption{Power spectrum of a Doppler modulated signal at frequencies  $f + 2f_{rot}$
of a source located at $(\pi /36 , \pi )$ with a resolution of $10^{-7}$.}
\label{fig:crp2frd}
\end{figure}

\begin{figure}[b]
\epsfig{file=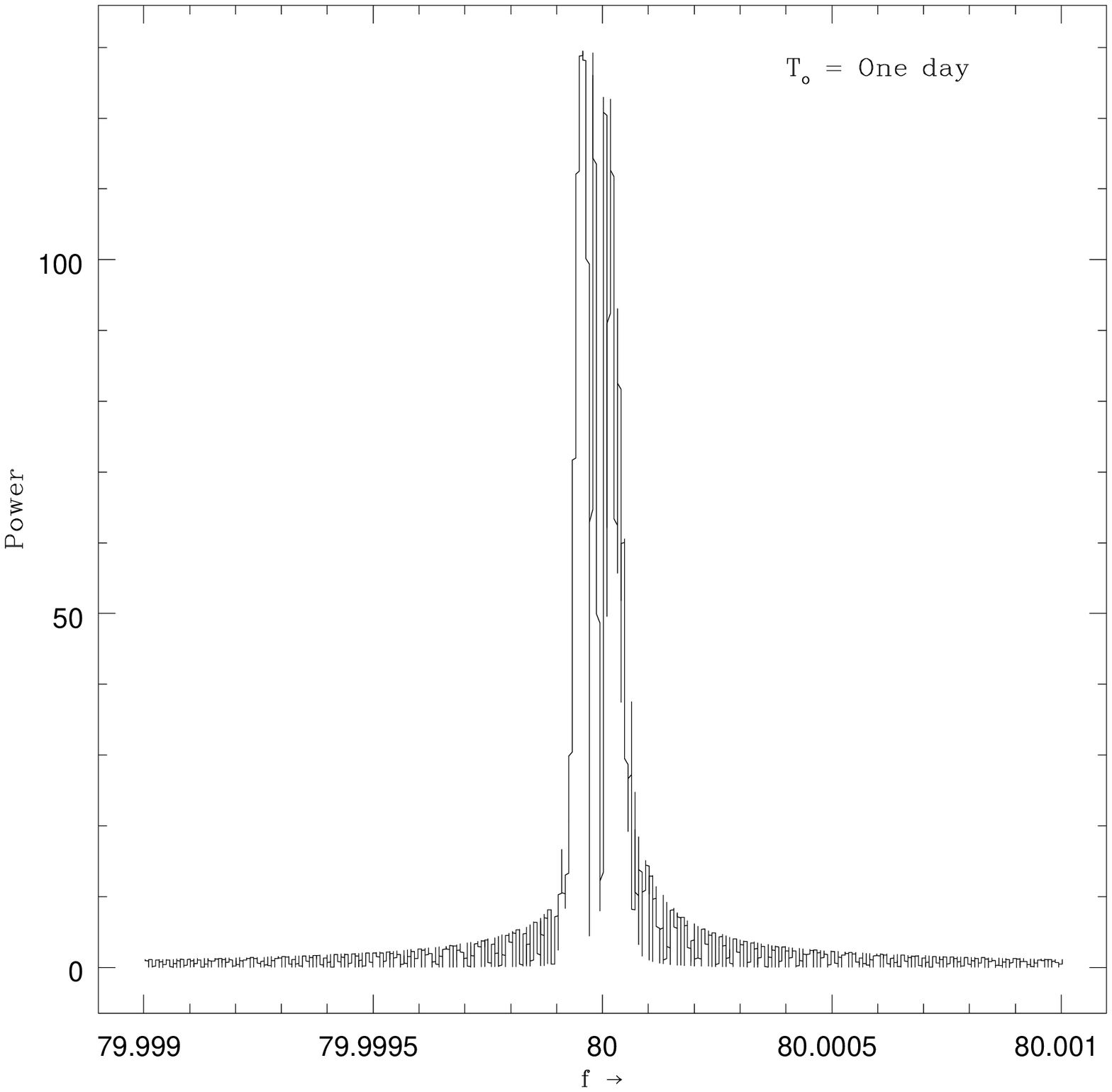,height=14.6cm}
\caption{Power spectrum of a Doppler modulated signal at frequencies  $f - 2f_{rot}$
of a source located at $(\pi /36 , \pi )$ with a resolution of $10^{-7}$.}
\label{fig:crm2frd}
\end{figure}

\begin{figure}[b]
\epsfig{file=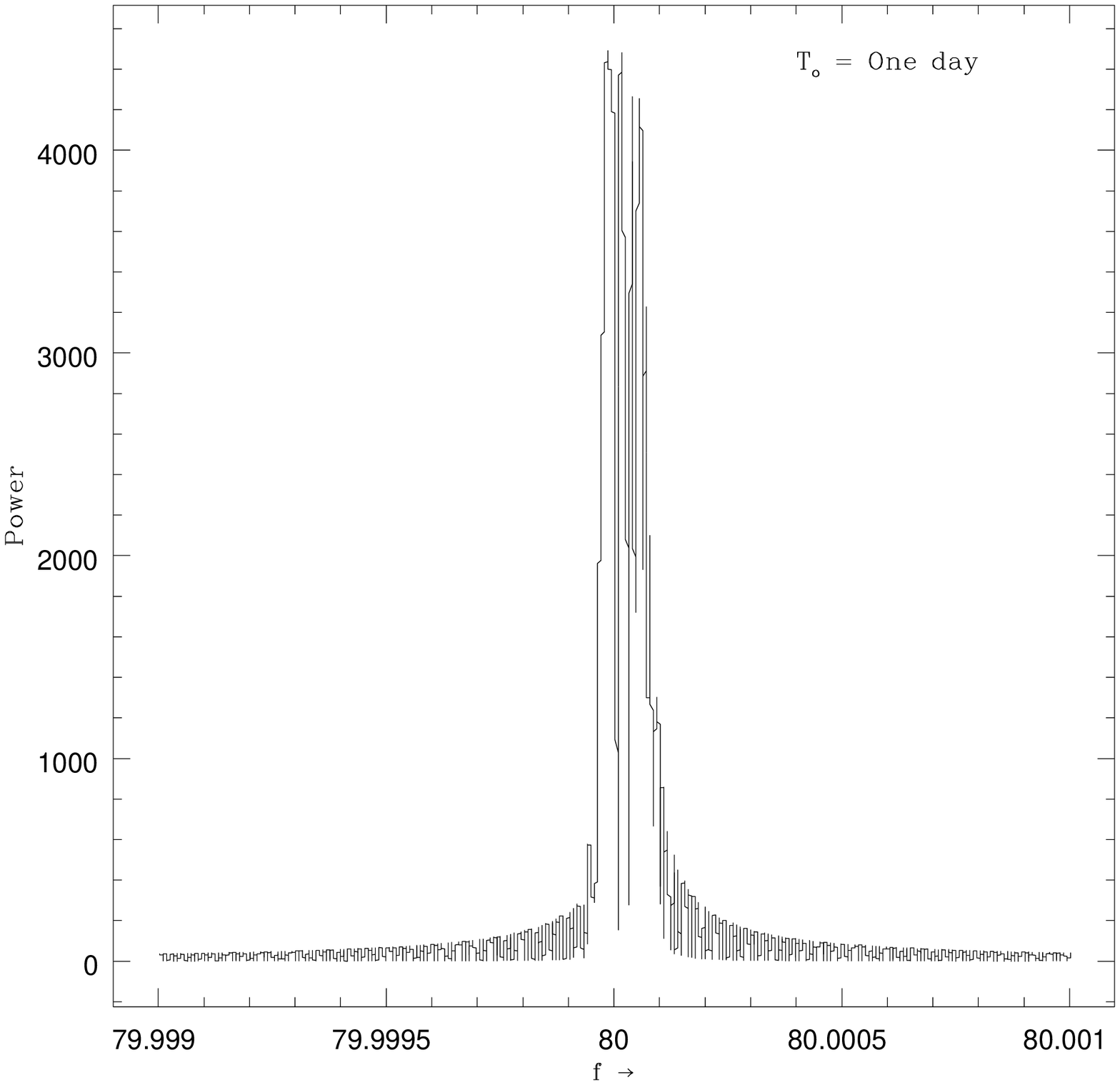,height=14.6cm}
\caption{Power spectrum of a Doppler modulated signal at frequencies  $f + f_{rot}$
of a source located at $(\pi /36 , \pi )$ with a resolution of $10^{-7}$.}
\label{fig:crpfrd}
\end{figure}

\begin{figure}[b]
\epsfig{file=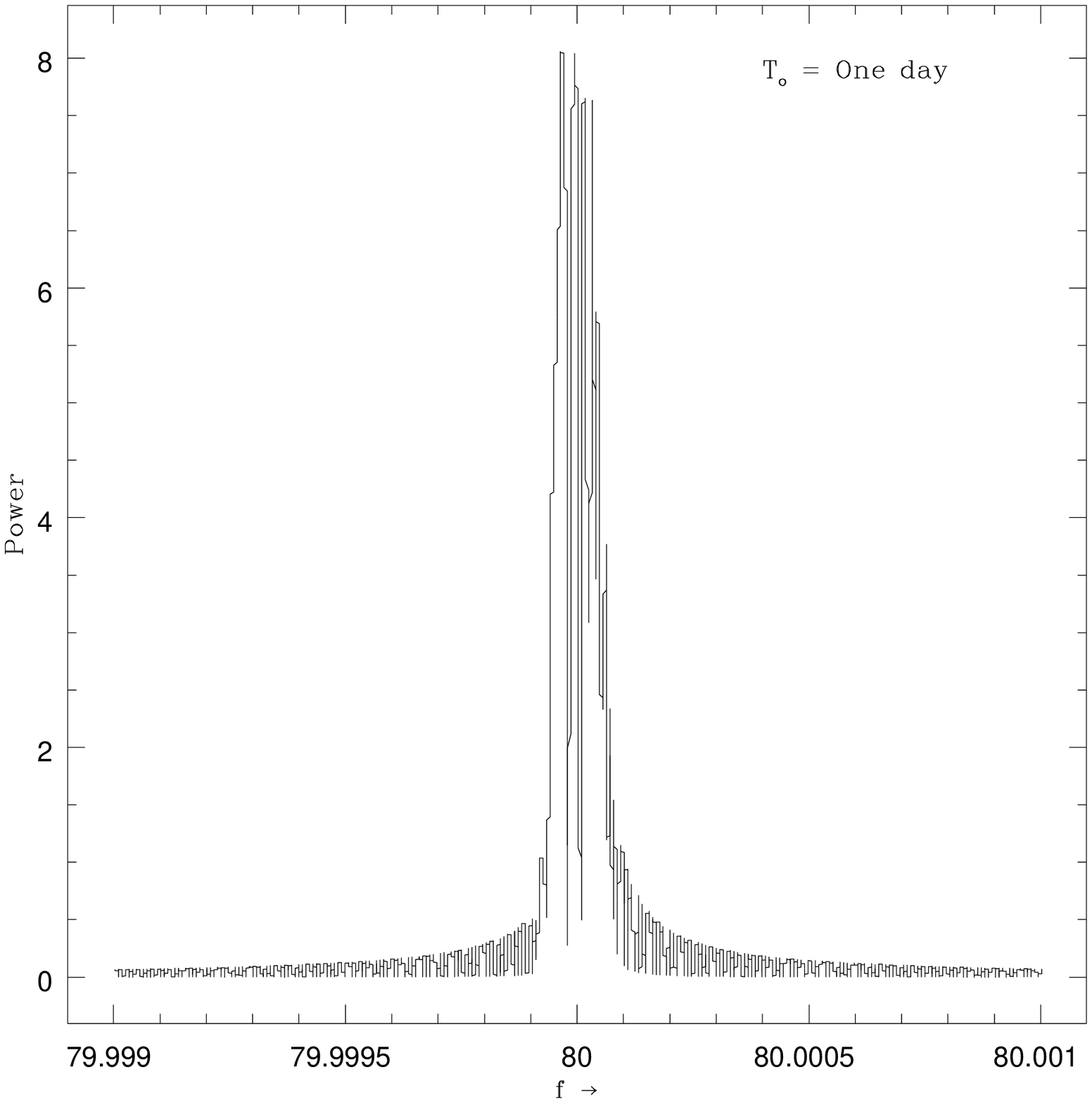,height=14.6cm}
\caption{Power spectrum of a Doppler modulated signal at frequencies  $f - f_{rot}$
of a source located at $(\pi /36 , \pi )$ with a resolution of $10^{-7}$.}
\label{fig:crmfrd}
\end{figure}

\begin{figure}[b]
\epsfig{file=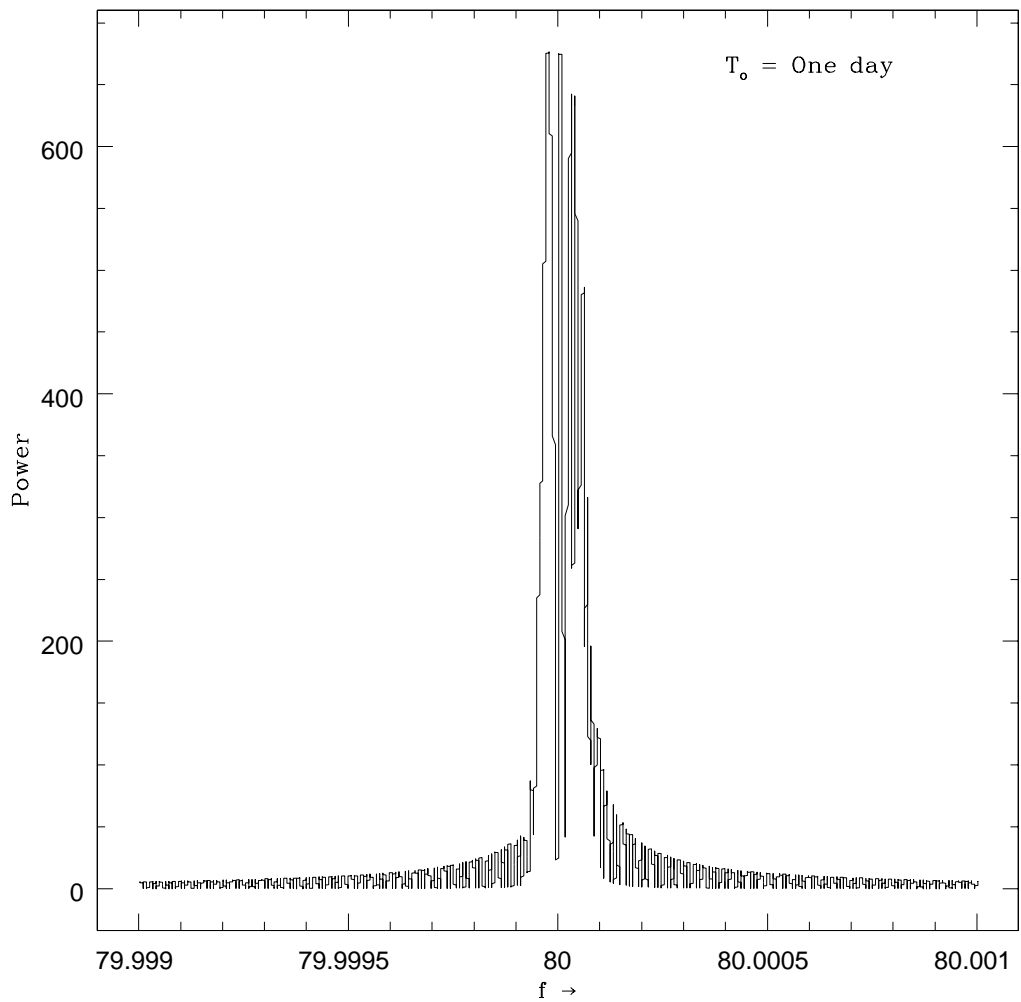,height=14.6cm}
\caption{Power spectrum of a Doppler modulated signal at frequencies  $f$ of a source
located at $(\pi /36 , \pi )$ with a resolution of $10^{-7}$.}
\label{fig:cr0shiftd}
\end{figure}

 We have plotted in Fig.~(\ref{fig:crd}) the power spectrum of the noise
free complete response of the signal for its various parameters as given
by~(\ref{eq:dayloc}). The contribution in the power spectrum of modulation
at frequencies $f + 2f_{rot}$, $f - 2f_{rot}$, $f + f_{rot}$
and $f- f_{rot}$ and $f$ are
represented in Figs.~(\ref{fig:crp2frd}),~(\ref{fig:crm2frd}),~(\ref{fig:crpfrd}),~(\ref{fig:crmfrd})
and~(\ref{fig:cr0shiftd}). It is observed that the most of the power will be at
$f + 2f_{rot}$ and least power will be in $f - f_{rot}$.

\clearpage

\vspace{0.65cm}

\section{Discussion}
\label{sec:sum3}
\indent In this Chapter we have considered the effect of Earth's motion 
on the response of the detector through FT analysis. It can easily
be inferred from Eqs.~(\ref{eq:rfd}) and~(\ref{eq:hfpd},~\ref{eq:hfcd})
that the splitting of frequencies (i) in AM arise explicitly due to rotational motion and
(ii) in FM arise due to rotational as well as orbital motion of earth.
In view of the fact that the data output at the detector is available in discrete
form, the analytical FT is not very convenient and one normally employs
the popular FFT. However, FFT has resolution limited 
to $1/T_o$. Further, it is important to understand for how much time one
can ignore the frequency shift arising due to Doppler effect. In fact, Schutz
(1991) has demonstrated that these effects due to rotational motion are
important after the time given by

\begin{equation}T_{max} = \left({\frac{2c}{\omega_{rot}^2 f_o R_e}}\right)^{1/2} \simeq
70\left( {\frac{f_o}{1 kHz}}\right)^{-1/2} min. \end{equation}

\vspace{0.5cm}

\noindent This means that for GW signal for frequency 80 Hz one has to take
into account these effects after data time $\simeq$ 4 hours. The
analytical FT studied in this Chapter leads to following inferences:
\begin{enumerate}[(i)]
\item FFT for one day observation data will not provide sufficient resolution as
to represent the correct picture of the frequency splitting.
\item The adequate resolution required for one day observation $\simeq 10^{-6}$ Hz.
\item The frequency split due to FM for frequency $f_o = 80$ Hz and source at
$(\theta , \phi )$ = $(\pi /36 ,\pi)$ is $\simeq$ 2 $\times 10^{-4}$ Hz and due
to AM is $\simeq$ $4.64 \times 10^{-5}$ Hz.
\item The drop in amplitude due to FM alone is about 56\%
\item The drop in amplitude due to AM alone is about 18\%.
\item The drop in amplitude for the complete response is about 74\%.
\item The maximum power due to AM is associated with $f_o + 2f_{rot}$.
\end{enumerate}

\noindent It is remarked that the drop of the amplitude in complete response is
severe both due to AM and FM as the relevant frequency range lie in the same region [see
conclusion (iii) above].

\par Finally, we would like to mention that we have presented the FT analysis
assuming the phase of the GW to be zero at that instant $t = 0$. However, one
may relax this condition and may obtain the results easily by taking into
consideration the effects of change of the time origin.

\chapter{Data analysis of continuous gravitational wave: Fourier transform-II}
\section{Introduction}
In previous Chapter we have presented Fourier analysis of one day observation
data set of the response of a laser interferometer. We have seen that the
amplitude and frequency modulations result into a large number of side bands
about the signal frequency $f_o$. Consequently the maximum power lies in the
frequency $f + 2f_{rot}$ with amplitude reduction by 74\% to what one would
have expected due to increased data interval. Hence, for GW detection it is
desirable to obtain FT for larger data observation time interval. To begin with
we present in the next section an analysis for data set of one year observation time. 
The analysis turns out to be simple in view of the observation that there exists correspondences
and identifications to the analysis of previous Chapter to make them identical.
We obtain in section 3 generalisations of the results for
arbitrary observation time. To facilitate analogous modifications we have
introduced corresponding quantities with tilde viz., $\tilde{\cal C}$ and $\tilde{\cal D}$ in place of ${\cal C}$ and
${\cal D}$. As an application of the results obtained here we consider spin
down and N-component signal analysis respectively in sections 4 and 5.
We present discussion and conclusions in the last section.

\vspace{0.65cm}

\section{Fourier transform for one year integration}
\label{sec:fty}
\subsection{Frequency modulation}
Let us rewrite the expression for phase of the GW signal referring to
Eqs.~(\ref{eq:phase}), ~(\ref{eq:daypq}) and~(\ref{eq:daydelta}) 
\begin{eqnarray}
\Phi(t) & = & 2\pi f_o t + {\cal Z}\cos (w_{orb}t - \phi ) + 
{\cal N}\cos (w_{rot}t - \delta ) - {\cal R} - {\cal Q} 
\end{eqnarray}

\vspace{0.1cm}

\noindent The FT for one year observation time $T_{obs}$ is given as

\begin{equation}
\left[\tilde{h}(f)\right]_y = \int_0^{\bar{a}T} \cos[\Phi (t)]e^{-i2\pi ft}dt\; ;
\label{eq:hfy1}
\end{equation}
\begin{equation}
\bar{a} = a^{-1} = w_{rot}/w_{orb}\; ;\quad T =\; one\; sideral\; day \; ;
\end{equation}
\begin{equation}
T_{obs} = \bar{a}T \simeq  3.14 \times 10^{7}\; sec.
\end{equation}

\vspace{0.1cm}

\noindent This may be splitted as usual into two terms as
\begin{equation}
\left[\tilde{h}(f)\right]_y = I_{\bar{\nu}_-} + I_{\bar{\nu}_+} \; ;
\label{eq:inuy}
\end{equation}
\begin{eqnarray}
\label{eq:inubar}
 I_{\bar{\nu}_-}& =& {1\over 2 w_{orb}}\int_0^{2\pi} e^{i \left[
 \bar{\xi}\bar{\nu}_- +
{\cal Z}\cos (\bar{\xi} - \phi ) + {\cal N}\cos (\bar{a}\bar{\xi} - \delta ) -{\cal R}  - {\cal 
Q} \right] } d\bar{\xi}\, ,\\
I_{\bar{\nu}_+}& = & {1\over 2 w_{orb}}\int_0^{2\pi} e^{- i \left[
\bar{\xi}\bar{\nu}_+
+ {\cal Z} \cos (\bar{\xi} - \phi ) + {\cal N}\cos (\bar{a}\bar{\xi} - \delta ) - {\cal R} - {
\cal Q} \right] } d\bar{\xi}\, , \\
\bar{\nu}_{\pm}& = & \frac{f_o \pm f}{f_{orb}} ; \quad \bar{\xi} \; = \; \xi_{orb} \; = \;
w_{orb}t  
\end{eqnarray}

\vspace{0.5cm}

\noindent Hereafter, we neglect the contribution of $I_{\bar{\nu}_+}$ to
$\left[\tilde{h}(f)\right]_y$ as it oscillates rapidly and contributes very
little and write $\bar{\nu}$ in place of $\bar{\nu}_-$. A careful comparison of
Eq.~(\ref{eq:inubar}) with~(\ref{eq:dayinu})
reveals that the integrand of the equations are identical with following
identifications and correspondences.

\begin{equation}
\left.\begin{array}{ccc}
\delta &\leftrightarrow &\phi\; ,\\
{\cal Z}&\leftrightarrow &{\cal N}\; , \\
a &\leftrightarrow & \bar{a}.
\end{array} \right\}
\end{equation}

\vspace{0.5cm}

\noindent Hence, we may employ the results obtained there by introducing obvious
corresponding quantities i.e. $\bar{\cal B}$, $\bar{\cal C}$, $\bar{\cal D}$
in place of ${\cal B}$, ${\cal C}$, ${\cal D}$ leaving ${\cal A}$ unchanged.
Thus

\begin{equation}
\label{eq:hfy}
\left[\tilde {h}(f)\right]_y  \simeq  {\bar{\nu} \over 2 w_{orb}} \sum_{k  =  - \infty}^{k =  
\infty} \sum_{m = - \infty}^{m =  \infty} e^{ i {\cal A}}\bar{{\cal B}}[
\bar{{\cal C}} - i\bar{{\cal D}} ]\; ;
\end{equation}  

\begin{equation}
\left.\begin{array}{lcl}
\vspace{0.2cm}
{\cal A}&  = &{(k + m)\pi\over 2} - {\cal R} - {\cal Q} \; ,\\
\vspace{0.2cm}
\bar{{\cal B}} & = & {J_k({\cal N}) J_m({\cal Z})\over {\bar{\nu}^2 - (\bar{a} k + m)^2}} \; , \\
\vspace{0.2cm}
\bar{{\cal C}} &= & \sin 2\bar{\nu}\pi \cos ( 2 \bar{a} k \pi - k \delta - m \phi ) -  \nonumber\\
\vspace{0.2cm}
&&{ \bar{a} k + m \over \bar{\nu}}\{\cos 2 \bar{\nu} \pi \sin ( 2 \bar{a} k \pi - k \delta - m \phi )
+ \sin ( k \delta + m \phi )\}\; , \\
\vspace{0.2cm}
\bar{{\cal D}} & = & \cos 2\bar{\nu}\pi \cos ( 2 \bar{a} k \pi - k \delta - m \phi ) + \nonumber\\
&& {k \bar{a} +m \over \bar{\nu}}\sin 2 \bar{\nu} \pi \sin ( 2 \bar{a} k \pi - k \delta - m \phi )
 - \cos ( k \delta + m \phi )  
\end{array}\right\}
\end{equation}

\vspace{0.5cm}

\noindent Now the FT of the two polarisation states can be written as
\begin{eqnarray}
\label{eq:hfpy}
\left[\tilde{h}_+(f)\right]_y&=&h_{o_+}\left[\tilde{h}(f)\right]_y \nonumber \\
& \simeq & {\bar{\nu} h_{o_+}\over 2 w_{orb}} \sum_{k  =  - \infty}^{k =  
\infty} \sum_{m = - \infty}^{m =  \infty} e^{ i {\cal A}}\bar{{\cal B}}[
\bar{{\cal C}} - i\bar{{\cal D}} ]\,\;
\end{eqnarray}
and
\begin{eqnarray}
\label{eq:hfcy}
\left[\tilde{h}_\times(f)\right]_y&=& - i h_{o_\times}\left[\tilde{h}(f)\right]_y \nonumber \\
& \simeq & {\bar{\nu} h_{o_\times}\over 2 w_{orb}} \sum_{k  =  - \infty}^{k =  
\infty} \sum_{m = - \infty}^{m =  \infty} e^{ i {\cal A}}\bar{{\cal B}}[
\bar{{\cal D}} - i\bar{{\cal C}} ]
\end{eqnarray}

\vspace{0.5cm}

\par The FT obtained contains double series of Bessel
functions of the order $k$ and $m$ ranging from $-\infty$ to $\infty$. It is 
well known that Bessel functions decrease rapidly as the order exceeds the
argument. Hence possible range of $k$ and $m$ over which the summation of the series is
to be considered depends on the arguments of Bessel functions i.e ${\cal Z}$ and 
${\cal N}$. Referring to Eq.~(\ref{eq:daypq}) it is found
that

\begin{equation}
\left.\begin{array}{ccl}
\vspace{0.2cm}
{\cal Z}_{max} & = &  3133215\left(\frac{f}{1\, KHz} \right)\\
{\cal N}_{max} & = &134\left(\frac{f}{1\, KHz} \right)
\end{array}\right\}
\end{equation}

\vspace{0.5cm}

\noindent The FT of a FM signal for

\begin{equation}
\left.\begin{array}{lll}
\vspace{0.2cm}
f_o = 50\; Hz\, , & h_o = h_\times = 1 & \\
\vspace{0.2cm}
\alpha = \pi /4\, , & \beta_o = 0\, , & \gamma =  \pi \, , \\
\theta = \pi /18\, , & \phi = 0\, , &  \psi = \pi /4.
\end{array} \right\}
\label{eq:yearloc}
\end{equation}

\begin{figure}[b]
\epsfig{file=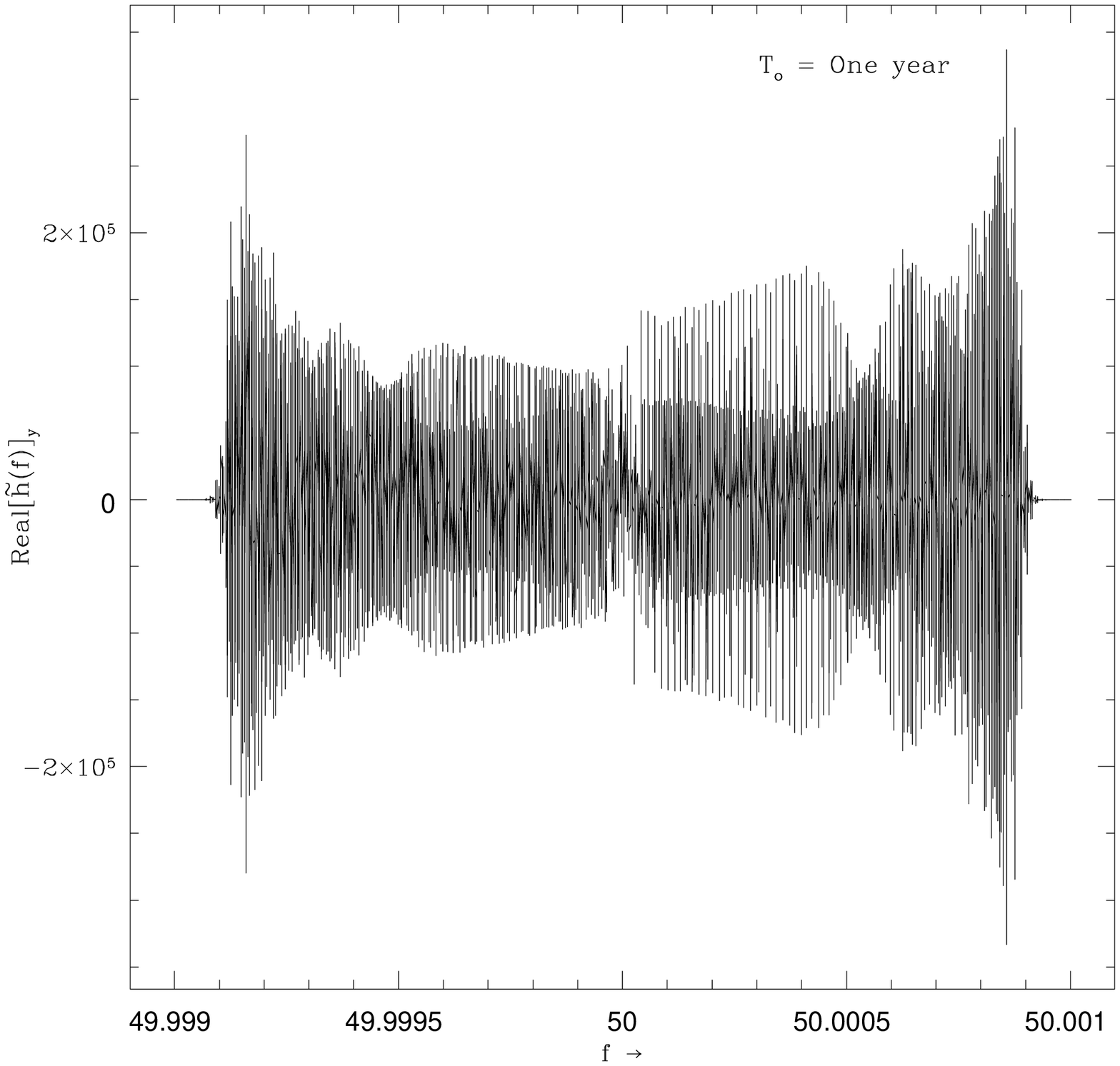,height=14.6cm}
\caption{FT of a FM signal of a source located at $(\pi /18 , 0)$
with a resolution of $3.17 \times 10^{-8}.$}
\label{fig:yrealfm}
\end{figure}

\begin{figure}[b]
\epsfig{file=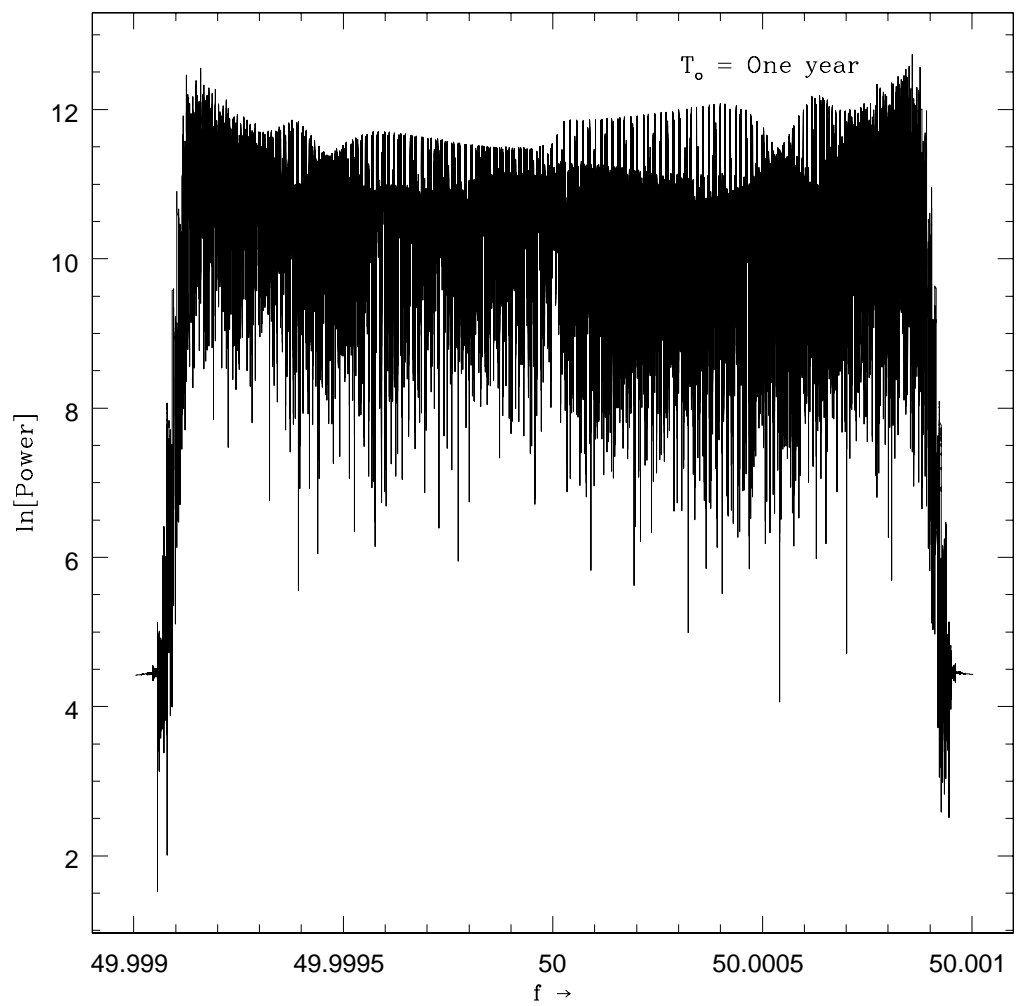,height=14.6cm}
\caption{Power spectrum of a FM signal of a source located at $(\pi /18 ,
0)$ with a resolution of $3.17 \times 10^{-8}.$}
\label{fig:yfmpower}
\end{figure}

\vspace{0.5cm}

\noindent is shown in Fig.~(\ref{fig:yrealfm}). The spectrum has a
resolution of $1/T_o \approx 3.17 \times 10^{-8} Hz$. The 
corresponding power spectrum is shown in Fig.~(\ref{fig:yfmpower}). We have
convinced ourselves by plotting the FT at higher resolutions that resolution
of $1/T_o$ is sufficient to represent relevant
peaks. We notice that the drop in amplitude is about 98\%. This may be
attributed to the presence of a very large number of side bands.

\clearpage
\subsection{Complete response}
The complete response of the detector may be obtained employing
Eqs.~(\ref{eq:hfpy}), ~(\ref{eq:hfcy}), ~(\ref{eq:rpt}) and~(\ref{eq:rct})
and one gets
 \begin{eqnarray}
\label{eq:rfpy}
\left[\tilde{R}_+(f)\right]_y & = & h_{o_+}\left[e^{-i 2 \beta_o}( F_{1_+} +
i F_{2_+} ) \left[\tilde{h} ( f + 2 f_{rot})/2\right]_y + \right. \nonumber \\
&& e^{i 2 \beta_o}( F_{1_+} - i F_{2_+} )\left[\tilde{h} ( f - 2 f_{rot})/2\right]_y
+ \nonumber \\
&& e^{-i \beta_o}( F_{3_+} + i F_{4_+} )\left[\tilde{h} ( f +  f_{rot})/2\right]_y + \nonumber \\
&& \left. e^{i \beta_o} ( F_{3_+} - i F_{4_+}) \left[\tilde{h} ( f -  f_{rot})/2\right]_y
+ F_{5_+}\left[\tilde{h}(f)\right]_y\right]\, ;
\end{eqnarray}
 \begin{eqnarray}
\label{eq:rfcy}
\left[\tilde{R}_\times(f)\right]_y & = & h_{o_\times}\left[e^{-i 2 \beta_o}(
F_{2_\times} -i F_{1_\times} )\left[\tilde{h}( f + 2 f_{rot})/2\right]_y - \right. \nonumber \\
&& e^{i 2 \beta_o}( F_{2_\times} + i F_{1_\times} )\left[\tilde{h}( f -
2 f_{rot})/2\right]_y + \nonumber \\
&& e^{-i \beta_o}( F_{4_\times} - i F_{3_\times} )\left[\tilde{h}( f +  f_{rot})/2 \right]_y - \nonumber \\
&& \left. e^{i \beta_o} ( F_{4_\times} + i F_{3_\times}) \left[\tilde{h}
( f -  f_{rot})/2\right]_y - i F_{5_\times}\left[\tilde{h}(f)\right]_y\right]
\end{eqnarray}

\noindent After rearranging these expressions we obtain
\begin{eqnarray}
\label{eq:rfy}
\left[\tilde{R}(f)\right]_y & = & e^{-i 2 \beta_o}\left[\tilde{h}( f
+ 2 f_{rot})/2\right]_y\left[ h_{o_+}( F_{1_+} + i F_{2_+} )
+ h_{o_\times} ( F_{2_\times} - i F_{1_\times} )\right] +\nonumber \\
&& e^{i2\beta_o}\left[\tilde{h}( f - 2 f_{rot})/2\right]_y\left[ h_{o_+}
( F_{1_+} - i F_{2_+} )
- h_{o_\times} ( F_{2_\times} + i F_{1_\times} )\right] +\nonumber \\
& & e^{-i\beta_o}\left[\tilde{h}( f + f_{rot})/2\right]_y\left[ h_{o_+}( F_{3_+} + i F_{4_+} )
+ h_{o_\times} ( F_{4_\times} - i F_{3_\times} )\right] +\nonumber \\
&& e^{i\beta_o}\left[\tilde{h}( f - f_{rot})/2\right]_y\left[ h_{o_+}( F_{3_+} - i F_{4_+} )
- h_{o_\times} ( F_{4_\times} + i F_{3_\times} )\right] +\nonumber \\
&&\left[\tilde{h}(f)\right]_y\left[ h_{o_+}F_{5_+} - i h_{o_\times}F_{5_\times}\right]
\end{eqnarray}

\par
\begin{figure}[b]
\epsfig{file=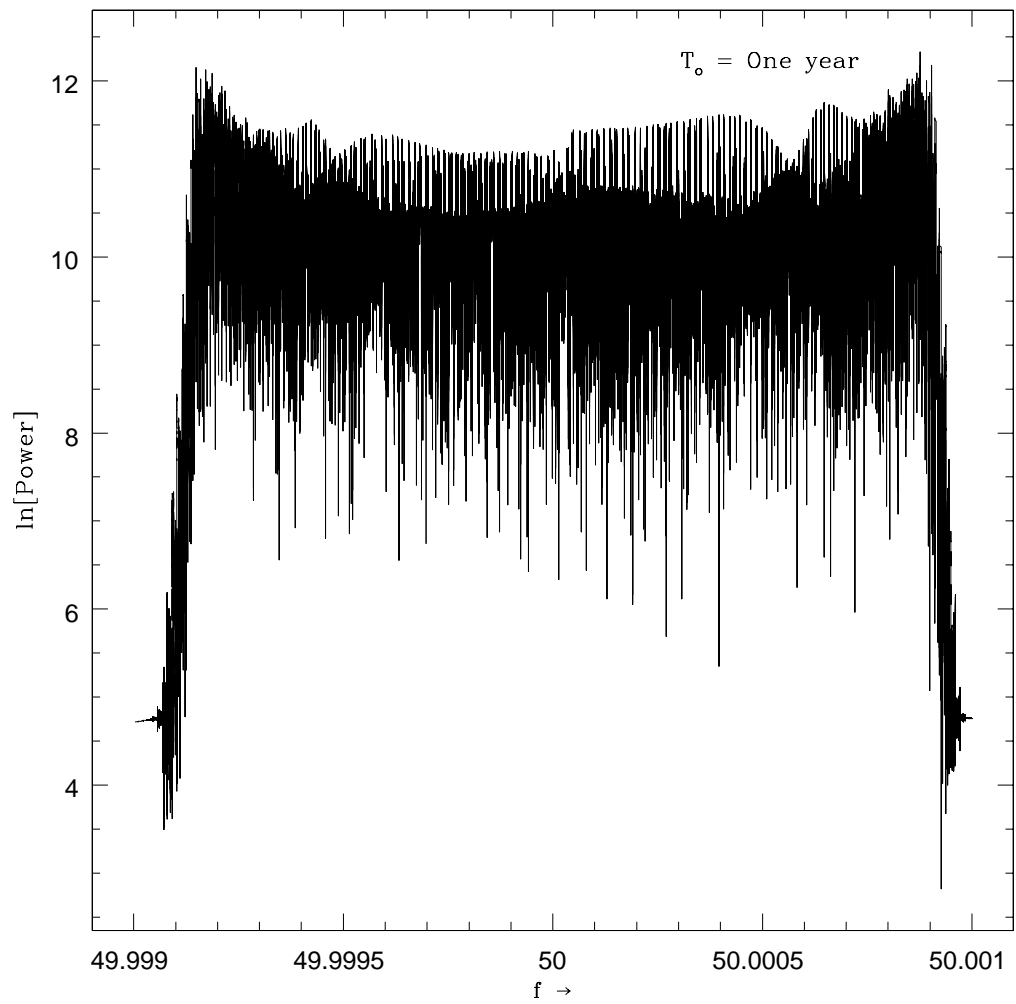,height=14.6cm}
\caption{Power spectrum of the complete response of a modulated signal of a
source located at $ (\pi /18 ,0)$ with a resolution of $3.17 \times 10^{-8}.$}
\label{fig:ycr}
\end{figure}

\begin{figure}[b]
\epsfig{file=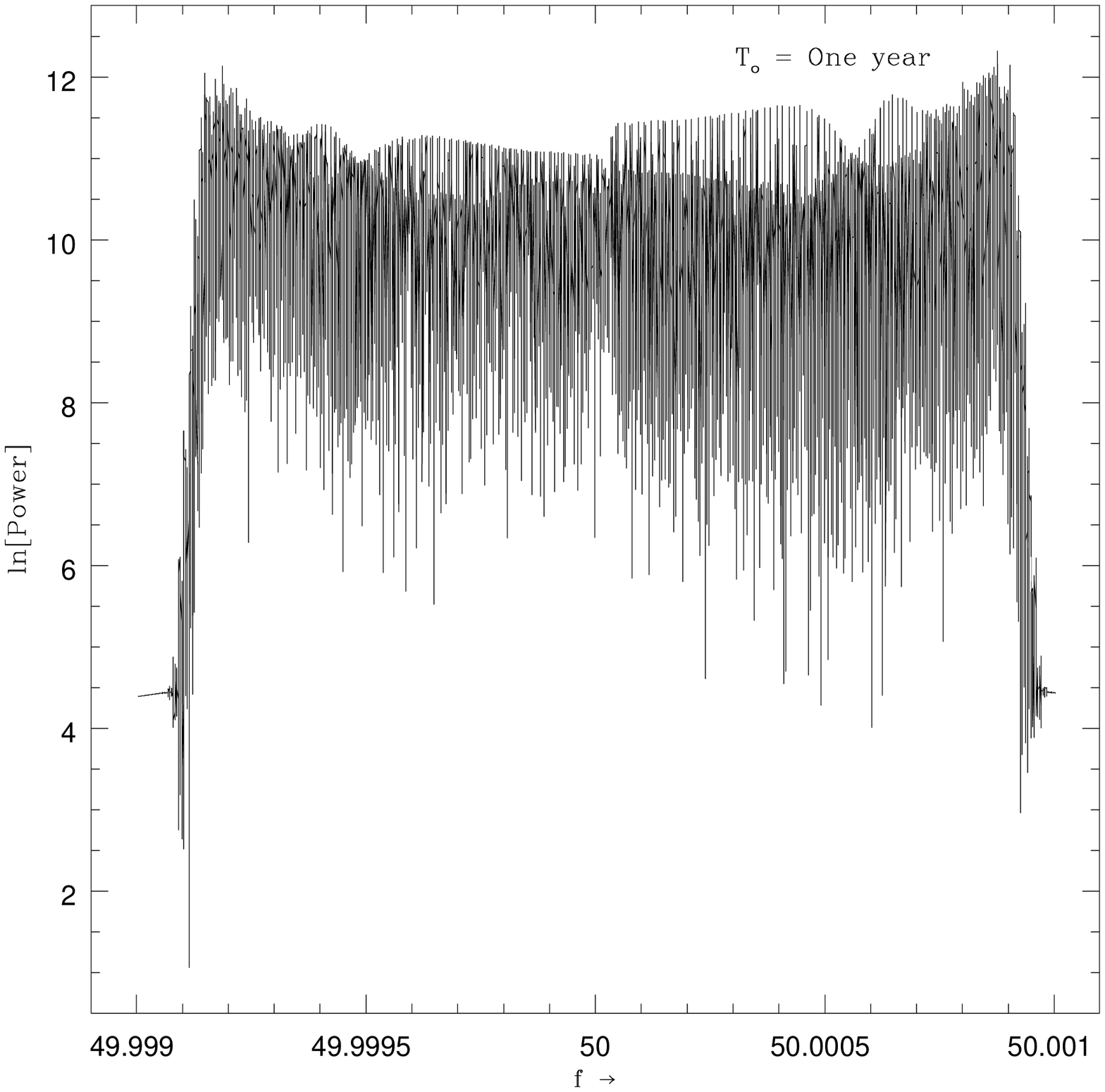,height=14.6cm}
\caption{Power spectrum of a Doppler modulated signal at frequencies $f + 2f_{rot}$  
of a source located at $(\pi /18 , 0)$ with a resolution of $3.17 \times 10^{-8}.$}
\label{fig:yp2frcr}
\end{figure}

\begin{figure}[b]
\epsfig{file=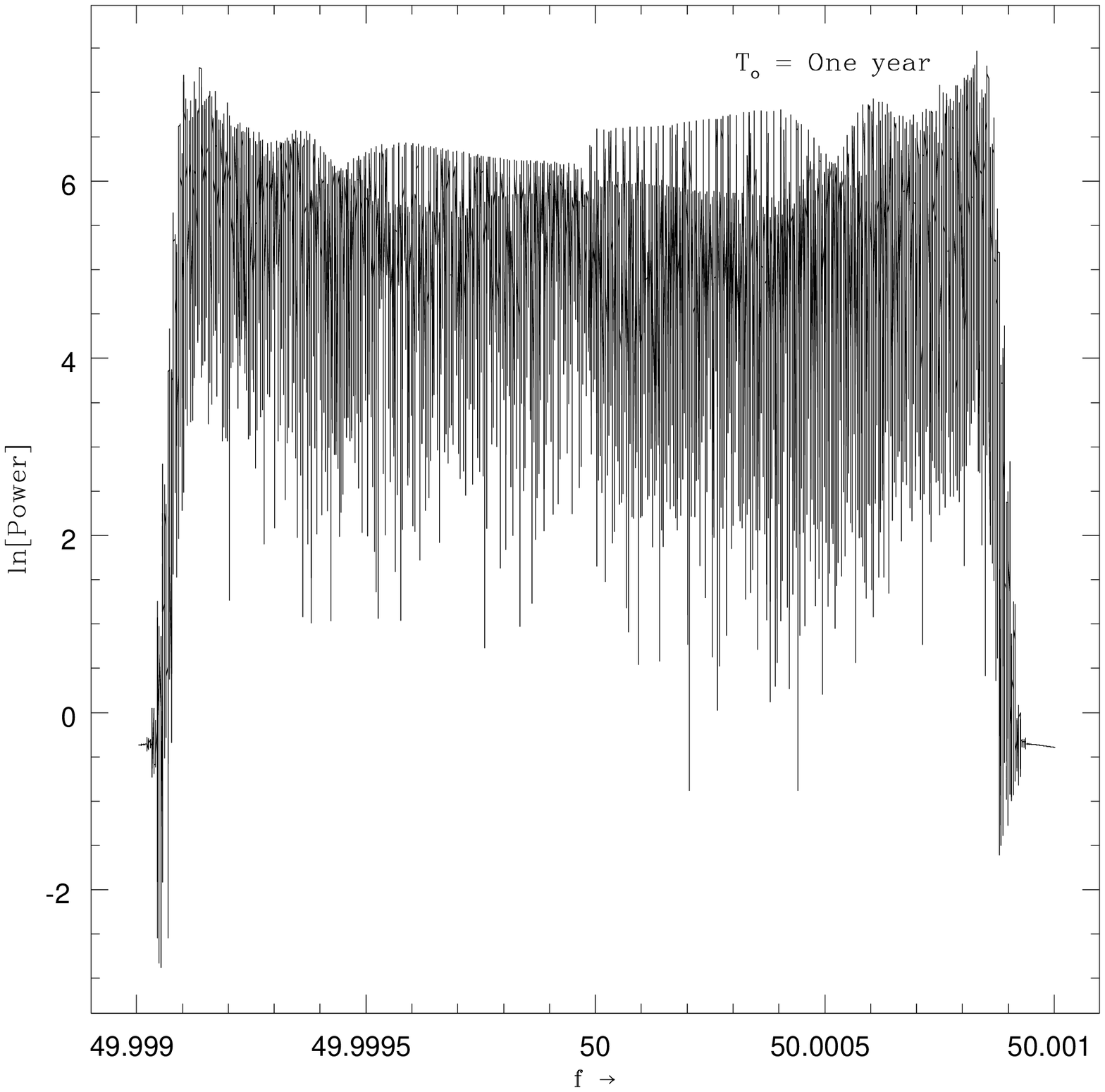,height=14.6cm}
\caption{Power spectrum of a Doppler modulated signal at frequencies $f - 2f_{rot}$
of a source located at $(\pi /18 , 0)$ with a resolution of $3.17 \times 10^{-8}.$}
\label{fig:ym2frcr}
\end{figure}

\begin{figure}[b]
\epsfig{file=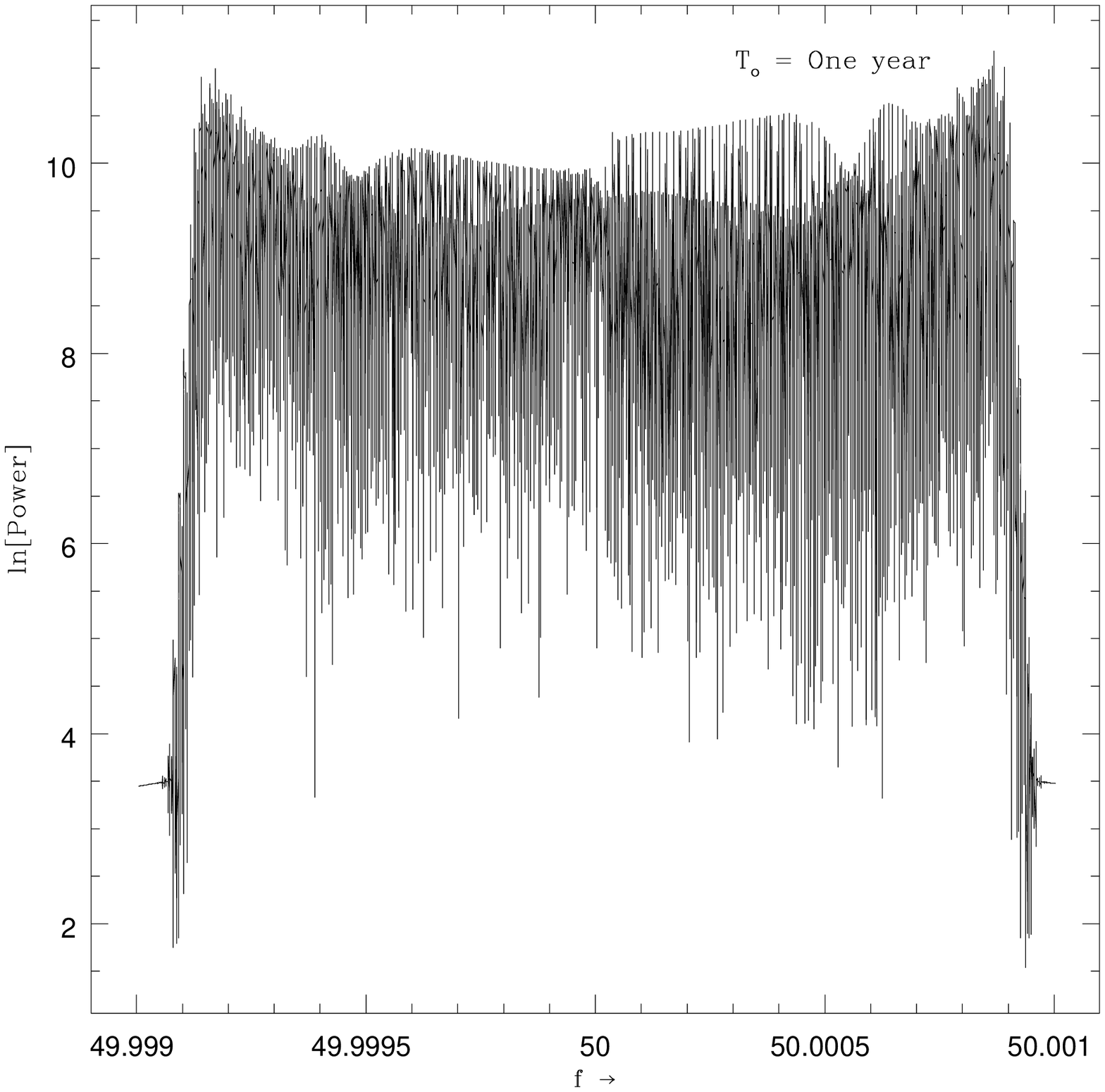,height=14.6cm}
\caption{Power spectrum of a Doppler modulated signal at frequencies $f + f_{rot}$
of a source located at $(\pi /18 , 0)$ with a resolution of $3.17 \times 10^{-8}.$}
\label{fig:ypfrcr}
\end{figure}

\begin{figure}[b]
\epsfig{file=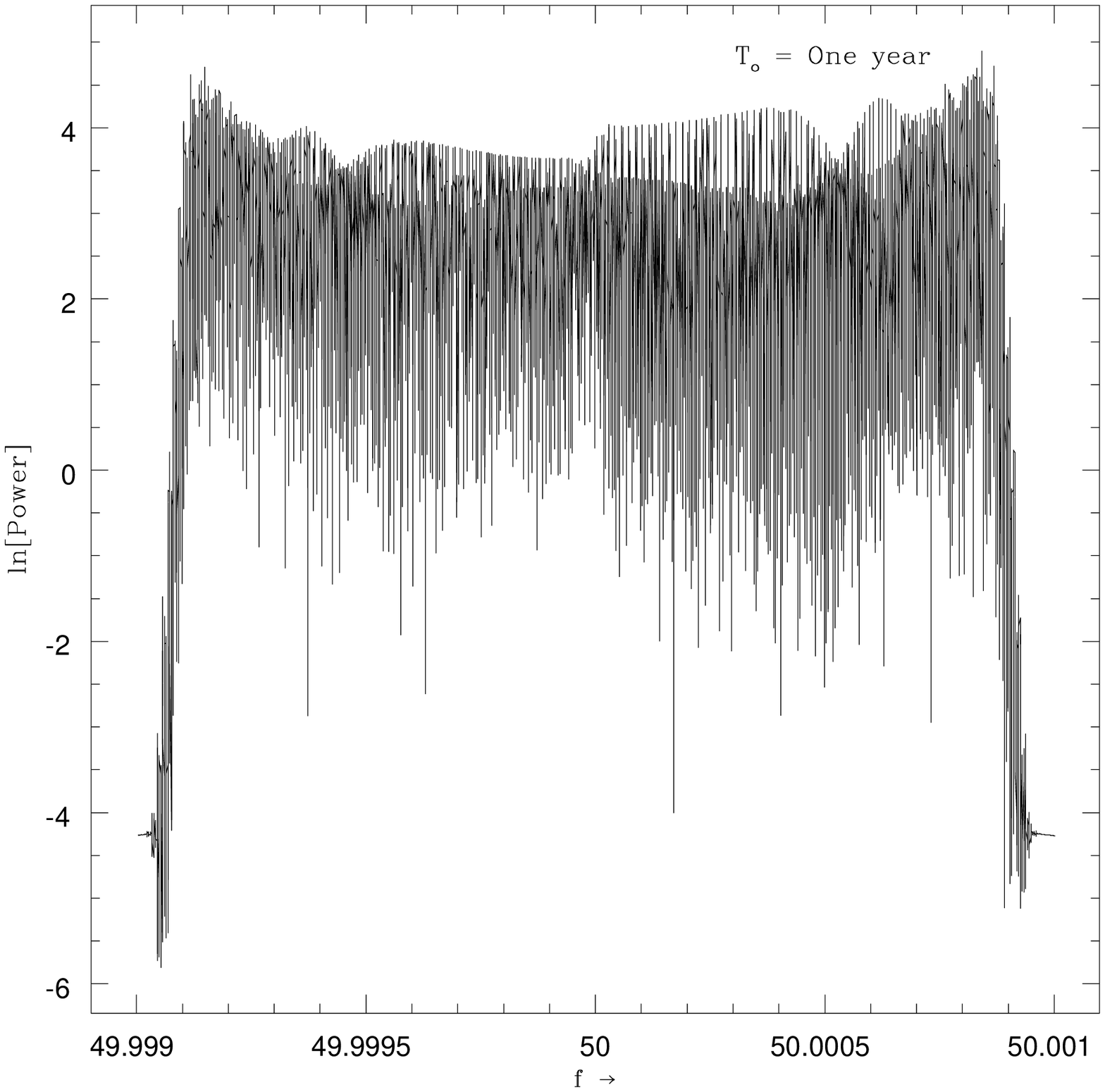,height=14.6cm}
\caption{Power spectrum of a Doppler modulated signal at frequencies $f - f_{rot}$
of a source located at $(\pi /18 , 0)$ with a resolution of $3.17 \times 10^{-8}.$}
\label{fig:ymfrcr}
\end{figure}

\begin{figure}[b]
\epsfig{file=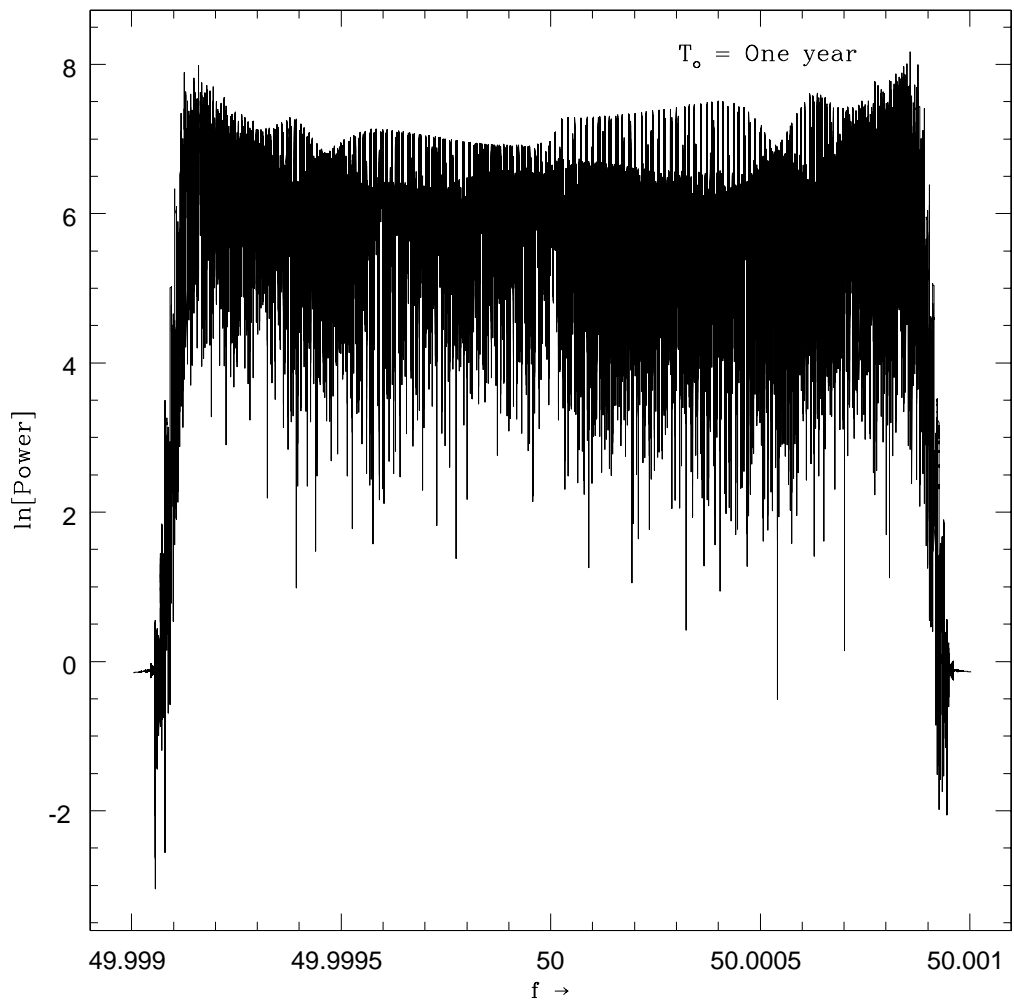,height=14.6cm}
\caption{Power spectrum of a Doppler modulated signal at frequencies  $f$
of a source located at $(\pi /18 , 0)$ with a resolution of $3.17 \times 10^{-8}.$}
\label{fig:y0shiftcr}
\end{figure}

 Figure~(\ref{fig:ycr}) shows the power spectrum of the complete response of
the Doppler modulated signal. We have kept here all parameters same as 
in FM. Figures~(\ref{fig:yp2frcr}), ~(\ref{fig:ym2frcr}), 
~(\ref{fig:ypfrcr}), ~(\ref{fig:ymfrcr}), and~(\ref{fig:y0shiftcr}) are the power 
spectrum  of complete response at frequencies $f + 2f_{rot}$, $f - 2f_{rot}$,
$f + f_{rot}$, $f - f_{rot}$ and $f$. In this case also we observe that the maximum  
power is associated with $f + 2f_{rot}$ and the least lies in $f - f_{rot}$.

\clearpage

\vspace{0.65cm}

\section{Fourier transform for arbitrary observation time}
\label{sec:ftt}
It is important to obtain the FT for arbitrary observation 
time. In this section we present this analysis. The results obtained will be
employed to outline how spin down of a pulsar due to gravitational
radiation back reaction or due to some other mechanism can be taken into account.

\noindent The FT for data of observation time $T_o$ is given via

\begin{equation}
\tilde{h}(f) \;= \;\int_0^{T_o}\cos [\Phi (t)]e^{-i2\pi f t} dt
\end{equation}

\vspace{0.5cm}

\noindent As usual, this splits into two terms as

\begin{equation} \tilde{h}(f)\ = I_{\nu_-} + I_{\nu_+} \; ;
 \label{eq:inut1}
 \end{equation}
 \begin{eqnarray}
\label{eq:inut}
 I_{\nu_-}& =& {1\over 2 w_{rot}}\int_0^{\xi_o} e^{i \left[
 \xi\nu_- +
{\cal Z}\cos (a\xi - \phi ) + {\cal N}\cos (\xi - \delta ) -{\cal R}  - {\cal 
Q} \right] } d\xi\, ,\\
I_{\nu_+}& = & {1\over 2 w_{rot}}\int_0^{\xi_o} e^{- i \left[
\xi\nu_+
+ {\cal Z} \cos (a\xi - \phi ) + {\cal N}\cos (\xi - \delta ) - {\cal R} - {
\cal Q} \right] } d\xi\, , \\
\nu_{\pm}& = & \frac{f_o \pm f}{f_{rot}} ; \quad \xi_o \; = \; w_{rot}T_o ;
\quad \xi \; = \; \xi_{rot} \; = \; w_{rot}t 
\end{eqnarray}

\vspace{0.5cm}

\noindent As $I_{\nu_+}$ contributes very little to $\tilde{h}(f)$
we drop $I_{\nu_+}$ and write $\nu$ in place of $\nu_-$. Using 
the identity~(\ref{eq:bessel}) we get 
\begin{eqnarray}
\tilde{h}(f) &\simeq & {1 \over 2 w_{rot}} e^{i ( 
- {\cal R} - {\cal Q})} \int_0^{\xi_o} e^{i\nu\xi} \left[ J_o( {\cal Z} ) + 2 
\sum_{k = 1}^{k =  \infty} J_k ({\cal Z}) i^k \cos k (a\xi - \phi )\right] 
 \nonumber \\
&& \times\,\left[J_o( {\cal N} ) + 2 \sum_{m = 1}^{m =  \infty} J_m ({\cal N}) i^m 
\cos m (\xi - \delta )\right] d\xi
\end{eqnarray}

\vspace{0.1cm}

\noindent After performing the integration and proceeding in a straight forward manner we have

\begin{equation}
\label{eq:hft}
\tilde{h}(f) \simeq  {\nu \over 2 w_{rot}} \sum_{k  =  - 
\infty}^{k = \infty} \sum_{m = - \infty}^{m =  \infty} e^{ i {\cal A}}{\cal 
B}[ \tilde{{\cal C}} - i\tilde{{\cal D}} ] \; ; \;
\end{equation}  

\vspace{0.1cm}

\begin{equation}
\left.\begin{array}{lcl}
\vspace{0.2cm}
{\cal A}&  = &{(k + m)\pi\over 2} - {\cal R} - {\cal Q}  \\
\vspace{0.2cm}
{\cal B} & = & {J_k({\cal Z}) J_m({\cal N})\over {\nu^2 - (a k + m)^2}} \\
\vspace{0.2cm}
\tilde{{\cal C}} &=& \sin \nu\xi_o \cos ( a k \xi_o + m\xi_o - k \phi - m \delta ) -  \\
\vspace{0.2cm}
&&{ a k + m \over \nu}\{\cos\nu\xi_o \sin ( a k \xi_o + m\xi_o - k \phi - m \delta )
+ \sin ( k \phi + m \delta )\}\nonumber \\
\vspace{0.2cm}
\tilde{{\cal D}} & = & \cos \nu\xi_o \cos ( a k \xi_o + m\xi_o - k \phi - m \delta ) + \\
&& {k a + m \over \nu}\sin \nu \xi_o \sin ( a k \xi_o + m\xi_o - k \phi - m \delta )
 - \cos ( k \phi + m \delta ) 
\end{array} \right\}
\end{equation}

\vspace{0.5cm}

\noindent The FT of the two polarisation states of the wave can now be written as
\begin{eqnarray}
\label{eq:hfpt}
{h}_+(f)&=&h_{o_+}\tilde{h}(f) \nonumber \\
&&\simeq {\nu h_{o_+}\over 2 w_{rot}} \sum_{k  =  - \infty}^{k =
\infty} \sum_{m = - \infty}^{m =  \infty} e^{ i {\cal A}}{\cal B}[ \tilde{{\cal C}}
- i\tilde{{\cal D}} ] \; ;
\end{eqnarray}
\begin{eqnarray}
\label{eq:hfct}
\tilde{h}_\times (f) &=&- i h_{o_\times}\tilde{h}(f) \nonumber \\
&&\simeq {\nu h_{o_\times}\over 2 w_{rot}} \sum_{k  =  - \infty}^{k =  
\infty} \sum_{m = - \infty}^{m =  \infty} e^{ i {\cal A}}{\cal B}[ \tilde{{\cal D}}
- i\tilde{{\cal C}} ]
\end{eqnarray}

\vspace{0.5cm}

\noindent Now it is simple matter to obtain the FT of complete response. One gets
\begin{eqnarray}
\label{eq:rft}
\tilde{R}(f) & = & e^{-i 2 \beta_o}\tilde{h}( f
+ 2 f_{rot})/2\left[ h_{o_+}( F_{1_+} + i F_{2_+} )
+ h_{o_\times} ( F_{2_\times} - i F_{1_\times} )\right] +\nonumber \\
&& e^{i2\beta_o}\tilde{h}( f - 2 f_{rot})/2\left[ h_{o_+}
( F_{1_+} - i F_{2_+} )
- h_{o_\times} ( F_{2_\times} + i F_{1_\times} )\right] +\nonumber \\
& & e^{-i\beta_o}\tilde{h}( f + f_{rot})/2\left[ h_{o_+}( F_{3_+} + i F_{4_+} )
+ h_{o_\times} ( F_{4_\times} - i F_{3_\times} )\right] +\nonumber \\
&& e^{i\beta_o}\tilde{h}( f - f_{rot})/2\left[ h_{o_+}( F_{3_+} - i F_{4_+} )
- h_{o_\times} ( F_{4_\times} + i F_{3_\times} )\right] +\nonumber \\
&&\tilde{h}(f)\left[ h_{o_+}F_{5_+} - i h_{o_\times}F_{5_\times}\right]
\end{eqnarray}

\begin{figure}[b]
\epsfig{file=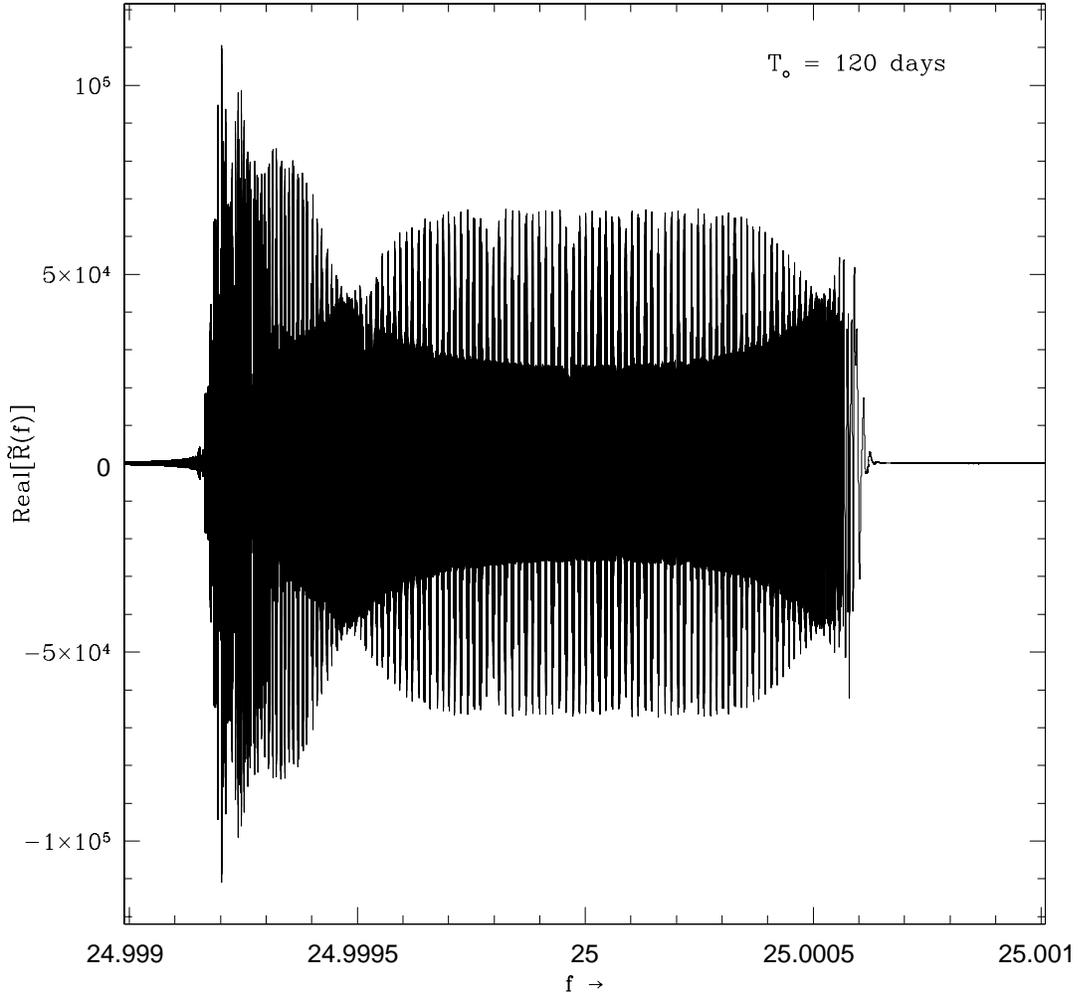,height=14.6cm}
\caption{FT of a FM signal of a source located at $(\pi /9 , \pi /4)$
with a resolution of $9.67 \times 10^{-8}.$}
\label{fig:120realfm}
\end{figure}

\begin{figure}[b]
\epsfig{file=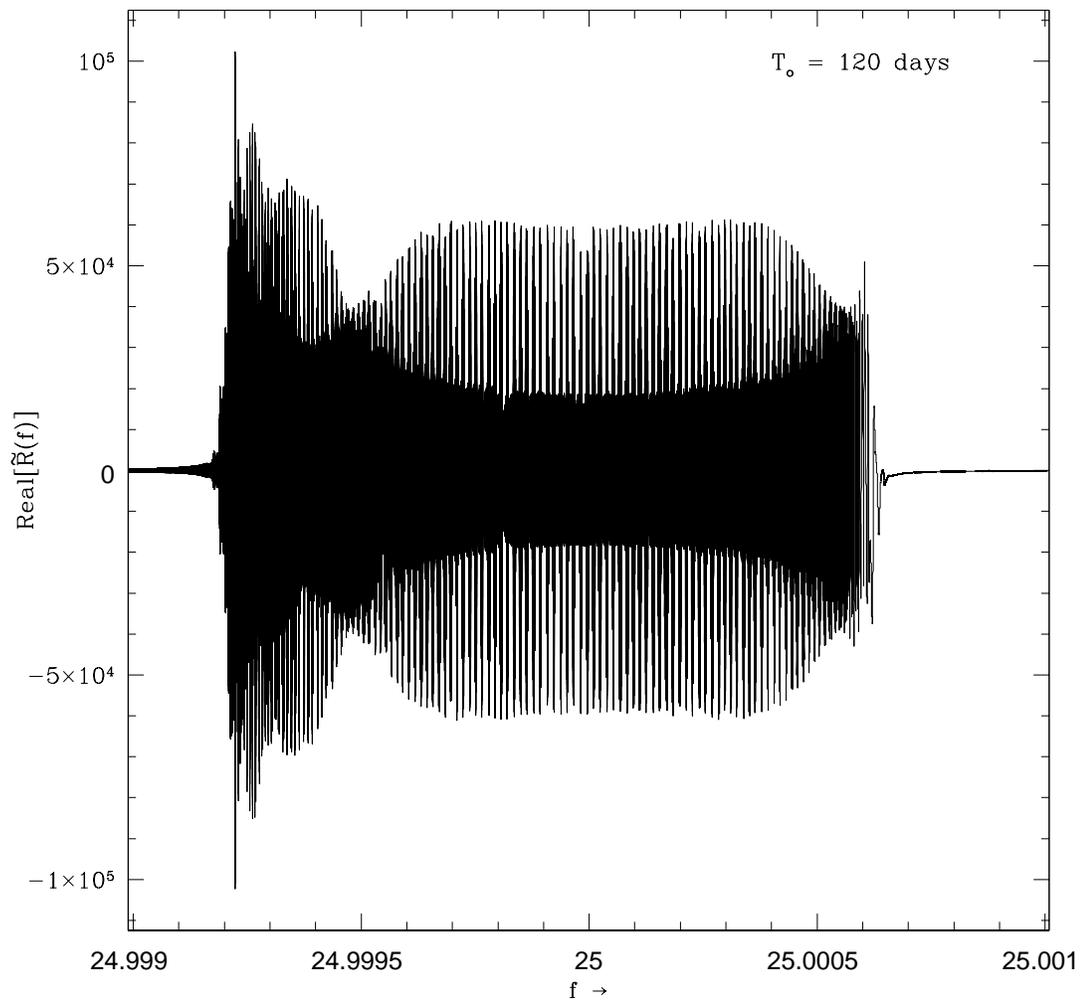,height=14.6cm}
\caption{FT of the complete response of a Doppler modulated of a 
source located at $(\pi /9 , \pi /4)$ with a resolution of $9.67
\times 10^{-8}.$}
\label{fig:120realcr}
\end{figure}

\begin{figure}[b]
\epsfig{file=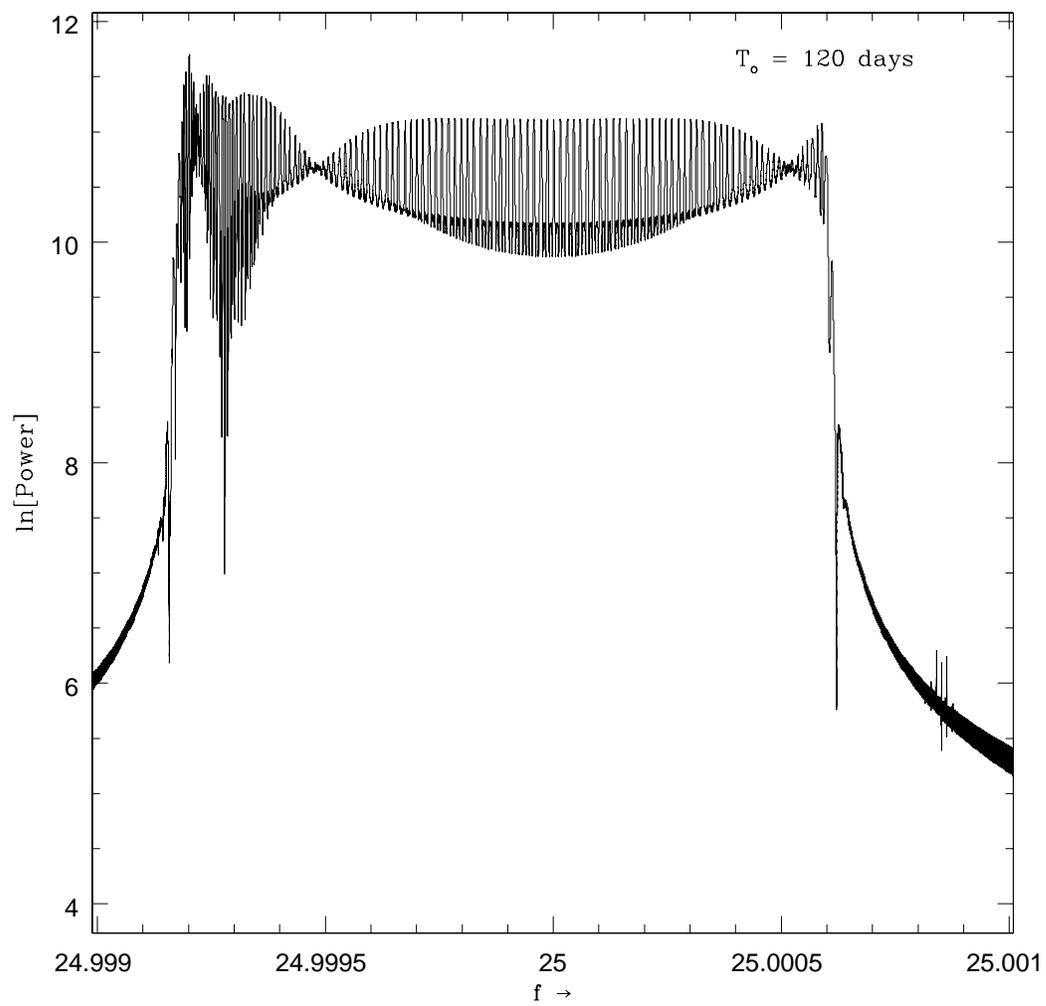,height=14.6cm}
\caption{Power spectrum of a FM signal of a source located at $(\pi /9 , \pi /4)$
with a resolution of $9.67 \times 10^{-8}.$}
\label{fig:120fmpower}
\end{figure}

\begin{figure}[b]
\epsfig{file=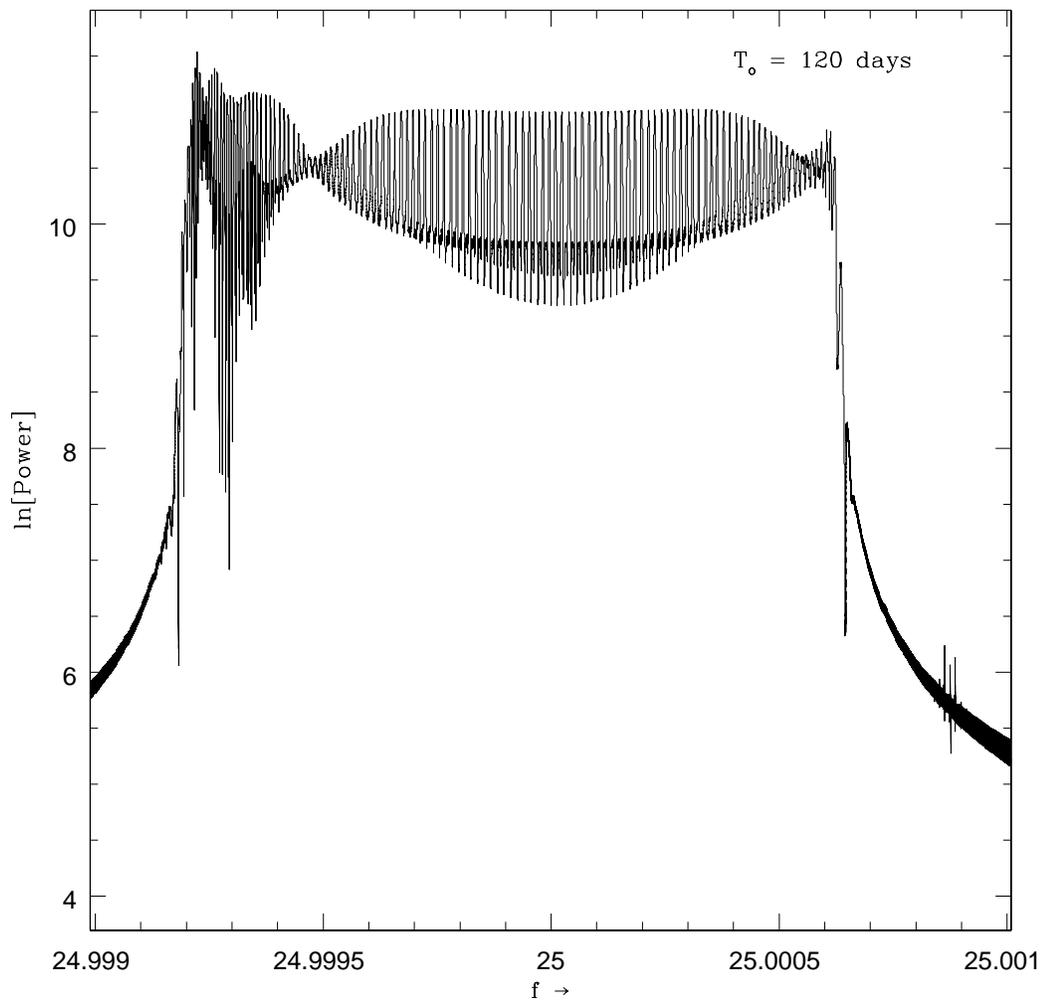,height=14.6cm}
\caption{Power spectrum of the complete response of a Doppler modulated signal
of a source located at $(\pi /9 , \pi /4)$ with a resolution of $9.67 \times 10^{-8}.$}
\label{fig:120cr}
\end{figure}

\begin{figure}[b]
\epsfig{file=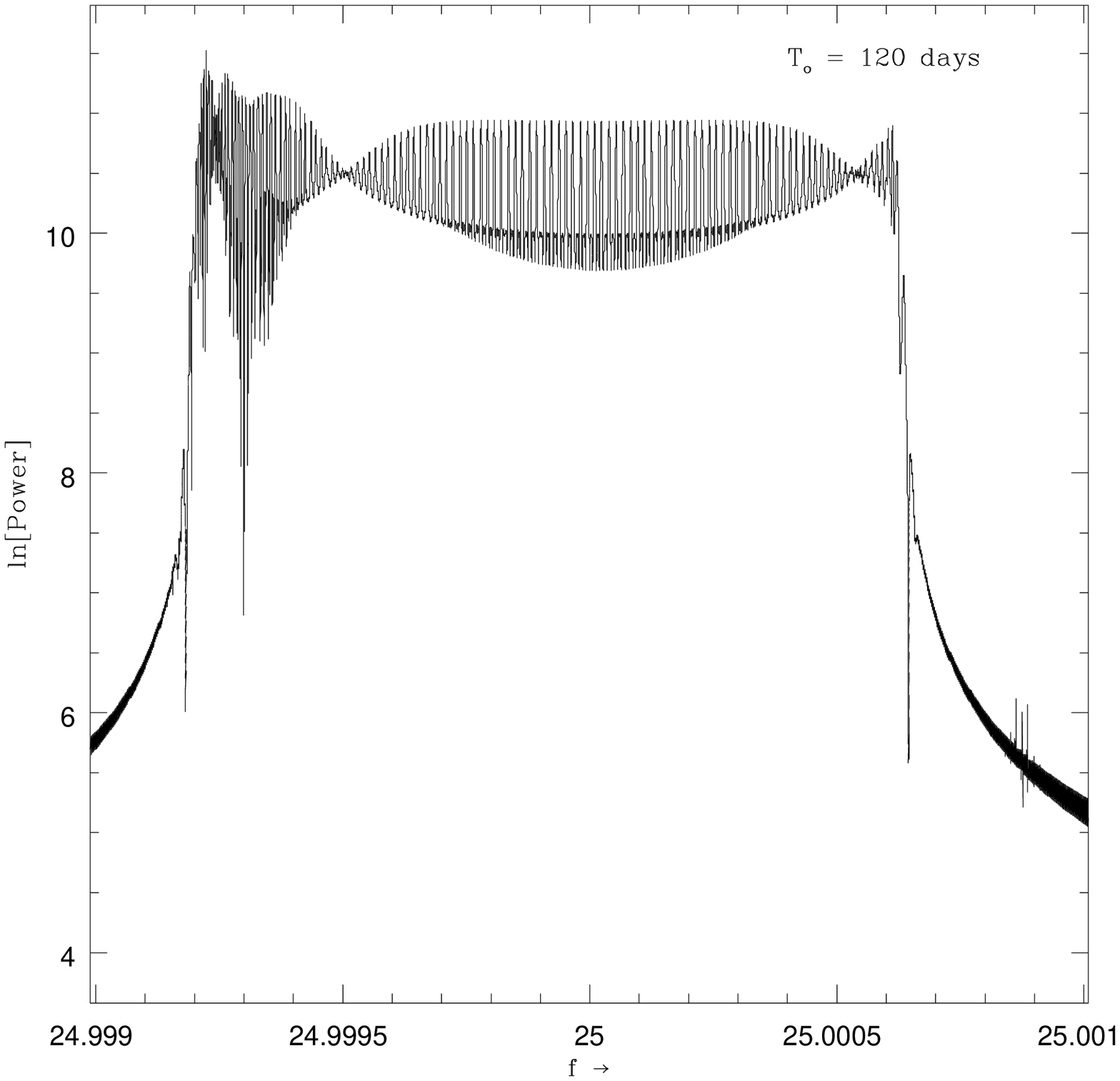,height=14.6cm}
\caption{Power spectrum of a Doppler modulated signal at frequencies  $f + 2f_{rot}$
of a source located at $(\pi /9 , \pi /4)$ with a resolution of $9.67 \times 10^{-8}.$}
\label{fig:120p2frcr}
\end{figure}

\begin{figure}[b]
\epsfig{file=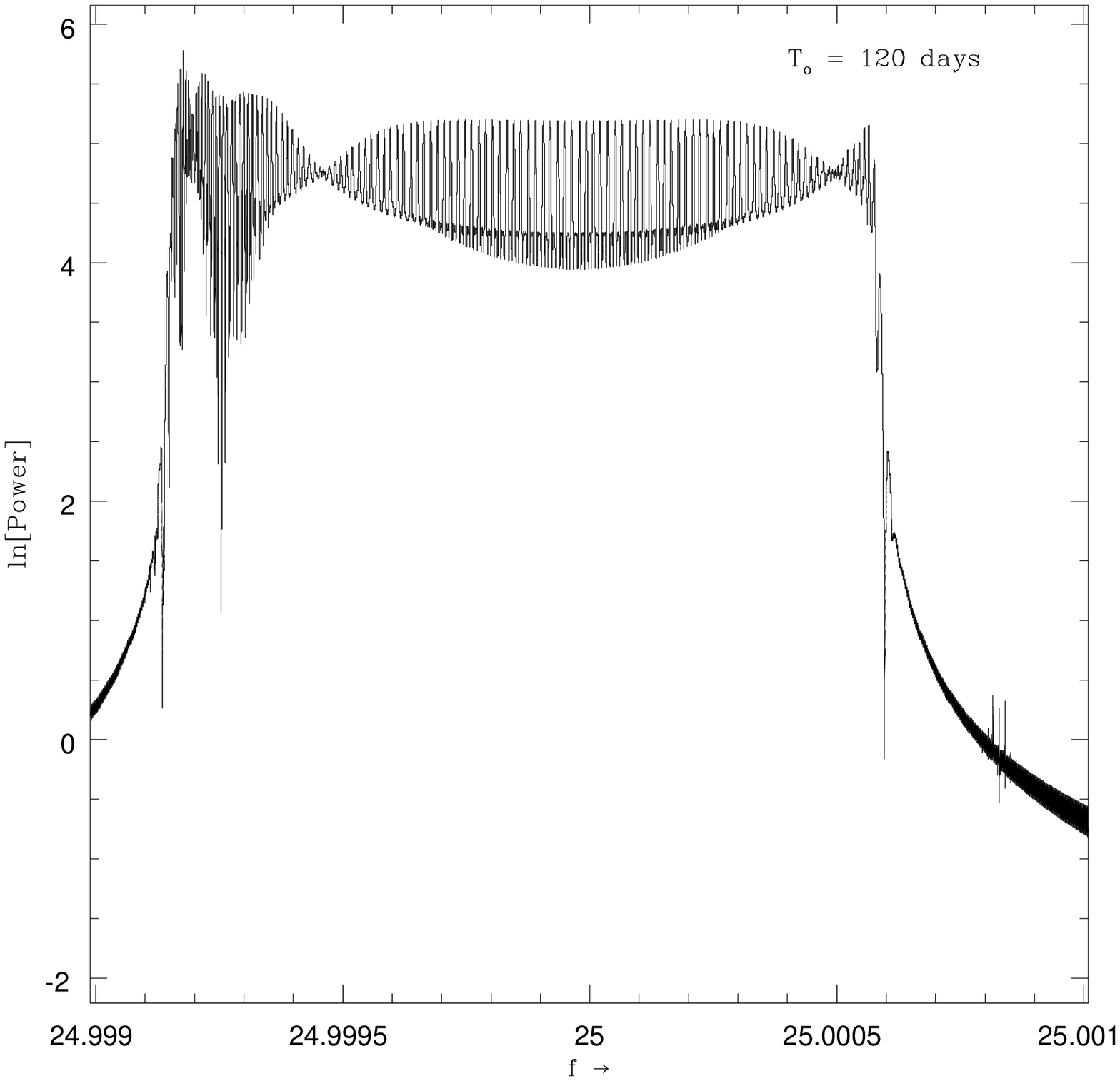,height=14.6cm}
\caption{Power spectrum of a Doppler modulated signal at frequencies  $f - 2f_{rot}$
of a source located at $(\pi /9 , \pi /4)$ with a resolution of $9.67 \times 10^{-8}.$}
\label{fig:120m2frcr}
\end{figure}

\begin{figure}[b]
\epsfig{file=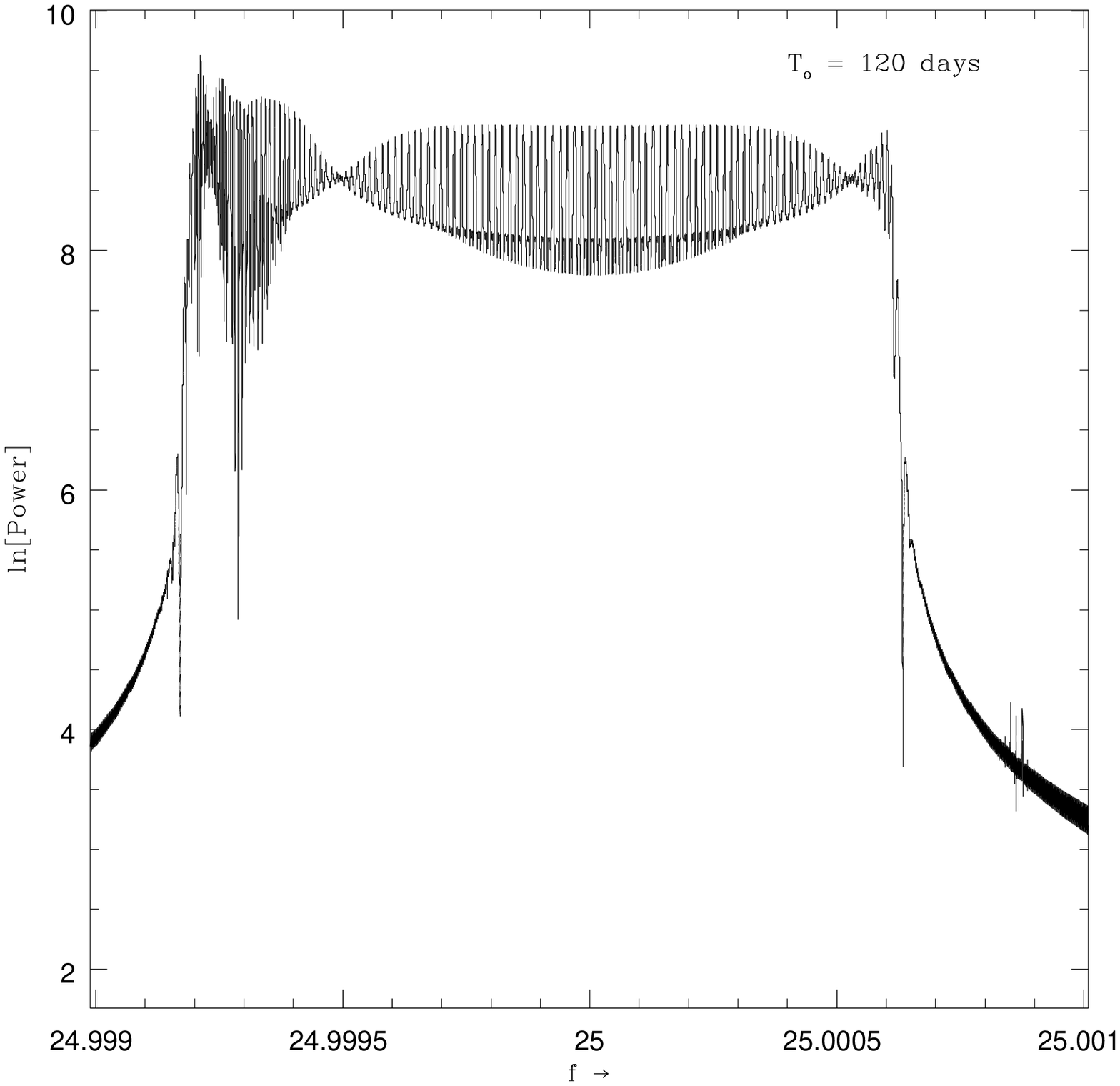,height=14.6cm}
\caption{Power spectrum of a Doppler modulated signal at frequencies $f + f_{rot}$
of a source located at $(\pi /9 , \pi /4)$ with a resolution of $9.67 \times 10^{-8}.$}
\label{fig:120pfrcr}
\end{figure}

\begin{figure}[b]
\epsfig{file=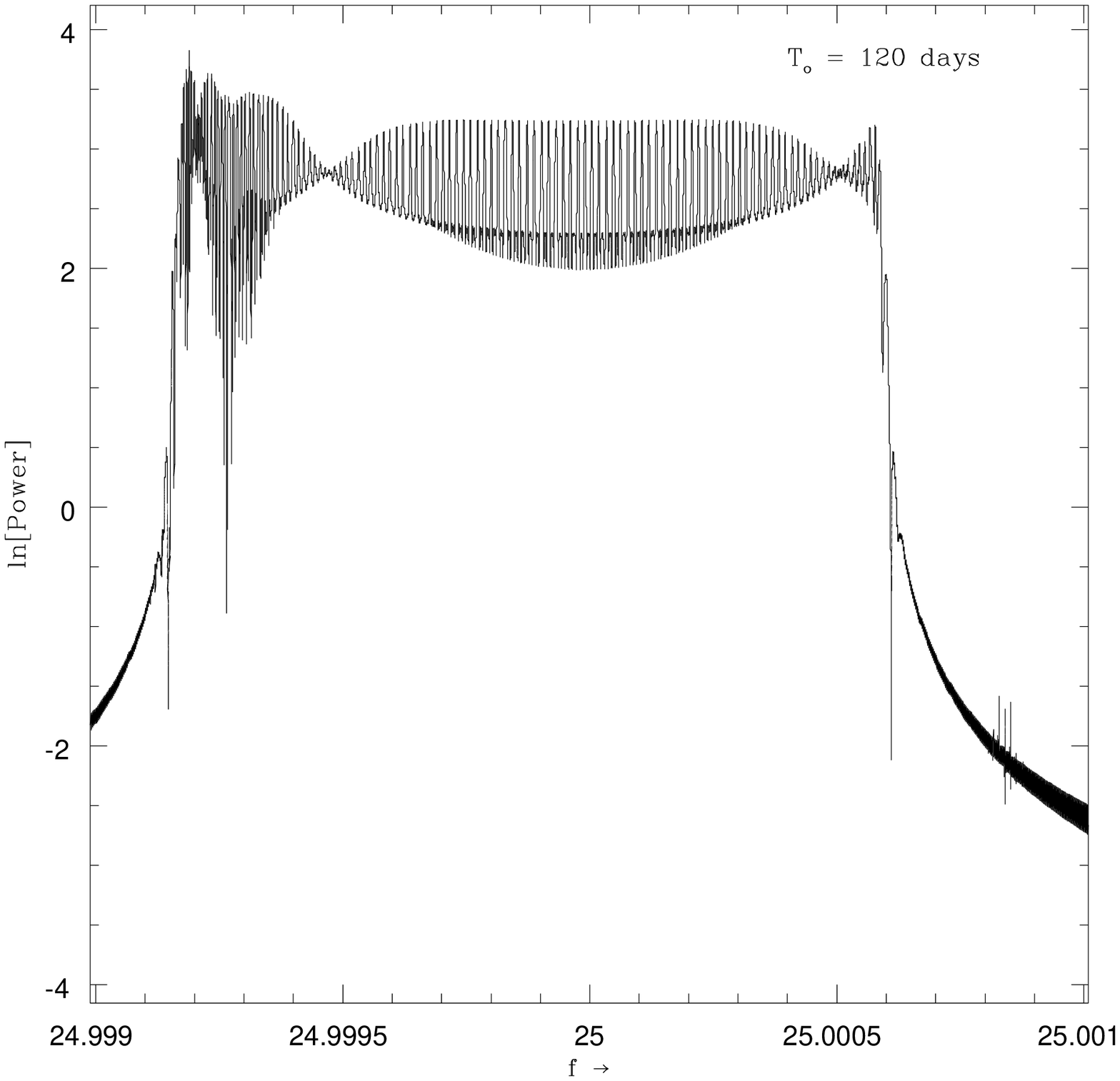,height=14.6cm}
\caption{Power spectrum of a Doppler modulated signal at frequencies $f - f_{rot}$
of a source located at $(\pi /9 , \pi /4)$ with a resolution of $9.67 \times 10^{-8}.$}
\label{fig:120mfrcr}
\end{figure}

\begin{figure}[b]
\epsfig{file=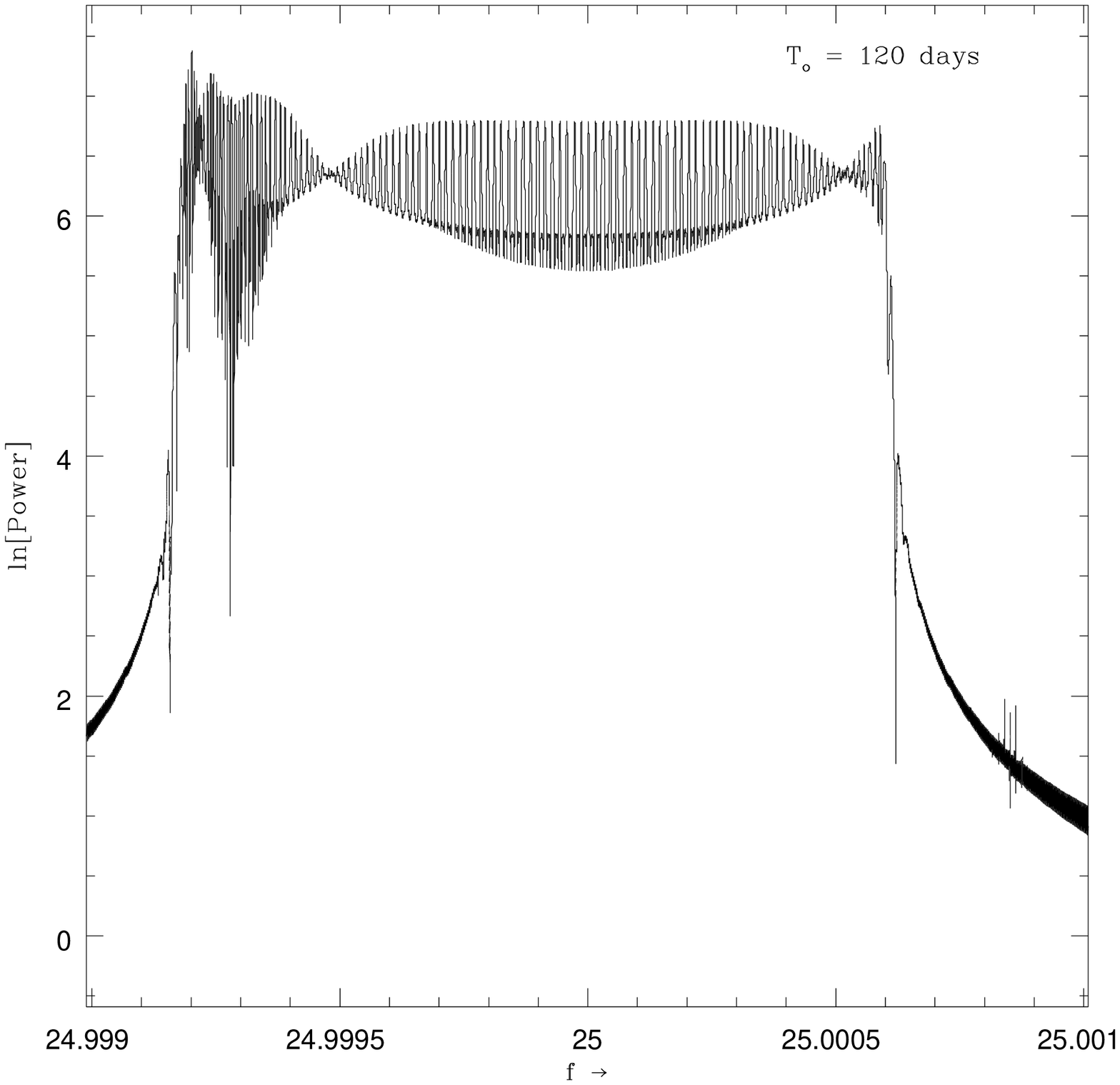,height=14.6cm}
\caption{Power spectrum of a Doppler modulated signal at frequencies $f$
of a source located at $(\pi /9 , \pi /4)$ with a
resolution of $9.67 \times 10^{-8}.$}
\label{fig:d0shiftcr}
\end{figure}
\clearpage

\noindent The FT of FM and complete response of the
detector for a data of 120 days and 

\begin{equation}
\left.\begin{array}{lll}
\vspace{0.2cm}
f_o = 25\; Hz\, , & h_o = h_\times = 1 & \\
\vspace{0.2cm}
\alpha = \pi /6\, , & \beta_o = \pi /3\, , & \gamma =  2\pi /3 \, , \\
\theta = \pi /9\, , & \phi = \pi /4\, , &  \psi = \pi /4. \\
\end{array} \right\}
\label{eq:120loc}
\vspace{0.4cm}
\end{equation}

\vspace{0.1cm}

\noindent are plotted in Figs.~(\ref{fig:120realfm}) and~(\ref{fig:120realcr})
with a resolution of $\,1/2T_o \approx 9.67 \times 10^{-8} Hz.$
The corresponding power spectra are plotted respectively in
Figs.~(\ref{fig:120fmpower}) and~(\ref{fig:120cr}). 
Figs.~(\ref{fig:120p2frcr}),~(\ref{fig:120m2frcr}),~(\ref{fig:120pfrcr})
and~(\ref{fig:120mfrcr}) represent the power spectra of complete response of the detector
at $f + 2f_{rot}$, $f - 2f_{rot}$, $f + f_{rot}$ and $f - f_{rot}$ whereas $\;$
Fig.~(\ref{fig:d0shiftcr}) $\;$ shows the
response of the detector at $f$. In this case also 
the most of the power of the signal lies in the frequency $f + 2f_{rot}$ 
and the least with $f - f_{rot}.$ The codes for computing the frequency
modulated signal and
complete response of the detector, written in FORTRAN, are given in appendices
A and B.

\section{Spin down}
\label{sec:fts}
Pulsars loose their rotational energy by processes like electro-magnetic breaking,
emission of particles and emission of GW. Thus, the rotational 
frequency is not completely stable, but varies over a time scale which is of the 
order of the age of the pulsar. Typically, younger pulsars have the largest spin
down rates. Current observations suggest that spin down is primarily due to
electro-magnetic breaking (Manchester, 1992 and Kulkarni, 1992).
Over the entire observing time $T_o$ the frequency drift would be small
but it may be taken into account for better sensitivity.   
To account this aspect we consider the evaluation of FT in a sequence of 
time windows by splitting the interval $0-T_o$ in M equal
 parts, each of interval $\bigtriangleup t$ $(T_o = M\bigtriangleup t$) 
such that the signal over a window may be treated as monochromatic. The
strategy is to evaluate the FT over the window and finally to add the result.
This process has been suggested by Brady and Creighton (2000) and Schutz (1998)
in numerical computing and
has been termed as {\it stacking\/} and {\it tracking\/}. For any such window
let the time interval of data under
consideration corresponds to $t = t_o + n\bigtriangleup t$ and $t = t_o + (n + 1)
\bigtriangleup t$ where $t_o$ is the instant representing the start of the data
set and $0 \le n \le M - 1$.
The window under consideration is the $n_{th}$ window. Now we have
\begin{eqnarray}
I& =& \int_{t_o + n\bigtriangleup t}^{t_o + (n+1)\bigtriangleup t} h(\bar{t})
e^{-i2\pi f\bar{t}}d\bar{t} \nonumber \\
\label{eq:s1}
& = &\int_0^{\bigtriangleup t} h(t+t_o\;+
n\bigtriangleup t ) e^{-i2\pi f (t+t_o+n\bigtriangleup t )}dt\; ;\\
\bar{t} & = & t + t_o + n\bigtriangleup t
\label{eq:s2}
\end{eqnarray}

\noindent Hence, the FT to account spin down is given via

\begin{equation}
\label{eq:s3}
\left[\tilde{h}(f)\right]_s \;= \;\int_0^{\bigtriangleup t}\cos [\Phi (t + t_o
+ n\bigtriangleup t )]e^{-i2\pi f (t + t_o + n\bigtriangleup t )} dt
\end{equation}

\vspace{0.5cm}

\noindent Taking the initial time of the data set
\begin{equation}
\label{eq:s4}
t_o = 0
\end{equation}

\vspace{0.1cm}

\noindent and proceeding in the same manner as in previous section the FT is
obtained as
\begin{eqnarray}
\left[\tilde{h}(f)\right]_s &\simeq & {1 \over 2 w_{rot}} e^{i \left[
2\pi n(f_o - f)\bigtriangleup t 
- {\cal R} - {\cal Q}\right]} \int_0^{\bigtriangleup t}
e^{i\nu\xi} \left[ J_o( {\cal Z} ) + \right.\nonumber \\
&&\left.2\sum_{k = 1}^{k =  \infty} J_k ({\cal Z}) i^k \cos k (a\xi - \lambda )\right] 
\;\times \nonumber \\
&& \left[J_o( {\cal N} ) + 2 \sum_{m = 1}^{m =  \infty} J_m ({\cal N}) i^m
\cos m (\xi - \zeta )\right] d\xi
\end{eqnarray}

\vspace{0.1cm}

\noindent After integration we get,
\begin{eqnarray}
\label{eq:hfs}
\left[\tilde {h}(f)\right]_s& =& {\nu \over 2 w_{rot}} \sum_{k  =  - \infty}^{k =  
\infty} \sum_{m = - \infty}^{m =  \infty} e^{ i {\cal A}_s}{\cal B} \left[
{\cal C }_s  -  i {\cal D}_s \right] 
\end{eqnarray}
\noindent where
\begin{equation}
\left.\begin{array}{lcl}
\vspace{0.2cm}
{\cal A}_s&  = &{(k + m)\pi\over 2} + 2\pi n\bigtriangleup t(f_o - f)  - {\cal R} - {\cal Q}\\
\vspace{0.2cm}
{\cal B} & = & {J_k({\cal Z}) J_m({\cal N})\over {\nu^2 - (a k + m)^2}} \, , \\
\vspace{0.2cm}
{\cal C}_s &= &  \sin (\nu \tau )\cos ( a k \tau  + m \tau - k \lambda - m \zeta ) \\
\vspace{0.2cm}
&&- { a k + m \over \nu}\{\cos (\nu \tau ) \sin ( a k \tau + m \tau - k \lambda  - m \zeta ) \\
\vspace{0.2cm}
&& + \sin ( k \lambda + m \zeta )\}\, , \\
 {\cal D}_s & = & \cos (\nu \tau )\cos (  a k \tau +  m \tau - k \lambda - m \zeta ) \\
\vspace{0.2cm}
&& + {a k + m \over \nu}\sin (\nu \tau ) \sin ( a k \tau + m \tau - k \lambda - m \zeta )\\
\vspace{0.2cm}
&&  - \cos ( k \lambda + m \zeta) \, , \\
\vspace{0.2cm}
\lambda & = & \phi -  a n \tau_o \, , \quad \zeta \; = \; \delta -  n\tau_o \, ,
\quad \tau \; = \; \xi_o \; = \; w_{rot}T_o\, ,\\
\tau_o &= &w_{rot}\bigtriangleup t \, , \quad n \; = \; 0, 1, 2, 3,......., M - 1.
\end{array} \right\}
\label{eq:sloc}
\end{equation}

\vspace{0.1cm}

\noindent The FT of the complete response would now be given via
\begin{eqnarray}
\label{eq:rfs}
\left[\tilde{R}(f)\right]_s & = & e^{-i 2 \beta_o}\left[\tilde{h}( f
+ 2 f_{rot})/2\right]_s\left[ h_{o_+}( F_{1_+} + i F_{2_+} )
+ h_{o_\times} ( F_{2_\times} - i F_{1_\times} )\right] +\nonumber \\
&& e^{i2\beta_o}\left[\tilde{h}( f - 2 f_{rot})/2\right]_s\left[ h_{o_+}
( F_{1_+} - i F_{2_+} )
- h_{o_\times} ( F_{2_\times} + i F_{1_\times} )\right] +\nonumber \\
& & e^{-i\beta_o}\left[\tilde{h}( f + f_{rot})/2\right]_s\left[ h_{o_+}( F_{3_+} + i F_{4_+} )
+ h_{o_\times} ( F_{4_\times} - i F_{3_\times} )\right] +\nonumber \\
&& e^{i\beta_o}\left[\tilde{h}( f - f_{rot})/2\right]_s\left[ h_{o_+}( F_{3_+} - i F_{4_+} )
- h_{o_\times} ( F_{4_\times} + i F_{3_\times} )\right] +\nonumber \\
&&\left[\tilde{h}(f)\right]_s\left[ h_{o_+}F_{5_+} - i h_{o_\times}F_{5_\times}\right]
\end{eqnarray}

\vspace{0.65cm}

\section{N-component signal}
\label{sec:ftn}
The FT in Eqs.~(\ref{eq:hft}) and~(\ref{eq:rft}) are for a
pulsar which emits GW signal at single frequency $f_o$. But there are known 
physical mechanisms which generate GW 
signals consisting of many components. An axially symmetric pulsar
undergoing free precession, emits quadrupole GW at
two frequencies, one equal to the sum of the spin frequency
and the precession frequency, and the other twice of it (Zimmermann, 1979, 1980). 
The quadrupole GW from a triaxial ellipsoid rotating
about one of its principal axes consists of one component only (Thorne, 1987). 
In this case the signal has frequency about 
twice the spin frequency of the star. In general, if a star is
non-axisymmetric and precesses, the GW signal consists
of more than two components. For the case of triaxial ellipsoid and
small wobble angle there is a third component with frequency equal
to twice the spin frequency of the star (Zimmermann, 1979, 1980). Recently,
new mechanisms e.g.
r-mode instability of spinning neutron stars (Anderson, 1998;
Lindblom et al., 1998; Owen, et al., 1998) and temperature asymmetry in the
interior of the neutron star with miss-aligned spin axis (Bildsten, 1998)
have been discussed in the literature.

\par In view of the above discussion CGW signal may 
consists of frequencies which are multiple of some basic frequencies. An
analysis of the GW data of N-component of signal has been made recently by
Jaranowski and Kr\'{o}lak (2000). We
in this section present Fourier analysis of N-component CGW signal. We model
the N-component signal as
\begin{eqnarray}
h(t)&=& \sum_{l = 1}^N h_l(t)\; ;\\
h_{l+}(t)&=& h_{o_{l+}}\cos [\Phi_l (t)] \\
h_{l\times} (t)&=& h_{o_{l\times}}\sin [\Phi_l (t)]\qquad \qquad l=1,2,.....,N\; ;
\end{eqnarray}
\begin{eqnarray}
\Phi_l(t) & = & 2\pi f_l \left[t + {\cal Z}_l\cos (a\xi_{rot} - \phi ) + 
{\cal N}_l\cos (\xi_{rot} - \delta_l ) - {\cal R}_l - {\cal Q}_l \right]
\end{eqnarray}

\noindent where $f_l$ represent the component frequency of the signal.
Let us write
\begin{equation}
 h_l(t) = \cos[\Phi_l (t)]
\end{equation}

\vspace{0.1cm}

\noindent It is trivial matter to obtain $\tilde {h}_l(f)$, $\tilde {R}_l(f)$
and to get
\begin{eqnarray}
\tilde {h}_N(f)& =& \sum_l \tilde{h}_l(f) \qquad and\\
\tilde {R}_N(f)& =& \sum_l \tilde{R}_l(f)
\end{eqnarray}

\vspace{0.1cm}

\noindent We do not continue this analysis as the requisite formalism is analogous
to what we have presented in Chapter 3 and earlier sections of the present
Chapter.

\vspace{0.65cm}

\section{Discussion and conclusions}
\label{sec:concl4}
The analysis and results obtained in earlier Chapter regarding FT of the
response of a Laser Interferometer have been generalised in the present
Chapter. In this context following points must be noted.
\renewcommand{\theenumii}{\roman{enumii}}
\begin{enumerate}
\item For longer observation time, say, 120 days the resolution provided by
FFT (equal to $1/T_o$) is sufficient to represent the structure of side bands.

\item Although in every case discussed, it turned out that the maximum power lies
in the frequency $f + 2f_{rot}$. However, this is not established conclusively
whether this result is generic. In any case, for any given detector location, we
can know in advance if similar type of behaviour is obtained.

\item The computer time required in computing analytical FT depends on the orientation
of the source and is independent of time interval of the observation data set.
In contrast to this the computation cost increases for FFT with increase of
the data set.

\item Throughout our analysis in Chapters 3 and 4 we have employed following
conditions.
\begin{enumerate}
\item The phase of the wave is zero at $t = 0$.

\item The observation time of the data set is from $t = 0$ to $t = T_o$.
\end{enumerate}

\item As remarked in Chapter 3 the requirement 4-(i) may be acheived by
translation of time-origin. Now let us see how the condition
4(ii) can be relaxed. Let the data set is taken for
\begin{equation}
t = t_i \quad to \quad  t = t_i + T_o
\end{equation}

\noindent Eqs.~(\ref{eq:s1}), ~(\ref{eq:s2}), ~(\ref{eq:s3}) and~(\ref{eq:s4})
reveal that the results obtained there corresponds to data set
\begin{equation}
t = n \bigtriangleup t \quad  to \quad  t = (n + 1)\bigtriangleup t
\end{equation}

\noindent To obtain the sought generalisation we set
\begin{equation}
t_i = n \bigtriangleup t \quad  and \quad  T_o = \bigtriangleup t
\end{equation}

\noindent and may use the results of section~(\ref{sec:ftt}) by noting that
for the present case we have
\begin{eqnarray}
\lambda & = & \phi -  n w_{orb} \bigtriangleup t \quad \Longrightarrow \quad
\lambda \; = \; \phi -  w_{orb}t_i\\
\zeta & = & \delta -  n w_{rot}\bigtriangleup t \quad \Longrightarrow \quad
\zeta \; = \; \delta -  w_{rot}t_i
\end{eqnarray}
\end{enumerate}

\chapter{All sky search: Study of templates}

\section{Introduction}

\indent Gravitational wave Laser Interferometer antennas are essentially omni - directional
with their response better than 50\% of the average over 75\% of the whole sky
(Grishchuk et al., 2000). Hence the data analysis systems will have to carry
out all sky searches for the sources. We know that the amplitude of intense GW
believed bathing Earth is very small, as compared to the sensitivity of GW
detectors and is further masked by the dominant noise. In these circumstances,
CGW sources are of prime importance because for such sources we can achieve
enhanced SNR by investigating longer observation data set. However, a long
observation time introduces modulation effects, arising due to relative
motion of detector and the source. As a consequence, there results distribution
of power in a forest of sidebands to such an extent that there is reduction
to the tune of 90\% of the expected power due to AM. The problem of all sky search gains
another dimension in view of the fact that there are reasons to believe the
presence of intense GW sources whose locations and even frequencies are not
known. Amongst such sources pulsars occupy an important position. Similar to
all sky search one will also have to do all frequency search. All sky all
frequency search is the holy grail of gravitation pulsar astronomy. In this
Chapter we confine ourselves to the problem of all sky search.

\par Search of CGW without a priori knowledge appears to be
computationally quite demanding even by the standard computers
expected to be available in near future. For example, in the case of
bandwidth $10^3$ Hz, observation time $10^7$ sec. and star's minimum
decay time  of $100$ years one would require $10^{14}\, Tflops$ computer (Frasca, 2000). Very fast
computer and large memories with ample amount of disk space seems inevitable.
However, choice of optimal data processing and clever programming is also
integral part of solution to this problem. Amongst these the pre-correction
of time series due to Doppler modulation before the data is processed may be
a method, which will reduce computational requirements. In reference to this, 
Schutz (1991) has introduced the concept of patches in the sky defined as the 
region of space for which similar Doppler corrections would be required. He
has also demonstrated that the number of patches required for $10^7$ sec.
observation data set and one 
KHz signal would be about $1.3 \times 10^{13}$ if one takes into account the
rotational motion of Earth. However, 
the size of the patch would also depend on the data analysis technique being
employed. 

\par Matched filtering is the most suitable technique for detection of signals
from sources viz., pulsars whose wave form is known. The wave forms are used
to construct a bank of templates, which represent expected signal wave form
with all possible ranges of its parameters. The time of arrival, source
location, frequency of the signal,
ellipticity of the source and its spin down represent important parameters of GW
emitted by a pulsar. For detection of GW we set a criterion to judge if the
cross correlation of the templates with the corresponding data set exceeds the
preassigned threshold. We would perform the data analysis 
employing the criterion of the {\it FF.\/}

\vspace{0.65cm}

\section{Matched filter analysis: Templates}
\label{sec:templates}
The bank of templates will be matched to only a discrete set of signals from
among the continuum of possible signals. Consequently, it is natural that all signals will
not get detected with equal probability. However, it is possible to 
choose judiciously the set of templates so that all signals of a given amplitude are
detected with a given minimum detection probability. The standard measure for
deciding what class of wave form is good enough is the {\it FF\/} (Apostolatos,
1995). It quantitatively describes the closeness of the true signals to the
template manifold in terms of the reduction of SNR due to cross correlation of
a signal outside the manifold with all the templates lying inside the manifold.
If the {\it FF\/} of a template family is unity the signal lies in the manifold.
If the {\it FF\/} is less than unity the signal lies outside manifold.

\par Even if the signal discrete templates lies within the template manifold
it would be unlikely that any of the actual templates used would correspond to the
signal. The parameters describing the search template
(source location, ellipticity, etc.) can vary continuously through
out a finite range of values. The set of templates characterised by
the continuously varying parameters is ofcourse infinite. However, in
practice the interferometer output must be cross correlated with a finite subset
of the templates whose parameter values vary in discrete steps from
one template to the next. This subset (``the discrete template
family'') has measure zero on the manifold of the full set of
possible templates (``the continuous template family''), so the
template which most closely match a signal will generally lie in 
between the signal and the nearest of the discrete template
family. The mismatch between the signal and the nearest of the 
discrete templates will cause some reduction in SNR. This would mean that
the members of the discrete template
family must be chosen so as to render acceptable loss of SNR.

\par The study of templates has been made by many research workers in time domain [Schutz
(1991), Kr\'olak (1997), Brady et. al. (1998), Brady and Creighton (2000),
Jaranowski et al. (1998) and Jaranowski and  Kr\'olak (2000)]. However,
the analysis in frequency domain has the advantage of incorporating
interferometer's spectral noise density. In order to determine the number
of templates required to perform matched filtering analysis let us rewrite
the formula expressing {\it FF\/} [Eq.~(\ref{eq:ff})] as

\begin{eqnarray}
FF(\theta , \phi) & = & \max\limits_{\theta , \phi} \frac{\langle h(f)|
h_T(f;\theta_T , \phi_T)\rangle}{\sqrt{\langle h_T(f;\theta_T , \phi_T )|h_T(f;
\theta_T , \phi_T )\rangle\langle h(f)|h(f)\rangle}}
\label{eq:ff1}
\end{eqnarray}
            
\vspace{0.6cm}

\noindent where, $h(f)$ and $h_T(f; \theta_T , \phi_T)$ represent respectively the
FTs of the actual signal wave form and the templates. The template parameter
$\theta_T$ and $\phi_T$ are chosen differing from the actual orientation of
the source in discrete steps. 

\par We have seen in earlier Chapters that the AM of
CGW data output results into redistribution of power at four additional frequencies
$f \pm 2f_{rot}$, $f \pm f_{rot}$ in accordance with the FM. Hence it is
sufficient for the analysis of {\it FF\/} to consider only the frequency
modulated FT. The results obtained in Chapter 4 regarding FT
of frequency modulated data output [Eq.~(\ref{eq:hft})] may be arranged using 
the symmetry property of the Bessel functions and one may write 

\begin{eqnarray}
\tilde{h}(f)&\simeq & { \nu \over w_{rot}}\left[ {J_o({\cal Z}) J_o
({\cal N}) \over 2\nu^2}\left[ \{ \sin ( {\cal R} + {\cal Q} ) - \sin
({\cal R} + {\cal Q} - \nu\xi_o )\}\; + \right. \right. \nonumber\\ &&
i \left. \{ \cos ( {\cal R} + {\cal Q} ) - \cos ({\cal R} + {\cal Q} - \nu
\xi_o )\} \right]\; + \nonumber \\
&& J_o ({\cal Z})\sum_{m = 1}^{m = \infty} {J_m({\cal N})\over 
\nu^2 - m^2} \left[ ( {\cal Y} {\cal U} -  {\cal X} {\cal V} ) - i ( 
{\cal X} {\cal U} + {\cal Y} {\cal V} ) \right]\; + \nonumber \\ 
&& \sum_{k = 1 }^{k = \infty}\sum_{m = -
\infty}^{m = \infty} e^{ i {\cal A}}{\cal B}\left[
\tilde{{\cal C}} - i\tilde{{\cal D}} \right]\; ;
\label{eq:fm_code}
\end{eqnarray} 

\begin{equation}
\left.\begin{array}{ccl}
{\cal X}& =& \sin ({\cal R}  + {\cal Q} - m \pi/2 )\\
{\cal Y}& =& \cos ({\cal R} +{\cal Q} - m \pi/2 )\\
{\cal U}& =& \sin \nu\xi_o \cos m ( \xi_o - \delta ) - {m\over \nu}\cos
\nu\xi_o \sin m ( \xi_o - \delta ) - \sin m\delta\\
{\cal V}& =& \cos \nu\xi_o \cos m ( \xi_o - \delta ) + {m\over \nu}\sin 
\nu\xi_o \sin m ( \xi_o - \delta ) - \cos m\delta\\
\end{array}\right\}
\end{equation}

\vspace{0.54cm}

\begin{figure}
\centering
\epsfig{file=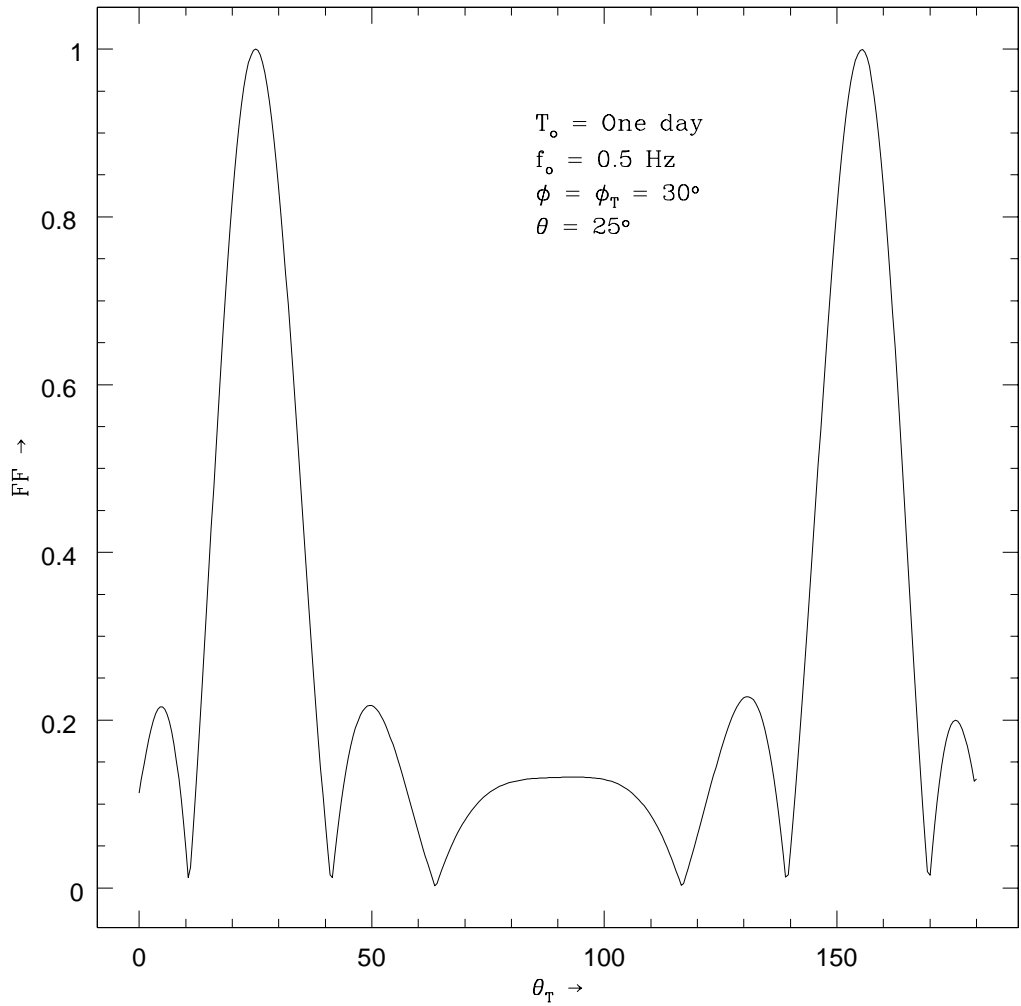,height=14.6cm}
\caption{ Variation of $FF$ with  $\theta_T.$}
\label{fig:symtheta}
\end{figure}

\begin{figure}
\centering
\epsfig{file=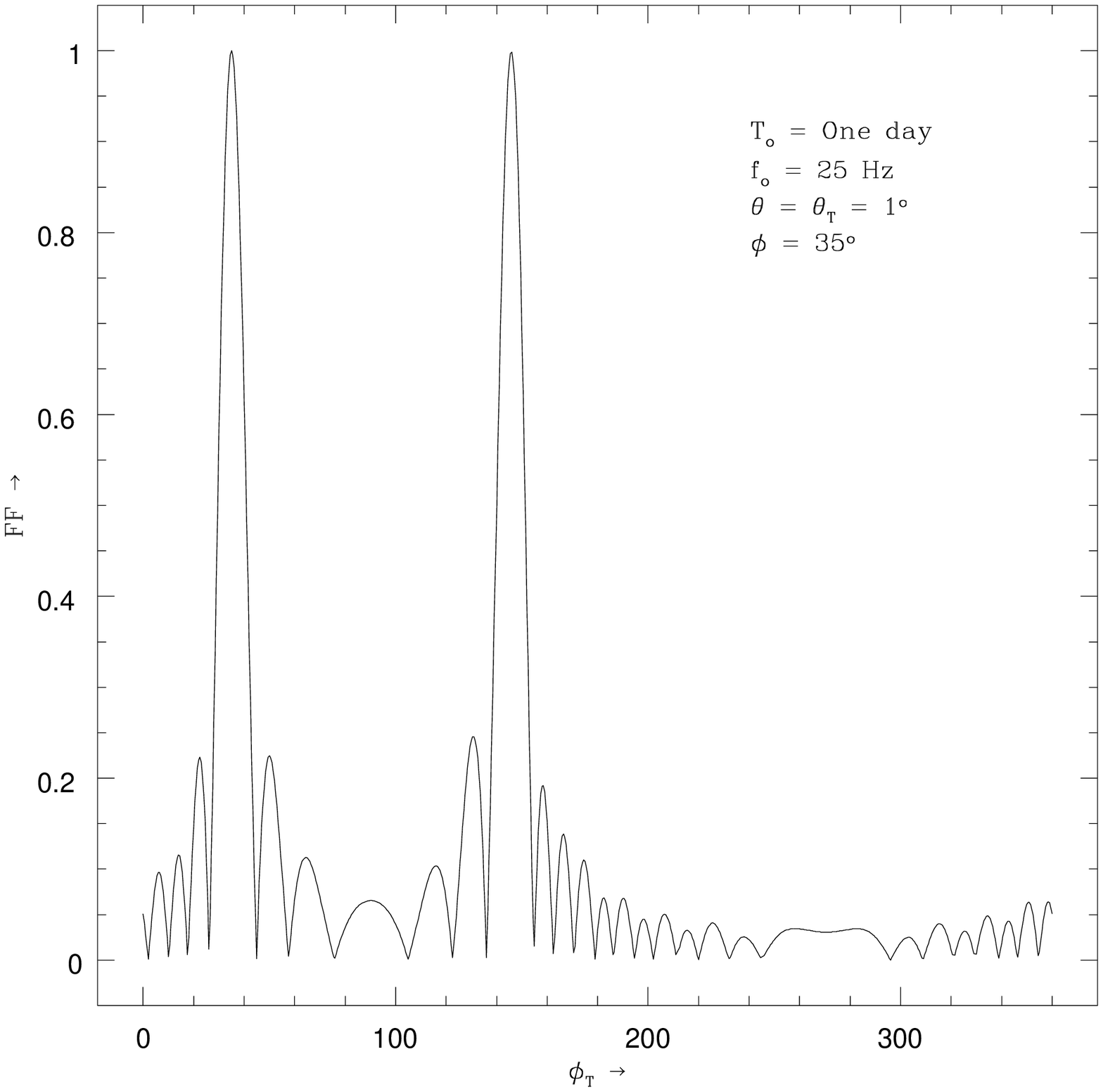,height=14.6cm}
\caption{Variation of $FF$ with $\phi_T.$}
\label{fig:symphi35}
\end{figure}

\begin{figure}
\centering
\epsfig{file=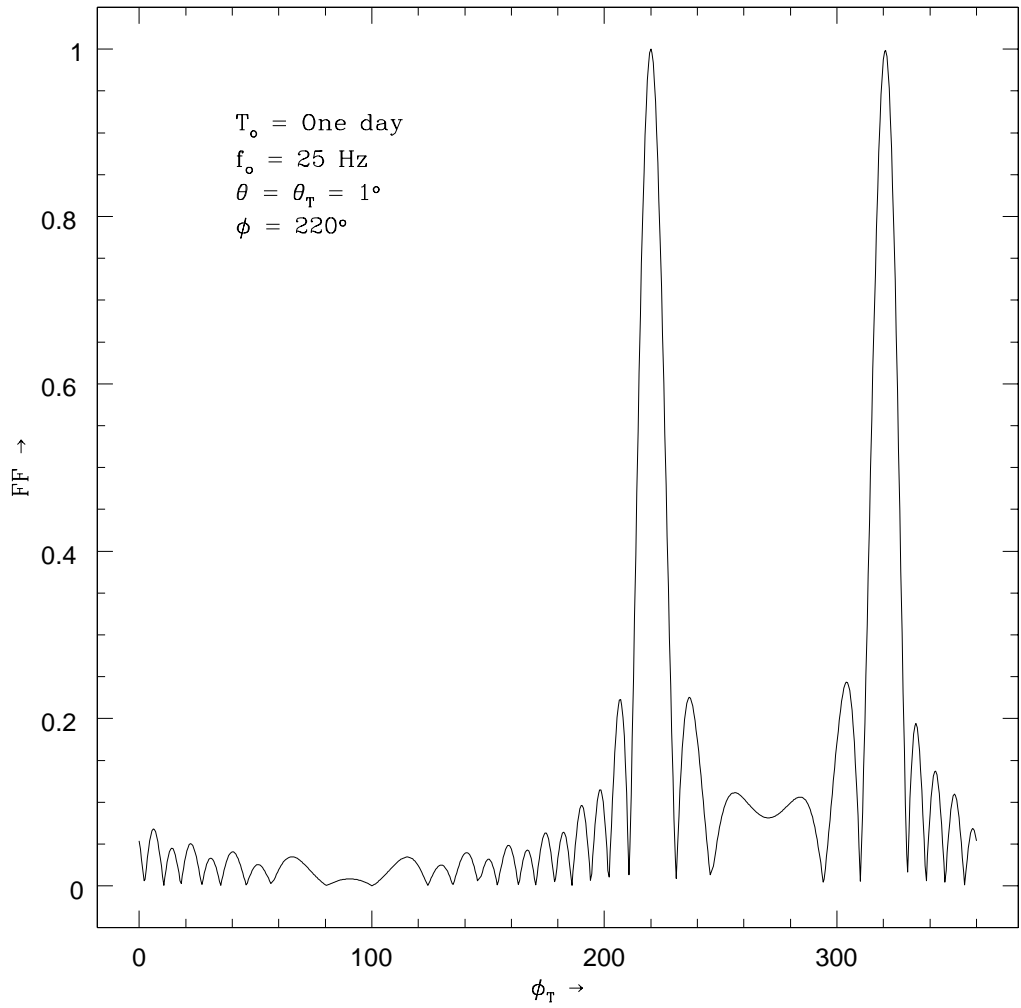,height=14.6cm}
\caption{Variation of $FF$ with $\phi_T.$}
\label{fig:symphi220}
\end{figure}

\noindent Now it is straight forward to compute {\it FF\/}. To understand the
procedure let us assume that there is a source at the location 
$(\theta , \phi ) = (25^o , 30^o)$ emitting
frequency $f_o = 0.5 $ Hz. We wish to analyze the data set for
$T_o =$ one sidereal day. We first fix the template parameter $\phi$ for the whole set given by
$\phi_T = \phi = 30^o$ and vary $\theta_T$ in discrete steps over its entire range
i.e. $0^o$ to $180^o$. The results obtained are plotted graphically in
Fig.~(\ref{fig:symtheta}).
It is remarked that in order to compute the inner product
defined via Eq.~(\ref{eq:ip}) one would require to integrate the expression over the band
width of Doppler modulated signal. This may be determined either analytically
by computing the maximum value of the Doppler shift in accordance with
Eq.~(\ref{eq:ds}) or may be taken as represented by the frequency spectrum of
the FT. In the present case we have taken the band width equal
to $0.002$ Hz. In a similar manner one may fix the $\theta-$parameter
of the template set and obtain the variation of {\it FF\/} with template
parameter $\phi_T$. Figures~(\ref{fig:symphi35}) and~(\ref{fig:symphi220})
represent respectively the behaviour for $f_o = 25$ Hz, $\theta = \theta_T = 1^o$,
$T_o = $ one sidereal day for source orientation $\phi = 35^o$ and $220^o.$\\

\clearpage

\clearpage

\indent\indent\indent The following points in reference to these plots may be noted.
\begin{enumerate}[(i)]
\item The {\it FF\/} is unity for $\theta_T = 25^o$, $155^o$ [Fig~(\ref{fig:symtheta}]
for $\phi_T = 35^o$, $145^o$ [Fig~(\ref{fig:symphi35})] and for $\phi_T = 220^o$,
$320^o$ [Fig~(\ref{fig:symphi220})].

\item It is observed that {\it FF \/} decreases exponentially with
template variables $\theta_T$ and $\phi_T$. The specific relations are found
numerically to be
\begin{equation}
FF = e^{0.047(\theta_T \;\sim \;\theta)^2}
\label{eq:thetaff}
\end{equation}
\begin{equation}
FF = e^{0.06(\phi_T\; \sim \;\phi)^2}
\label{eq:phiff}
\end{equation}

\item The oscillatory behaviours do not represent any real situation as it
arise because of the improper choice of numerical integration technique.
However, we are content with the technique we have employed as the region
of such artificial facets fall in the region of the $FF < 0.25$.
\end{enumerate}

\vspace{0.3cm}
\noindent Finally, we conclude this section by noting the symmetry property of the
template parameters. A closer look of the graphs and the remark (i)
above reveal the following symmetry property. The $FF$ is symmetrical under 
following transformations.
\begin{equation}
\theta_T \longrightarrow  \pi - \theta_T  \qquad 0 \le \theta_T \le \pi
\end{equation}
\begin{equation}
\phi_T \longrightarrow \pi - \phi_T \qquad 0 \le \phi_T \le \pi
\end{equation}
\begin{equation}
\phi_T \longrightarrow 3\pi - \phi_T \qquad \pi \le \phi_T \le 2\pi 
\end{equation}

\noindent Let us note that these symmetry properties are based on our results
for one sidereal day observation time. The generic nature of the symmetries
may be established only after studying the variation of $FF$ with $T_o$.
Unfortunately, we could not make this analysis because of our limitations on
computational facilities. 

\vspace{0.65cm}

\section{Number of Templates}
It is important to study the problem of number of templates for all sky
search in the light of $FF$.
The results of previous section reveal that the grid
spacing $\bigtriangleup \theta$ in the $\theta-$parameter of templates may be
expressed symbolically as a function of $FF$, $f_o$, and $T_o$ i.e.  
\begin{equation}
\bigtriangleup\theta = {\cal F}(FF, f_o, T_o)
\end{equation}
\noindent Similarly, we have
\begin{equation}
\bigtriangleup\phi = {\cal G}(FF, f_o, T_o)
\end{equation}

\vspace{0.3cm}

\noindent In view of this, Equations~(\ref{eq:thetaff}) and~(\ref{eq:phiff}) may be equivalently expressed as
\begin{equation}
{\cal F}(FF, 0.5, T) = (0.047)^{-1}\ln (FF) 
\end{equation}
\begin{equation}
{\cal G}(FF, 25, T) = (0.06)^{-1}\ln (FF)
\end{equation}

\noindent For any chosen value of $FF$ one can determine $\bigtriangleup\theta$ and
$\bigtriangleup\phi$. But there is no unique choice for it. 
Our interest would be in the assignment of
$\bigtriangleup\theta$ and  $\bigtriangleup\phi$ such that the 
spacing are maximum resulting into the least number of
templates. As we have mentioned earlier 
there is stringent requirement on reducing computer time. Accordingly, 
there is serious
need of adopting some procedure/formalism to achieve this. For example, one
may adopt the method of
hierarchical search given by Mohanty and Dhurandhar (1996) and 
Mohanty (1998). This
search is carried out in two steps. At the first level one would start with
template bank with a coarse spacing in the
parameter space but with a lower threshold. In the next level a more
finely spaced set of templates and a higher threshold would be used
but only around those templates of the previous level which crossed the
previous threshold.

\par
\begin{figure}
\centering
\epsfig{file=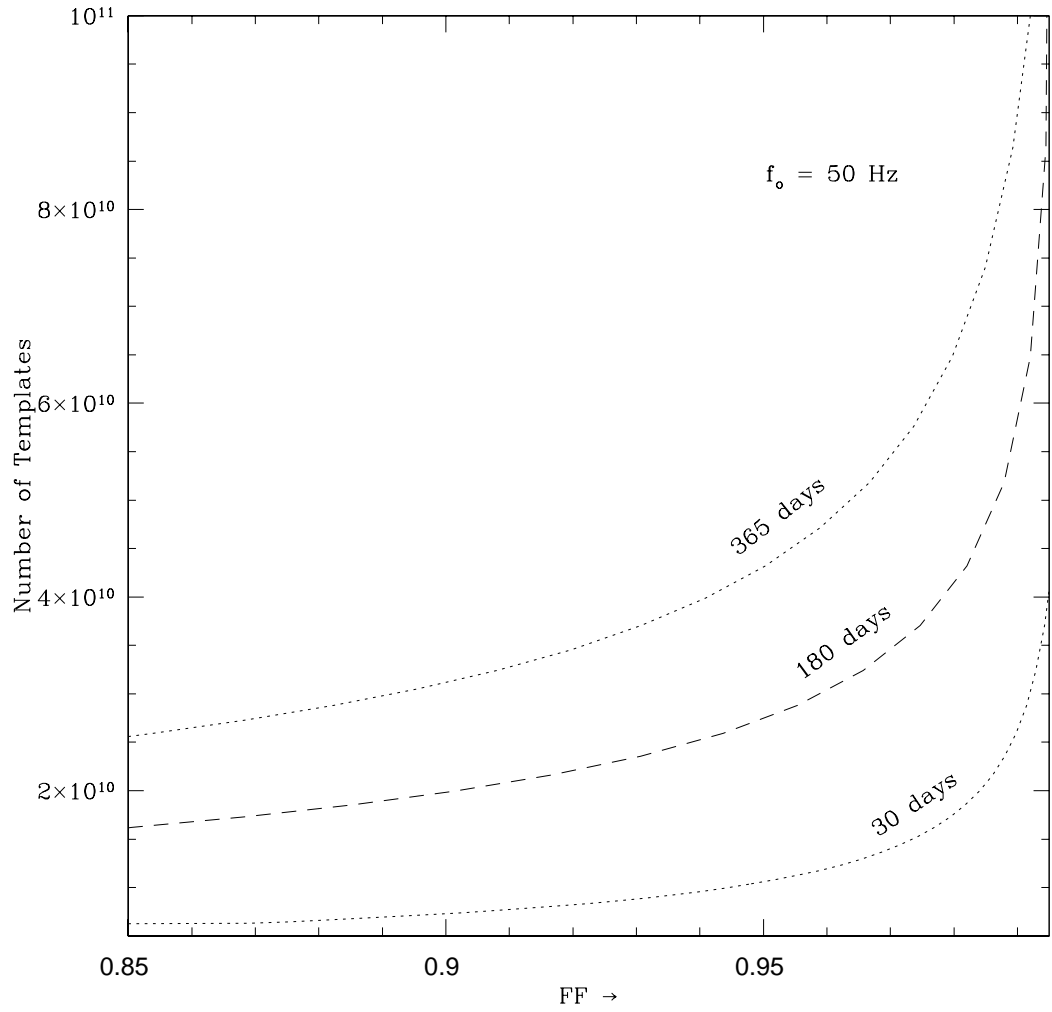,height=14.6cm}
\caption{Variation of number of templates with $FF$ for fixed $f_o$ at
different $T_o$.}
\label{fig:temtime}
\end{figure}

\begin{figure}
\centering
\epsfig{file=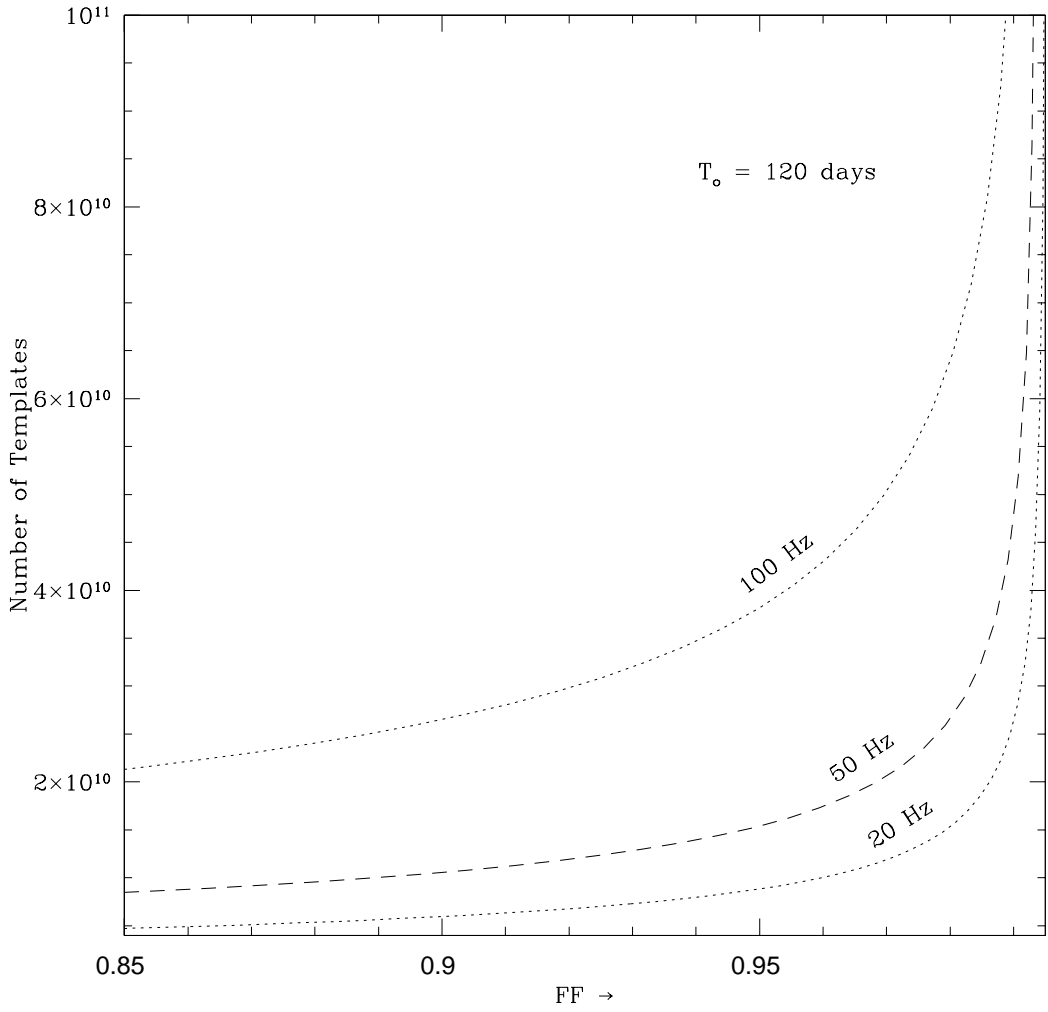,height=14.6cm}
\caption{Variation of number of templates with $FF$ for fixed $T_o$ and of
frequencies.}
\label{fig:temfreq}
\end{figure}

 However, an important issue related to the problem of number of templates
is regarding the study of the 
behaviour of number of templates with $FF$ for different
$f_o$ and $T_o$.  We have made investigations of this aspect. We assume a source
location $(\theta , \phi ) = (1^o , 30^o )$. We choose some value of $FF$, say
$0.995$. Now taking $\phi_T = 30^o$ we determine the spacing $\bigtriangleup
\theta$ to yield the selected $FF$. In the case under investigation
$\bigtriangleup\theta$ is found to equal $4.5 \times 10^{-5}$. Thereafter, 
we introduce spacing
$\bigtriangleup\phi$ in the so obtained bank of templates and determine the resulting $FF$.
The results obtained may be expressed in the form of a graph such as shown in
Figs.~(\ref{fig:temtime}) and~(\ref{fig:temfreq}). Interestingly the nature
of these curves are similar. We have obtained a best fit to the graphs and obtain
the relation.

\begin{equation}
N_{Templates} = \exp [a -bx + cx^2 -dx^3 + ex^4]\; ;\quad x = FF\, ,
\end{equation}
$$ 0.85 \le x \le 0.99$$

\noindent where $a$, $b$, $c$, $d$ and $e$ are constants. The values of these
constants are given in Table~(\ref{table:coefficients}). 

\par Let us note from the graphs, for sake of comparison, that the number of
templates required for $FF$ equal to $0.97$ are respectively $1.44 \times
10^{10}$, $3.5 \times 10^{10}$ and $5.5 \times
10^{10}$ for observation data set of 30, 120 
and 365 days and $f_o = 50$ Hz. Similarly the number of templates required
to analyse the observation data set
of 120 days of GW frequencies 20, 50 and 100 Hz 
are respectively $1.22 \times 10^{10}$,
$2.16 \times 10^{10}$ and $5 \times 10^{10}$. It is observed that higher
$FF$ requires exponentially increasing large number of templates.

\begin{table}
\centering
\begin{tabular}{|c|c|c|c|c|c|c|}
\hline
&&&&&&\\
$f_o$ & $T_o$ & a & b & c & d & e\\
(Hz)&(days)&$\times 10^{- 2}$&$\times 10^{- 4}$&$\times 10^{- 5}$&$\times 10^{- 6}$&$\times 10^{- 7}$\\
\hline
&&&&&&\\
& 30 & $2138.05 $ & $2071.43$ & $7225.73 $
& $6239.43$ & $2036.14$\\ 
&&&&&&\\ \cline{2-7}
&&&&&&\\
50 & 180 & $2317.05 $ & $- 71.3155$ & $1746.61 $
&$ 2146.55$  &$ 944.931$\\ 
&&&&&&\\ \cline{2-7}
&&&&&&\\
& 365 &$ 2382.96 $ & $216.917 $ &$ 2464.42 $
&$ 2464.42$ & $1031.54$\\ 
&&&&&&\\
\hline
&&&&&&\\
20 & & $2047.55$ & $- 794.473 $ &$ 3564.56 $
&$ 4650.68$ &$ 1945.87$\\ 
&&&&&&\\ \cline{1-1} \cline{3-7}
&&&&&&\\
50 & 120 & $2266.59$ & $4269.44$ & $15655.0$
&$ 17509.5$ & $6484.51$\\
&&&&&&\\ \cline{1-1} \cline{3-7}
&&&&&&\\
100& &$ 2360.23$ & $- 206.906$ & $1158.27$
&$ 1491.01$ & $733.520 $\\ 
&&&&&&\\
\hline
\end{tabular}

\vspace{0.6cm}

\caption{Coefficients of the best fit graphs obtained for the 
number of templates.}
\label{table:coefficients}
\end{table}

\clearpage

\clearpage

\section{Discussion}
\label{sec:concl5}

\par In view of the complexity of the FT which
contains exponential as well as Bessel functions; one has to be careful in
computing $FF$. We have found useful to employ the 
Romberg integration using Pad\'e approximation. We have used
(i) QROMO of numerical recipes instead of QROMB as the former takes care of
singularities, and (ii) RATINT routine for Pad\'e approximation. The basic structure of 
the code is given in the appendix A and C (FM.F \& TEM.F).

\par To compute $FF$ the code has been written in FORTRAN. The Bessel functions
are computed using the routine BESSJ0(X), BESSJ1(X) and BESSJ(N,X) of numerical
recipes (Press et al. 1986). The BESSJ0(X) and BESSJ1(X) compute
the Bessel function of first kind of zeroth and first order
respectively for any value of $X$, while BESSJ(N,X) computes the
Bessel function of first kind of arbitrary integeral order $N$ for any
value of $X$.

\par We  have noticed marked symmetries in all sky search in both $\theta$ and
$\phi$ space for one day observation time.
It has been found that any $FF$ corresponds to two values each in $\theta_T$ and
$\phi_T$. Accordingly, computation burden will be reduced by a factor of
four. However, it is not clear whether the symmetry
property can be established analytically as well. The source
location, because of the symmetry, is uncertain. Some other 
analysis is to be adopted for getting the exact location.

\par We have computed the number of templates required for all sky search for
matched filtering assuming the noise power spectral density 
$S_n(f)$ to be flat. However, for realistic situation, the effect of the noise
have to be taken into consideration.

\par The issues of optimum template parameterization and placement, and the
related computational burden have been discussed in literature by several
authors notably by Sathyaprakash and Dhurandhar (1991), Dhurandhar and Sathyaprakash
(1994), Owen (1996), Apostolatos (1995, 1996), Mohanty and 
Dhurandhar (1996), Mohanty (1998), Owen and Sathayaprakash
(1999). The question of possible efficient interpolated representation
of the correlators is a problem of current interest and remains still unsolved.

\appendix
\addcontentsline{toc}{chapter}{\numberline{}Appendices}{}

\baselineskip=0.660cm

\chapter{Source code of the frequency modulated signal: {\it FM.F\/}}

C\hspace{27pt} {\bf FT of the frequency modulated signal}\\
\indent\hspace{19pt} FUNCTION hfm(fnu,dtheta,dphi)\\
\indent \hspace{19pt} IMPLICIT none\\
\indent \hspace{19pt} INTEGER k,m,ms,ks \\
\indent \hspace{19pt} REAL t,f0,fnu,dtheta,dphi,dalpha,dbeta0 \\
C\hspace{30pt} [t = $T_o$, f0 = $f_o$, fnu = $\nu$, dtheta = $\theta^o$,
dphi = $\phi^o$,  dalpha = $\alpha^o$,\\
\indent \hspace{9.2pt} *\,dbeta0 = $\beta_o^o$]\\
\indent \hspace{19pt} PARAMETER (t = 86164.0, f0 = 25.0, dalpha = 75.0,\\ 
\indent \hspace{19pt} dbeta0 = 60.0, ks = 720, ms = 8)\\
C\hspace{30pt} [`ks' \& `ms' are the values of `k' \& `m' respectively]\\
\indent \hspace{19pt} REAL h,z,bessj0,bessj1,bessj,bsh(-ms:ms),bsz(0:ks)  \\
\indent \hspace{19pt} REAL$^*$8 pi,fr,pfr,a,u,rc,rad,theta,phi,epsilon,alpha,beta0,\\
\indent \hspace{9.2pt} *\,sth,sal,sbe0,sph,sep,cth,cal,cbe0,cph,cep,p,q,r,delta,crq,\\
\indent \hspace{9.2pt} *\,srq,bs0,fnn,cfnn,sfnn,rqf,fnu2,rplusq               \\
\indent \hspace{19pt} REAL$^*$8 dm(1:ms),dmu(1:ms),rqm(1:ms),cmud(1:ms), smud(1:ms),         \\
\indent \hspace{9.2pt} *\,smd(1:ms),cmd(1:ms),crqm(1:ms),srqm(1:ms),wmn(1:ms),vmn(1:ms)     \\
\indent \hspace{9.2pt} *\,akd(1:ks,-ms:ms),rqk(1:ks,-ms:ms),skm(1:ks,-ms:ms),              \\
\indent \hspace{9.2pt} *\,pmd(1:ks,-ms:ms),akm(1:ks,-ms:ms),ckmu(1:ks,-ms:ms),\\
\indent \hspace{9.2pt} *\,skmu(1:ks,-ms:ms),ckmr(1:ks,-ms:ms),ckm(1:ks,-ms:ms),             \\
\indent \hspace{9.2pt} *\,skmr(1:ks,-ms:ms)                                \\
\indent \hspace{19pt} COMPLEX sum1,sum2,hfm                        \\
C\hspace{30pt} [hfm $\longrightarrow$ Spectrum of the frequency modulated signal.]\\
\indent \hspace{19pt} EXTERNAL bessj0,bessj1,bessj\\ \\
\indent \hspace{19pt}         pi = 4.d0*datan(1.d0)  \\
\indent \hspace{19pt}         fr = 1.d0/86164.d0\\
\indent \hspace{19pt}         pfr = 2.d0*pi*fr    \\
\indent \hspace{19pt}         a = 1.d0/365.2572d0   \\
\indent \hspace{19pt}         u = 2.d0*pi*fr*t        \\
\indent \hspace{19pt}         rc = 2.d0*pi*f0*0.0637/3.d0\\
\indent \hspace{19pt}         rad = pi/180.d0              \\
\indent \hspace{19pt}         theta = rad*dtheta              \\
\indent \hspace{19pt}         phi = rad*dphi                    \\
\indent \hspace{19pt}         epsilon = rad*23.45d0              \\
\indent \hspace{19pt}         alpha = rad*75.d0                    \\
\indent \hspace{19pt}         beta0 = rad*60.d0                      \\
\indent \hspace{19pt}         sth = dsin(theta)                         \\
\indent \hspace{19pt}         sal = dsin(alpha)                           \\
\indent \hspace{19pt}         sbe0 = dsin(beta0)                            \\
\indent \hspace{19pt}         sph = dsin(phi)                                 \\
\indent \hspace{19pt}         sep = dsin(epsilon)                               \\
\indent \hspace{19pt}         cth = dcos(theta)                                   \\
\indent \hspace{19pt}         cal = dcos(alpha)                                     \\
\indent \hspace{19pt}         cbe0 = dcos(beta0)                                      \\
\indent \hspace{19pt}         cph = dcos(phi)                                           \\
\indent \hspace{19pt}         cep = dcos(epsilon)                                         \\
\indent \hspace{19pt}         p = rc*sal*(cbe0*(sth*cep*sph + cth*sep) - sbe0*sth*cph)      \\
\indent \hspace{19pt}         q = rc*sal*(sbe0*(sth*cep*sph + cth*sep) + cbe0*sth*cph)        \\
\indent \hspace{19pt}         z = 2.d0*pi*f0*1496.d0/3.d0*sth                                   \\
\indent \hspace{19pt}         r = z*cph                                                           \\
\indent \hspace{19pt}         rplusq = r + q                                                      \\
\indent \hspace{19pt}         delta = datan(p/q)\\
\indent \hspace{19pt}         h = dsqrt(p*p + q*q)\\
\indent \hspace{19pt}         crq = dcos(rplusq)    \\
\indent \hspace{19pt}         srq = dsin(rplusq)      \\
\indent \hspace{19pt}         bsz(0) = bessj0(z)        \\
\indent \hspace{19pt}         bsh(0) = bessj0(h)          \\
\indent \hspace{19pt}         bsz(1) = bessj1(z)            \\
\indent \hspace{19pt}         bsh(1) = bessj1(h)              \\
\indent \hspace{19pt}         bsh(-1) = - bsh(1)                \\
\indent \hspace{19pt}         bs0 = bsz(0)*bsh(0)                 \\
\indent\indent\indent DO 10 m = 2,ms                        \\
\indent \hspace{19pt}             bsh(m) = bessj(m,h,bsh(0),bsh(1))                 \\
\indent \hspace{19pt}             bsh(-m) = (-1)**m*bsh(m)              \\
$\left. \right.$\hspace{5pt} 10 \hspace{12pt} CONTINUE                                  \\
\indent\indent\indent DO 20 k = 2,ks                               \\
\indent \hspace{19pt}            bsz(k) = bessj(k,z,bsz(0),bsz(1))                          \\
$\left. \right.$\hspace{5pt} 20 \hspace{12pt} CONTINUE                                        \\
\indent\indent\indent DO 30 m = 1,ms                                     \\
\indent \hspace{19pt}            dm(m) = m*delta                                    \\
\indent \hspace{19pt}            dmu(m) = m*(u-delta)                                \\
\indent \hspace{19pt}            rqm(m) = rplusq - m*pi/2.d0                            \\
\indent \hspace{19pt}            cmud(m) = dcos(dmu(m))                                   \\
\indent \hspace{19pt}            smud(m) = m*dsin(dmu(m))                                   \\
\indent \hspace{19pt}            smd(m) = m*dsin(dm(m))                                       \\
\indent \hspace{19pt}            cmd(m) = dcos(dm(m))                                           \\
\indent \hspace{19pt}            crqm(m) = dcos(rqm(m))                                           \\
\indent \hspace{19pt}            srqm(m) = dsin(rqm(m))                                             \\
$\left. \right.$\hspace{5pt} 30 \hspace{12pt} CONTINUE\\ 
\indent\indent\indent DO 40 m = -ms,ms\\
\indent\indent\indent\indent DO 50 k = 1,ks   \\
\indent \hspace{19pt}               akd(k,m) = a*k*u+m*u-k*phi - m*delta\\
\indent \hspace{19pt}               rqk(k,m) = (k + m)*pi/2.d0 - rplusq    \\
\indent \hspace{19pt}               pmd(k,m) = k*phi+m*delta               \\
\indent \hspace{19pt}               akm(k,m) = k*a + m                         \\
\indent \hspace{19pt}               ckmu(k,m) = dcos(akd(k,m))               \\
\indent \hspace{19pt}               skm(k,m) = akm(k,m)*dsin(pmd(k,m))         \\
\indent \hspace{19pt}               skmu(k,m) = akm(k,m)*dsin(akd(k,m))          \\
\indent \hspace{19pt}               ckmr(k,m) = dcos(rqk(k,m))                     \\
\indent \hspace{19pt}               ckm(k,m) = dcos(pmd(k,m))                        \\
\indent \hspace{19pt}               skmr(k,m) = dsin(rqk(k,m))                         \\
$\left. \right.$\hspace{5pt} 50 \hspace{5pt} \indent CONTINUE                                                 \\
$\left. \right.$\hspace{5pt} 40 \hspace{12pt} CONTINUE \\
\indent \hspace{19pt}         fnn = fnu*u                                                     \\
\indent \hspace{19pt}         fnu2 = fnu*fnu                                                     \\
\indent \hspace{19pt}         cfnn = dcos(fnn)                                                     \\
\indent \hspace{19pt}         sfnn = dsin(fnn)                                                       \\
\indent \hspace{19pt}         rqf = rplusq - fnn\\
\indent \hspace{19pt}         sum1 = cmplx(0.0,0.0)     \\
\indent \hspace{19pt}         sum2 = cmplx(0.0,0.0)       \\
\indent\indent\indent DO 60 m = 1,ms                \\
\indent \hspace{19pt}         wmn(m) = sfnn*cmud(m) - (cfnn*smud(m) + smd(m))/fnu\\
\indent \hspace{19pt}         vmn(m) = cfnn*cmud(m) + sfnn*smud(m)/fnu - cmd(m)    \\
\indent \hspace{19pt}         sum1 = sum1 + bsz(0)*bsh(m)/(fnu2 - m*m)*              \\
\indent \hspace{9.2pt} *\,cmplx((wmn(m)*crqm(m) - vmn(m)*srqm(m)),            \\
\indent \hspace{9.2pt} *\,- (wmn(m)*srqm(m) + vmn(m)*crqm(m)))  \\
$\left. \right.$\hspace{5pt} 60 \hspace{12pt} CONTINUE   \\
\indent\indent\indent DO 70 m = - ms,ms                                           \\
\indent\indent \indent\indent DO 80 k = 1,ks                                                \\
\indent \hspace{19pt}         sum2 = sum2 + bsz(k)*bsh(m)/(fnu2 - (a*k+m)**2)*                   \\
\indent \hspace{9.2pt} *\,cmplx((sfnn*ckmu(k,m) - (skm(k,m) + skmu(k,m)*cfnn)/fnu),           \\
\indent \hspace{9.2pt} *\,- (cfnn*ckmu(k,m) - ckm(k,m) + skmu(k,m)*                             \\
\indent \hspace{9.2pt} *\,sfnn/fnu))*cmplx(ckmr(k,m),skmr(k,m))                               \\
$\left. \right.$\hspace{5pt} 80 \hspace{5pt} \indent  CONTINUE\\
$\left. \right.$\hspace{5pt} 70 \hspace{5pt}   CONTINUE\\
\indent \hspace{19pt}        hfm = fnu/pfr*(bs0/(2.0*fnu2)*cmplx((srq - dsin(rqf)),\\
\indent \hspace{9.2pt} *\,(crq - dcos(rqf))) + sum1 + sum2)                     \\
\indent\indent\indent          RETURN                                                 \\
\indent\indent\indent\indent  END                                              

\chapter{Source code of the noise free complete response of the detector:
{\it CR.F\/}}

C\hspace{27pt} {\bf FT of the nosie free complete response of the detector}\\
\indent \hspace{19pt}        IMPLICIT none  \\
\indent \hspace{19pt}        INTEGER k,m,n,ms,ks,j,l                                         \\
\indent \hspace{19pt}        PARAMETER (ks = 27225,ms = 15,j = 174) \\
C\hspace{27pt} [ j $\longrightarrow$ Number of frequencies bins in the spectrum\\ 
C\hspace{27pt} of the complete response.]\\
\indent \hspace{19pt}        REAL t,f0,f,dtheta,dphi,dalpha,dbeta0,dgamma,dpsi,h0\_p,  \\
\indent \hspace{9.2pt} *\,h0\_c,res\_n,z,h,bessj0,bessj1,bessj\\
\indent \hspace{19pt}        PARAMETER (t = 86164.0, f0 = 50.0, f = 49.9995, dtheta = 10.0,\\
\indent \hspace{9.2pt} *\,dphi = 0.0, dalpha = 46.45, dbeta0 = 0.0, dgamma = 171.8,   \\
\indent \hspace{9.2pt} *\,dpsi = 45.0, h0\_p = 1.0, h0\_c = 1.0, res\_n = 2.0*t)  \\
C\hspace{30pt} [dgamma = $\gamma^o$, dpsi = $\psi^o$, h0\_p \& h0\_c $ \longrightarrow $ Amplitude of the plus  \\
C\hspace{27pt} and cross polarisation of the signal respectively and \\
C\hspace{27pt} res\_n $\longrightarrow$ resolution of the spectrum.]\\
\indent \hspace{19pt}        REAL$^*$8 pi,rad,fr,pfr,a,u,rc,theta,phi,epsilon,alpha,\\
\indent \hspace{9.2pt} *\,psi,beta0,sth,sal,sbe0,sph,sep,cth,cal,cbe0,cph,cep,gamma,ssi,csi\\
\indent \hspace{9.2pt} *\,,s2ep,c2ep,s2gm,c2gm,al,am,an,ap,aq,ar,ag,ah,ai,aj,f1\_p,f2\_p,f3\_p  \\
\indent \hspace{9.2pt} *\,,f4\_p,f5\_p,sph\_c,cph\_c,al\_c,am\_c,an\_c,ap\_c,aq\_c,ar\_c,ag\_c,ah\_c,      \\
\indent \hspace{9.2pt} *\,ai\_c,aj\_c,f1\_c,f2\_c,f3\_c,f4\_c,f5\_c,p,q,r,rplusq,delta,crq,srq,fn(0:j)\\
C\hspace{27pt} [fn $\longrightarrow$ frequency bins of the complete response.]  \\
\indent \hspace{19pt}        REAL$^*$8 bsh(-ms:ms),bsz(0:ks),bs0,dm(1:ms),dmu(1:ms),   \\
\indent \hspace{9.2pt} *\,rqm(1:ms),cmud(1:ms),smud(1:ms),smd(1:ms),cmd(1:ms),crqm(1:ms),    \\
\indent \hspace{9.2pt} *\,srqm(1:ms),akd(1:ks,-ms:ms),rqk(1:ks,-ms:ms),pmd(1:ks,-ms:ms),       \\
\indent \hspace{9.2pt} *\,akm(1:ks,-ms:ms),ckmu(1:ks,-ms:ms),skm(1:ks,-ms:ms),\\
\indent \hspace{9.2pt} *\,skmu(1:ks,-ms:ms),ckmr(1:ks,-ms:ms), ckm(1:ks,-ms:ms),\\
\indent \hspace{9.2pt} *\,skmr(1:ks,-ms:ms),fnu(-2:2,0:j),fnn(-2:2,0:j),  \\
\indent \hspace{9.2pt} *\,cfnn(-2:2,0:j),sfnn(-2:2,0:j),rqf(-2:2,0:j),wmn(1:ms,-2:2),\\
\indent \hspace{9.2pt} *\,vmn(1:ms,-2:2)   \\
\indent \hspace{19pt}        COMPLEX fp2fr,fm2fr,fpfr,fmfr,sum1(-2:2,0:j),sum2(-2:2,0:j),     \\
\indent \hspace{9.2pt} *\,hfm(-2:2,0:j),crp2f(0:j),crp1f(0:j),cr0f(0:j),crm1f(0:j),          \\
\indent \hspace{9.2pt} *\,crm2f(0:j),cr(0:j)   \\
C\hspace{27pt} [cr $\longrightarrow$ complete response of the detector and crp2f, crp1f, cr0f, \\
C\hspace{27pt} crm2f \& crm1f $\longrightarrow$ complete response at $f + 2f_r$, $f + f_r$, $f$, \\ 
C\hspace{27pt} $f - 2f_r$ \& $f - f_r$ respectively.]\\
\indent \hspace{19pt}        EXTERNAL bessj0,bessj1,bessj\\ \\
\indent \hspace{19pt}         pi = 4.d0*datan(1.d0)   \\
\indent \hspace{19pt}         rad = pi/180.d0\\
\indent \hspace{19pt}         fr = 1.d0/86164.d0\\
\indent \hspace{19pt}         pfr = 2.d0*pi*fr    \\
\indent \hspace{19pt}         a = 1.d0/365.2572d0   \\
\indent \hspace{19pt}         u = 2.d0*pi*fr*t        \\
\indent \hspace{19pt}         rc = 2.d0*pi*f0*0.0637d0/3.d0\\
\indent \hspace{19pt}         theta = rad*dtheta      \\
\indent \hspace{19pt}         phi = rad*dphi            \\
\indent \hspace{19pt}         epsilon = rad*23.45d0        \\
\indent \hspace{19pt}         alpha = rad*dalpha            \\
\indent \hspace{19pt}         psi = rad*dpsi                  \\
\indent \hspace{19pt}         beta0 = rad*dbeta0                \\
\indent \hspace{19pt}         sth = dsin(theta)                           \\
\indent \hspace{19pt}         sal = dsin(alpha)                             \\
\indent \hspace{19pt}         sbe0 = dsin(beta0)                            \\
\indent \hspace{19pt}         sph = dsin(phi)                                   \\
\indent \hspace{19pt}         sep = dsin(epsilon)                                 \\
\indent \hspace{19pt}         cth = dcos(theta)                                     \\
\indent \hspace{19pt}         cal = dcos(alpha)                                       \\
\indent \hspace{19pt}         cbe0 = dcos(beta0)                                        \\
\indent \hspace{19pt}         cph = dcos(phi)                                             \\
\indent \hspace{19pt}         cep = dcos(epsilon)                                           \\
\indent \hspace{19pt}         gamma = rad*dgamma                                       \\
\indent \hspace{19pt}         ssi = dsin(psi)                                                   \\
\indent \hspace{19pt}         csi = dcos(psi)                                                     \\
\indent \hspace{19pt}         s2ep = dsin(2.d0*epsilon)                                             \\
\indent \hspace{19pt}         c2ep = dcos(2.d0*epsilon)\\
\indent \hspace{19pt}         s2gm = dsin(2.d0*gamma)    \\
\indent \hspace{19pt}         c2gm = dcos(2.d0*gamma)      \\ \\
C\hspace{27pt} {\bf Beam pattern for plus polarisation}\\ \\
\indent \hspace{19pt}        al = csi*cph - cth*sph*ssi              \\
\indent \hspace{19pt}        am = csi*sph + cth*cph*ssi                \\
\indent \hspace{19pt}        an = - ssi*cph - cth*sph*csi                 \\
\indent \hspace{19pt}        ap = - ssi*sph + cth*cph*csi                  \\
\indent \hspace{19pt}        aq = sth*sph                                    \\
\indent \hspace{19pt}        ar = sth*cph                                      \\
\indent \hspace{19pt}        ag = ((al*aq + am*ar)*sep - (al*an - am*ap)*cep)/2.d0 \\
\indent \hspace{19pt}        ah = ((an*an - ap*ap)*cep*cep - (al*al - am*am) +          \\
\indent \hspace{9.2pt} *\,(aq*aq - ar*ar)*sep*sep - (an*aq + ap*ar)*s2ep)/2.d0      \\
\indent \hspace{19pt}        ai = ((aq*aq - ar*ar)*(s2ep) - (an*an - ap*ap)*s2ep -         \\
\indent \hspace{9.2pt} *\,2.d0*(an*aq + ap*ar)*c2ep)/2.d0                           \\
\indent \hspace{19pt}        aj = ((al*aq + am*ar)*cep + (al*an - am*ap)*sep)/2.d0             \\
\indent \hspace{19pt}        f1\_p = h0\_p*(ah*s2gm*(cal*cal + 1) - 4.d0*ag*cal*c2gm)/2.d0      \\
\indent \hspace{19pt}        f2\_p = h0\_p*(ah*cal*c2gm + ag*s2gm*((cal)**2 + 1.d0))                 \\
\indent \hspace{19pt}        f3\_p = h0\_p*(ai*sal*c2gm + aj*dsin(2.d0*alpha)*s2gm)                \\
\indent \hspace{19pt}        f4\_p = h0\_p*(4.d0*aj*sal*c2gm - ai*dsin(2.d0*alpha)*s2gm)/2.d0        \\
\indent \hspace{19pt}        f5\_p = h0\_p*sal*sal*s2gm*(ah + al*al - am*am)*3.d0/2.d0                 \\ \\
C\hspace{27pt} {\bf Beam pattern for cross polarisation}\\
                                            \\
\indent \hspace{19pt}        sph\_c = dsin(phi - pi/4.d0)             \\
\indent \hspace{19pt}        cph\_c = dcos(phi - pi/4.d0)                \\
\indent \hspace{19pt}        al\_c = csi*cph\_c - cth*sph\_c*ssi             \\
\indent \hspace{19pt}        am\_c = csi*sph\_c + cth*cph\_c*ssi               \\
\indent \hspace{19pt}       an\_c = - ssi*cph\_c - cth*sph\_c*csi            \\
\indent \hspace{19pt}        ap\_c = -ssi*sph\_c + cth*cph\_c*csi                 \\
\indent \hspace{19pt}        aq\_c = sth*sph\_c                                     \\
\indent \hspace{19pt}        ar\_c = sth*cph\_c                                       \\
\indent \hspace{19pt}        ag\_c = ((al\_c*aq\_c + am\_c*ar\_c)*sep - (al\_c*an\_c - \\
\indent \hspace{9.2pt} *\,am\_c*ap\_c)*cep)/2.d0\\
\indent \hspace{19pt}        ah\_c = ((an\_c*an\_c - ap\_c*ap\_c)*cep*cep -        \\
\indent \hspace{9.2pt} *\,(al\_c*al\_c - am\_c*am\_c) +(aq\_c*aq\_c - ar\_c*ar\_c)*sep*sep -    \\
\indent \hspace{9.2pt} *\,(an\_c*aq\_c + ap\_c*ar\_c)*s2ep)/2.d0     \\
\indent \hspace{19pt}        ai\_c = ((aq\_c*aq\_c - ar\_c*ar\_c)*(s2ep) - (an\_c*an\_c - ap\_c*ap\_c)*      \\
\indent \hspace{9.2pt} *\,s2ep - 2.d0*(an\_c*aq\_c + ap\_c*ar\_c)*c2ep)/2.d0\\
\indent \hspace{19pt}        aj\_c = ((al\_c*aq\_c + am\_c*ar\_c)*cep + (al\_c*an\_c - \\
 \indent \hspace{9.2pt} *\,am\_c*ap\_c)*sep)/2.d0\\
\indent \hspace{19pt}        f1\_c = h0\_c*(ah\_c*s2gm*(cal*cal + 1) - 4.d0*ag\_c*cal*c2gm)/2.d0       \\
\indent \hspace{19pt}        f2\_c = h0\_c*(ah\_c*cal*c2gm+ag\_c*s2gm*((cal)**2 + 1.d0))                  \\
\indent \hspace{19pt}        f3\_c = h0\_c*(ai\_c*sal*c2gm + aj\_c*dsin(2.d0*alpha)*s2gm)                 \\
\indent \hspace{19pt}        f4\_c = h0\_c*(4.d0*aj\_c*sal*c2gm - \\
\indent \hspace{9.2pt} *\,ai\_c*dsin(2.d0*alpha)*s2gm)/2.d0         \\
\indent \hspace{19pt}        f5\_c = h0\_c*sal*sal*s2gm*(ah\_c + al\_c*al\_c - am\_c*am\_c)*3.d0/2.d0\\
                                                                        \\
C\hspace{27pt} {\bf Both polarisations are suitably arrange to compute \\ 
C\hspace{27pt} complete response}\\                       \\
\indent \hspace{19pt}        fp2fr = cmplx((f1\_p + f2\_c), (f2\_p - f1\_c))/2.d0\\
\indent \hspace{19pt}        fm2fr = cmplx((f1\_p - f2\_c), (- f2\_p - f1\_c))/2.d0\\
\indent \hspace{19pt}        fpfr = cmplx((f3\_p + f4\_c), (f4\_p - f3\_c))/2.d0     \\
\indent \hspace{19pt}        fmfr = cmplx((f3\_p - f4\_c), (- f4\_p - f3\_c))/2.d0     \\
                                                             \\
C\hspace{27pt} {\bf Computation for frequncy modulation}                      \\
                                                                 \\
\indent \hspace{19pt}         p = rc*sal*(cbe0*(sth*cep*sph + cth*sep) - sbe0*ar)          \\
\indent \hspace{19pt}         q = rc*sal*(sbe0*(sth*cep*sph + cth*sep) + cbe0*ar)            \\
\indent \hspace{19pt}         z = 2.d0*pi*f0*1496.d0/3.d0*sth                                  \\
\indent \hspace{19pt}         r = z*cph                                                          \\
\indent \hspace{19pt}         rplusq = r + q                                                     \\
\indent \hspace{19pt}         delta = datan(p/q)                                                     \\
\indent \hspace{19pt}         h = dsqrt(p*p + q*q)                                                     \\
\indent \hspace{19pt}         crq = dcos(rplusq)                                                         \\
\indent \hspace{19pt}         srq = dsin(rplusq)\\
\indent \hspace{19pt}         bsz(0) = bessj0(z)  \\
\indent \hspace{19pt}         bsh(0) = bessj0(h)    \\
\indent \hspace{19pt}         bsz(1) = bessj1(z)      \\
\indent \hspace{19pt}         bsh(1) = bessj1(h)        \\
\indent \hspace{19pt}         bsh(-1) = - bsh(1)          \\
\indent \hspace{19pt}         bs0 = bsz(0)*bsh(0)           \\
\indent\indent\indent       DO 10 m = 2,ms  \\
\indent \hspace{19pt}             bsh(m) = bessj(m,h,bsh(0),bsh(1))           \\
\indent \hspace{19pt}             bsh(-m) = (-1)**m*bsh(m)        \\
$\left. \right.$\hspace{5pt} 10 \hspace{5pt} CONTINUE                            \\
\indent\indent\indent         DO 20 k = 2,ks                          \\
\indent \hspace{19pt}            bsz(k) = bessj(k,z,bsz(0),bsz(1))                    \\
$\left. \right.$\hspace{5pt} 20 \hspace{5pt}   CONTINUE                                  \\
\indent\indent\indent       DO 30 m = 1,ms                                \\
\indent \hspace{19pt}            dm(m) = m*delta                              \\
\indent \hspace{19pt}            dmu(m) = m*(u-delta)                          \\
\indent \hspace{19pt}            rqm(m) = rplusq - m*pi/2.d0                      \\
\indent \hspace{19pt}            cmud(m) = dcos(dmu(m))                             \\
\indent \hspace{19pt}            smud(m) = m*dsin(dmu(m))                             \\
\indent \hspace{19pt}            smd(m) = m*dsin(dm(m))                                 \\
\indent \hspace{19pt}            cmd(m) = dcos(dm(m))                                     \\
\indent \hspace{19pt}            crqm(m) = dcos(rqm(m))                                     \\
\indent \hspace{19pt}            srqm(m) = dsin(rqm(m))                                       \\
$\left. \right.$\hspace{5pt} 30 \hspace{5pt}   CONTINUE                                               \\
\indent\indent\indent         DO 40 m = - ms,ms                     \\
\indent\indent\indent\indent             DO 50 k = 1,ks                  \\
\indent \hspace{19pt}               akd(k,m) = a*k*u+m*u - k*phi - m*delta                                \\
\indent \hspace{19pt}               rqk(k,m) = (k + m)*pi/2.d0 - rplusq                                 \\
\indent \hspace{19pt}               pmd(k,m) = k*phi + m*delta                                              \\
\indent \hspace{19pt} \indent \hspace{19pt}   akm(k,m) = k*a + m\\
\indent \hspace{19pt}               ckmu(k,m) = dcos(akd(k,m))\\
\indent \hspace{19pt}               skm(k,m) = akm(k,m)*dsin(pmd(k,m))\\
\indent \hspace{19pt}               skmu(k,m) = akm(k,m)*dsin(akd(k,m)) \\
\indent \hspace{19pt}               ckmr(k,m) = dcos(rqk(k,m))            \\
\indent \hspace{19pt}               ckm(k,m) = dcos(pmd(k,m))               \\
\indent \hspace{19pt}               skmr(k,m) = dsin(rqk(k,m)) \\
$\left. \right.$\hspace{5pt} 50 \hspace{5pt} \indent       CONTINUE                                        \\
$\left. \right.$\hspace{5pt} 40 \hspace{5pt}  CONTINUE                                              \\
\indent\indent\indent      DO 60 n = 0,j                                        \\
\indent \hspace{19pt}         fn(n) = f + n/res\_n                                     \\
$\left. \right.$\hspace{5pt} 60 \hspace{5pt}    CONTINUE                                                  \\
\indent\indent\indent  DO 70 n = 0,j         \\
\indent\indent\indent\indent DO 80 l = - 2,2               \\
\indent \hspace{19pt}         fnu(l,n) = (f0 - fn(n))/fr + l                                     \\
\indent \hspace{19pt}         fnn(l,n) = fnu(l,n)*u                                              \\
\indent \hspace{19pt}         cfnn(l,n) = dcos(fnn(l,n))                                            \\
\indent \hspace{19pt}         sfnn(l,n) = dsin(fnn(l,n))                                              \\
\indent \hspace{19pt}         rqf(l,n) = rplusq - fnn(l,n)                                              \\
\indent \hspace{19pt}         sum1(l,n) = cmplx(0.0,0.0)                                                  \\
\indent \hspace{19pt}         sum2(l,n) = cmplx(0.0,0.0)\\
\indent\indent\indent          DO 90 m=1,ms              \\
\indent \hspace{19pt} wmn(m,l) = sfnn(l,n)*cmud(m) - (cfnn(l,n)*smud(m) + \\ 
\indent \hspace{9.2pt} *\,smd(m))/fnu(l,n)\\
\indent \hspace{19pt}         vmn(m,l) = cfnn(l,n)*cmud(m) + sfnn(l,n)*\\
\indent \hspace{9.2pt} *\,smud(m)/fnu(l,n) - cmd(m)    \\
\indent \hspace{19pt}         sum1(l,n) = sum1(l,n) + bsz(0)*bsh(m)/(fnu(l,n)**2 - m*m)*              \\
\indent \hspace{9.2pt} *\,cmplx((wmn(m,l)*crqm(m) - vmn(m,l)*srqm(m)), -   \\
\indent \hspace{9.2pt} *\,(wmn(m,l)*srqm(m) + vmn(m,l)*crqm(m)))                                                     \\
$\left. \right.$\hspace{5pt} 90 \hspace{5pt}     CONTINUE                                                                \\
\indent\indent\indent         DO 100 m = - ms,ms                                                            \\
\indent\indent\indent\indent DO 110 k = 1,ks\\
\indent \hspace{19pt}         sum2(l,n) = sum2(l,n) + bsz(k)*bsh(m)/(fnu(l,n)**2 - \\
\indent \hspace{9.2pt} *\,(a*k + m)**2)*\\
\indent \hspace{9.2pt} *\,cmplx((sfnn(l,n)*ckmu(k,m) - (skm(k,m) + skmu(k,m)*cfnn(l,n))/\\
\indent \hspace{9.2pt} *\,fnu(l,n)), - (cfnn(l,n)*ckmu(k,m) - ckm(k,m)+skmu(k,m)*                   \\
\indent \hspace{9.2pt} *\,sfnn(l,n)/fnu(l,n)))*cmplx(ckmr(k,m),skmr(k,m))                     \\
$\left. \right.$\hspace{5pt} 110 \hspace{5pt} \indent   CONTINUE                                                               \\
$\left. \right.$\hspace{5pt} 100 \hspace{5pt}   CONTINUE                                                                 \\
\indent \hspace{19pt}        hfm(l,n) = fnu(l,n)/pfr*(bs0/(2.d0*fnu(l,n)**2)*                            \\
\indent \hspace{9.2pt} *\,cmplx((srq - dsin(rqf(l,n))),(crq - dcos(rqf(l,n)))) +                          \\
\indent \hspace{9.2pt} *\,sum1(l,n) + sum2(l,n))\\
$\left. \right.$\hspace{5pt} 80 \hspace{5pt} \indent  CONTINUE       \\
$\left. \right.$\hspace{5pt} 70 \hspace{5pt}  CONTINUE                 \\
                                 \\
C\hspace{27pt} {\bf Computation for complete response}\\
                                         \\
\indent\indent\indent    DO 120 n=0,j                         \\
\indent \hspace{19pt}        crp2f(n) = fp2fr*hfm(2,n)               \\
\indent \hspace{19pt}        crp1f(n) = fpfr*hfm(1,n)                  \\
\indent \hspace{19pt}        cr0f(n) = (cmplx(f5\_p,f5\_c))*hfm(0,n)       \\
\indent \hspace{19pt}        crm1f(n) = fm2fr*hfm(-1,n)                    \\
\indent \hspace{19pt}        crm2f(n) = fmfr*hfm(-2,n)                       \\
\indent \hspace{19pt}        cr(n) = crp2f(n) + crp1f(n) + cr0f(n) + crm1f(n) + crm2f(n)\\
\indent \hspace{19pt}  write(10,*) fn(n),cabs(cr(n)),cabs(crp2f(n))),cabs(crp1f(n))                                           \\
\indent \hspace{19pt}        write(20,*) fn(n),cabs(crm2f(n)),cabs(crm1f(n))),cabs(cr0f(n))                                               \\
$\left. \right.$\hspace{5pt} 120 \hspace{5pt}   CONTINUE\\
\indent\indent\indent      STOP       \\
\indent\indent\indent\indent      END

\chapter{Source code for the estimation of templates for all sky search:
{\it TEM.F\/}}
C\hspace{27pt} {\bf Estimation of templates for all sky search}\\
\indent \hspace{19pt}        IMPLICIT none\\
\indent \hspace{19pt}        INTEGER n,nt   \\
\indent \hspace{19pt}        PARAMETER (nt = 721)\\
\indent \hspace{19pt}        REAL$^*$8 fr,pi\\
\indent \hspace{19pt}        REAL x,y,f0,st,sig,dtheta,dphi,tdtheta,tdphi(1:nt),sigsig,ff(1:nt),\\
\indent \hspace{9.2pt} *\,sigtem(1:nt),temtem(1:nt),tem\_theta,tem\_phi(1:nt),skytem(1:nt)\\
C\hspace{30pt} [ff = Fitting Factor, skytem = Number of templates for all sky search]\\
C\hspace{30pt}        FREQUENCY AND SOURCE LOCATION \\
\indent \hspace{19pt}  PARAMETER (f0 = 1.0, dtheta = 1.0, tdtheta = 0.999955,\\
\indent \hspace{9.2pt} *\,dphi = 35.0)  \\
\indent \hspace{19pt}         EXTERNAL st,qromo,sig     \\ \\
\indent \hspace{19pt}         pi = 4.d0*datan(1.d0)              \\
\indent \hspace{19pt}         fr = 1.d0/86164.d0\\
\indent \hspace{19pt}         x = (f0 - 0.999998d0)/fr\\
\indent \hspace{19pt}         y = (f0 - 1.000002d0)/fr  \\
\indent \hspace{19pt}         CALL qromo(sig,x,y,dtheta,dphi,dtheta,dphi,sigsig) \\
\indent \hspace{19pt}         tem\_theta = 180.0/(dtheta - tdtheta)                 \\
\indent\indent\indent           DO 10 n = 0,nt                                         \\
\indent \hspace{19pt}         tdphi(n) = 0.5*float(n)\\
\indent \hspace{19pt}         tem\_phi(n) = abs(360.d0/(dphi + 0.1e-4 - tdphi(n)))        \\
\indent \hspace{19pt}         CALL qromo(sig,x,y,tdtheta,tdphi(n),tdtheta,tdphi(n),temtem(n))\\
\indent \hspace{19pt}         CALL qromo(st,x,y,dtheta,dphi,tdtheta,tdphi(n),sigtem(n))        \\
\indent \hspace{19pt}         ff(n) = abs(sigtem(n)/(sqrt(sigsig*temtem(n))))                    \\
\indent \hspace{19pt}         skytem(n) = tem\_theta*tem\_phi(n)                                  \\
\indent \hspace{19pt}         write(10) sngl(ff(n)),skytem(n)\\
 $\left. \right.$\hspace{5pt} 10 \hspace{5pt}     CONTINUE                                      \\
\indent\indent\indent        STOP                                            \\
\indent\indent\indent\indent        END                                               \\
                                                            \\
                                                              \\
\indent \hspace{19pt}       SUBROUTINE qromo(func,a,b,dtheta,dphi,tdtheta,tdphi,ss)   \\
\indent \hspace{19pt}       INTEGER JMAX,JMAXP,K,KM                                     \\
\indent \hspace{19pt}       REAL a,b,func,ss,EPS,dtheta,dphi,tdtheta,tdphi                \\
\indent \hspace{19pt}       EXTERNAL func                                                   \\
\indent \hspace{19pt}       PARAMETER (EPS = 1.e-6, JMAX = 14, JMAXP = JMAX+1,\\
\indent \hspace{9.2pt} *\,K = 5, KM = K-1)         \\
C\hspace{27pt}      USES ratint,midpnt                                                  \\
\indent \hspace{19pt}       INTEGER j                                                             \\
\indent \hspace{19pt}       REAL dss,h(JMAXP),s(JMAXP)                                              \\
\indent \hspace{19pt}       h(1) = 1.\\
 \indent\indent\indent      DO 10 j = 1,JMAX\\
\indent \hspace{19pt}         CALL midpnt(func,a,b,dtheta,dphi,tdtheta,tdphi,s(j),j)\\
\indent\indent\indent            IF (j.ge.K) THEN                                        \\
\indent \hspace{19pt}           CALL ratint(h(j - KM),s(j - KM),K,0.,ss,dss)                \\
\indent\indent\indent     IF (abs(dss).le.EPS*abs(ss)) RETURN                       \\
\indent\indent\indent         ENDIF      \\
\indent \hspace{19pt}         s(j+1) = s(j)                                                     \\
\indent \hspace{19pt}         h(j+1) = h(j)/9.                                                    \\
$\left. \right.$\hspace{5pt} 10 \hspace{5pt}    CONTINUE                                                              \\
\indent \hspace{19pt}       PAUSE 'too many steps in qromo'                                         \\
\indent\indent\indent      END\\
           \\
             \\
\indent \hspace{19pt}          SUBROUTINE midpnt(func,a,b,dtheta,dphi,tdtheta,tdphi,s,n)\\
\indent \hspace{19pt}       INTEGER n                                                     \\
\indent \hspace{19pt}       REAL a,b,s,func,dtheta,dphi,tdtheta,tdphi                       \\
\indent \hspace{19pt}       EXTERNAL func                                                     \\
\indent \hspace{19pt}       INTEGER it,j                                                        \\
\indent \hspace{19pt}       REAL ddel,del,sum,tnm,x                                               \\
\indent\indent\indent       IF (n.eq.1) THEN                                                        \\
\indent \hspace{19pt}         s = (b - a)*func(0.5*(a + b),dtheta,dphi,tdtheta,tdphi)\\
\indent\indent\indent      ELSE                                                 \\
\indent \hspace{19pt}         it = 3**(n - 2)                                          \\
\indent \hspace{19pt}         tnm = it                                                 \\
\indent \hspace{19pt}         del = (b - a)/(3.*tnm)                                       \\
\indent \hspace{19pt}         ddel = del + del                                               \\
\indent \hspace{19pt}         x = a + 0.5*del                                                  \\
\indent \hspace{19pt}         sum = 0.                                                         \\
\indent\indent\indent         DO 10 j=1,it                                                     \\
\indent \hspace{19pt}           sum = sum + func(x,dtheta,dphi,tdtheta,tdphi)                        \\
\indent \hspace{19pt}           x = x + ddel                                                           \\
\indent \hspace{19pt}           sum = sum + func(x,dtheta,dphi,tdtheta,tdphi)                            \\
\indent \hspace{19pt}           x = x + del\\
$\left. \right.$\hspace{5pt} 10 \hspace{5pt}       CONTINUE   \\
\indent \hspace{19pt}         s = (s + (b - a)*sum/tnm)/3.\\
\indent\indent\indent      ENDIF                     \\
\indent\indent\indent      RETURN                      \\
\indent\indent\indent\indent      END                           \\
                                      \\
                                        \\
\indent \hspace{19pt}       SUBROUTINE ratint(xa,ya,n,x,y,dy)   \\
\indent \hspace{19pt}       INTEGER n,NMAX                        \\
\indent \hspace{19pt}       REAL dy,x,y,xa(n),ya(n),TINY            \\
\indent \hspace{19pt}       PARAMETER (NMAX = 10,TINY = 1.e-25)           \\
\indent \hspace{19pt}       INTEGER i,m,ns                              \\
\indent \hspace{19pt}       REAL dd,h,hh,t,w,c(NMAX),d(NMAX)              \\
\indent \hspace{19pt}       ns = 1                                            \\
\indent \hspace{19pt}       hh = abs(x - xa(1))                                   \\
\indent\indent\indent       DO 10 i = 1,n                                         \\
\indent \hspace{19pt}         h = abs(x - xa(i))                                      \\
\indent\indent\indent         IF (h.eq.0.) THEN                                     \\
\indent \hspace{19pt}           y = ya(i)                                               \\
\indent \hspace{19pt}           dy = 0.0                                                  \\
\indent\indent\indent          RETURN                                                    \\
\indent\indent\indent        ELSE IF (h.lt.hh) THEN                                        \\
\indent \hspace{19pt}           ns = i                                                          \\
\indent \hspace{19pt}           hh = h                                                            \\
\indent\indent\indent        ENDIF                                                               \\
\indent \hspace{19pt}         c(i) = ya(i)                                                            \\
\indent \hspace{19pt}         d(i) = ya(i) + TINY                                                         \\
$\left. \right.$\hspace{5pt} 10 \hspace{5pt}   CONTINUE\\
\indent \hspace{19pt}       y = ya(ns)  \\
\indent \hspace{19pt}       ns = ns - 1     \\
\indent\indent\indent       DO 20 m = 1,n - 1 \\
\indent\indent\indent\indent         DO 30 i = 1,n - m \\
\indent \hspace{19pt}           w = c(i + 1) - d(i) \\
\indent \hspace{19pt}           h = xa(i + m) - x     \\
\indent \hspace{19pt}           t = (xa(i) - x)*d(i)/h\\
\indent \hspace{19pt}           dd = t - c(i + 1)         \\
\indent\indent\indent          IF(dd.eq.0.)PAUSE 'failure in ratint'\\
\indent \hspace{19pt}           dd = w/dd                                \\
\indent \hspace{19pt}           d(i) = c(i + 1)*dd                           \\
\indent \hspace{19pt}           c(i) = t*dd                                  \\
$\left. \right.$\hspace{5pt} 30 \hspace{5pt}\indent       CONTINUE                                       \\
\indent\indent\indent        IF (2*ns.lt.n - m) THEN                             \\
\indent \hspace{19pt}           dy = c(ns + 1)                                       \\
\indent\indent\indent        ELSE                                                 \\
\indent \hspace{19pt}           dy = d(ns)                                             \\
\indent \hspace{19pt}           ns = ns - 1                                                \\
\indent\indent\indent        ENDIF                                                      \\
\indent \hspace{19pt}         y = y + dy                                                       \\
$\left. \right.$\hspace{5pt} 20 \hspace{5pt}    CONTINUE                                                         \\
\indent\indent\indent      RETURN                                                             \\
\indent\indent\indent\indent      END                                                                  \\
                                                                             \\
\indent \hspace{19pt}       FUNCTION sig(fnu,dtheta,dphi)                                            \\
\indent \hspace{19pt}         REAL sig,fnu,dtheta,dphi                                                 \\
\indent \hspace{19pt}         COMPLEX hfm\\
\indent \hspace{19pt}         EXTERNAL hfm \\
\indent \hspace{19pt}         sig = real(hfm(fnu,dtheta,dphi)*conjg(hfm(fnu,dtheta,dphi)))\\
\indent\indent\indent      RETURN                                                          \\
\indent\indent\indent\indent      END                                                               \\
                                                                            \\
\indent \hspace{19pt}       FUNCTION st(fnu,dtheta,dphi,tdtheta,tdphi)                          \\
\indent \hspace{19pt}         REAL st,fnu,dtheta,dphi,tdtheta,tdphi                                \\
\indent \hspace{19pt}         COMPLEX hfm                                                           \\
\indent \hspace{19pt}          EXTERNAL hfm\\
\indent \hspace{19pt}         st = real(hfm(fnu,dtheta,dphi)*conjg(hfm(fnu,tdtheta,tdphi)))\\
\indent\indent\indent      RETURN                                                         \\
\indent\indent\indent\indent      END

\end{document}